\documentclass[longauth]{aa}
\usepackage[varg]{txfonts}
\usepackage{adjustbox}
\usepackage{booktabs}
\usepackage{CJKutf8}
\usepackage{color}

\bibpunct{(}{)}{;}{a}{}{,}

\begin{document}

\title{Wild behaviour near the finish line: the Type Ibn SN~2020able and its stratified environment}

\author{L.~Tartaglia\inst{\ref{inst:oaab}} \and
E.~Cappellaro\inst{\ref{inst:oapd}} \and
G.~Valerin\inst{\ref{inst:oapd}} \and
A.~Pastorello\inst{\ref{inst:oapd}} \and
S.~Benetti\inst{\ref{inst:oapd}} \and
A.~Reguitti\inst{\ref{inst:oapd},\ref{inst:oabr}} \and
L.~Tomasella\inst{\ref{inst:oapd}} \and
P.~Lundqvist\inst{\ref{inst:okc}} \and
K.~Misra\inst{\ref{inst:arya}} \and
A.~Gangopadhyay\inst{\ref{inst:okc}} \and
R.~Dastidar\inst{\ref{inst:oabr}} \and
B.~Ailawadhi\inst{\ref{inst:nvrangpura}} \and
Dimple\inst{\ref{inst:ubirm},\ref{inst:igwa}} \and
B.~Kumar\inst{\ref{inst:yunnanun},\ref{inst:yunnanunlab}} \and
V.~K.~Jha\inst{\ref{inst:ncra}} \and
M.~Singh\inst{\ref{inst:kora}} \and
N.~Elias-Rosa\inst{\ref{inst:oapd},\ref{inst:iss}} \and
C.~P.~Guti\'errez\inst{\ref{inst:iss}} \and
D.~A.~Howell\inst{\ref{inst:lcogt},\ref{inst:ucsb}} \and
D.~Hiramatsu\inst{\ref{inst:uflo}} \and
C.~Pellegrino\inst{\ref{inst:goddard}} \and
J.~Andrews\inst{\ref{inst:gemini}} \and
D.~J.~Sand\inst{\ref{inst:uaz}} \and
N.~Smith\inst{\ref{inst:uaz}} \and
J.~Pearson\inst{\ref{inst:uaz}} \and
S.~Wyatt\inst{\ref{inst:goddard}} \and
P.~Ochner\inst{\ref{inst:unipd},\ref{inst:oapd}} \and
Y.-Z.~Cai (\begin{CJK*}{UTF8}{gbsn}蔡永志\end{CJK*})\inst{\ref{inst:oapd},\ref{inst:yunnanobs},\ref{inst:yunnanlab}} \and
J.~Burke\inst{} \and
C.~McCully\inst{\ref{inst:lcogt}} \and
M.~Newsome\inst{\ref{inst:austin}} \and
E.~Padilla Gonzalez\inst{\ref{inst:balt}}
}

\institute{
INAF - Osservatorio Astronomico d'Abruzzo, Via Mentore Maggini s.n.c., I-64100 Teramo, Italy \\ \email{leonardo.tartaglia@inaf.it}\label{inst:oaab} \and
INAF - Osservatorio Astronomico di Padova, Vicolo dell'Osservatorio 5, I-35122 Padova, Italy\label{inst:oapd} \and
INAF - Osservatorio Astronomico di Brera, Via E. Bianchi 46, I-23807 Merate (LC), Italy\label{inst:oabr} \and
Oskar Klein Centre, Department of Astronomy, Stockholm University, AlbaNova, SE-106 91 Stockholm, Sweden\label{inst:okc} \and
Aryabhatta Research Institute of Observational Sciences, Nainital-263001, India\label{inst:arya} \and
Physical Research Laboratory, Navrangpura, Ahmedabad, Gujarat 380009, India\label{inst:nvrangpura} \and
School of Physics and Astronomy, University of Birmingham, Edgbaston, Birmingham B15 2TT, UK\label{inst:ubirm} \and
Institute for Gravitational Wave Astronomy, University of Birmingham, Birmingham B15 2TT, UK\label{inst:igwa} \and
South-Western Institute for Astronomy Research, Yunnan University, Kunming, Yunnan 650504, People’s Republic of China\label{inst:yunnanun} \and
Yunnan Key Laboratory of Survey Science, Yunnan University, Kunming, Yunnan 650500, People’s Republic of China\label{inst:yunnanunlab} \and
National Centre for Radio Astrophysics, Tata Institute of Fundamental Research, Post Bag 3, Ganeshkhind, Pune, 411007, India\label{inst:ncra} \and
Indian Institute of Astrophysics, Koramangala 2nd Block, Bangalore 560034, India\label{inst:kora} \and
Institute of Space Sciences (ICE, CSIC), Campus UAB, Carrer de Can Magrans, s/n, E-08193 Barcelona, Spain\label{inst:iss} \and
Las Cumbres Observatory, 6740 Cortona Drive, Suite 102, Goleta, CA 93117-5575, USA \label{inst:lcogt} \and
Department of Physics, University of California, Santa Barbara, CA 93106-9530, USA \label{inst:ucsb} \and
Department of Astronomy, University of Florida, Bryant Space Science Center, Gainesville, FL 32611-2055, USA \label{inst:uflo} \and
NASA Goddard Space Flight Center, 8800 Greenbelt Road, Greenbelt, MD 20771, USA\label{inst:goddard} \and
Università degli Studi di Padova, Dipartimento di Fisica e Astronomia, Vicolo dell'Osservatorio 3, 35122, Padova, Italy\label{inst:unipd} \and
Gemini Observatory/NSF’s National Optical-Infrared Astronomy Research Laboratory, 670 N. Aohoku Place, Hilo, HI 96720, USA \label{inst:gemini} \and
Steward Observatory, University of Arizona, 933 North Cherry Avenue, Tucson, AZ 85721-0065, USA \label{inst:uaz} \and
Yunnan Observatories, Chinese Academy of Sciences, Kunming 650216, P.R. China\label{inst:yunnanobs} \and
International Centre of Supernovae, Yunnan Key Laboratory, Kunming 650216, P.R. China\label{inst:yunnanlab} \and
Department of Astronomy, The University of Texas at Austin, 2515 Speedway, Stop C1400, Austin, TX 78712, USA \label{inst:austin} \and
Department of Physics and Astronomy, Johns Hopkins University, Baltimore, MD 21218, USA \label{inst:balt}
}

\date{Accepted 7 August, 2026}

\abstract{We present spectrophotometric observations of the Type Ibn supernova SN~2020able, along with a comprehensive analysis and modelling of its evolution. 
The transient occurred in SDSS J092602.93+243115.1, a faint host galaxy characterised by significantly sub-solar metallicity ($12+\log(\rm{O/H})\simeq8.28\,\rm{dex}$). 
The early photometric evolution and high peak luminosity ($L_{peak}\simeq4.4\times10^{43}\,\rm{erg}\,\rm{s^{-1}}$) indicate an efficient conversion of ejecta kinetic energy into radiation via interaction with the circumstellar medium. 
Light-curve modelling suggests that this interaction involves ejecta with a mass of $M_{\rm{ej}}\simeq1.3\,M_{\sun}$ impacting a shell of $M_{\rm{CSM}}\simeq0.6\,M_{\sun}$.
However, interaction is found to be geometrically and temporally confined within a few days after explosion, with the later evolution of the transient being dominated by the free expansion of the ejecta.
Spectroscopic analysis reveals a complex, stratified circumstellar environment. 
In addition to the innermost shocked region, we identify two additional He-rich shells: a dense inner one ($n_{\rm{e}}\simeq1.4\times10^{9}\,\rm{cm^{-3}}$) located at $R\gtrsim9.3\times10^{14}\,\rm{cm}$, and a more extended outer component reaching $R_{out}\simeq3.3\times10^{15}\,\rm{cm}$ and expanding at approximately $1.2\times10^3\,\rm{km}\,\rm{s^{-1}}$. 
The physical parameters of the former suggest it was produced by a discrete pre-supernova eruptive event similar to that of SN~2006jc, which expelled approximately $1.8\times10^{-2}\,M_{\sun}$ of material. 
Conversely, the expansion velocity of the outermost shell is consistent with the escape velocity of WR stars, pointing towards a late WN to WC-type progenitor as the most viable candidate for SN~2020able.
}

\titlerunning{The stratified environment of SN~2020able}
\authorrunning{L.~Tartaglia et al.}

\maketitle
\nolinenumbers

\section{Introduction} \label{sec:intro}
Massive stars are expected to end their lives with spectacular explosions, giving rise to core-collapse supernovae (CC SNe) following the collapse of their iron cores. The observational variety of CC SNe is determined by the composition of the residual envelope at the time of explosion, which is a direct consequence of the progenitor mass-loss history. 
Consequently, the mechanisms driving the mass-loss events and their efficiency during the late evolutionary stages are key parameters in determining the classification of the resulting transient \citep[see][]{2003ApJ...591..288H}.
Type Ib/c SNe are also labelled ``stripped-envelope" \citep{1996ApJ...459..547C} since their spectra lack H lines, with the Ib and Ic classes being characterised by the presence or absence of strong helium features, respectively \citep[see e.g.][]{1997ARA&A..35..309F,2019MNRAS.482.1545S}.
Specific mass-loss mechanisms are required to efficiently remove the outer layers of progenitor stars thereby leading to stripped envelope supernovae.
Candidate progenitors include massive Wolf-Rayet (WR) stars and, more predominantly according to observed fractions \citep{2011MNRAS.412.1522S}, lower-mass helium stars stripped in binaries via Roche-lobe overflow or common-envelope evolution.
As for the more common Type IIn SNe, emission lines with relatively narrow profiles (full-width-at-half-maximum - FWHM - velocities of approximately $10^3\,\rm{km}\,\rm{s^{-1}}$) are key features in classifying SNe Ibn \citep{2008MNRAS.389..113P,2017ApJ...836..158H,2019ApJ...871L...9H}, where the lower-case suffix stands for ``narrow".
Although narrow lines are historically associated with the ongoing dynamical interaction between SN ejecta and the circumstellar medium (CSM), \citet{2025A&A...703A.177T} recently showed that they are not necessarily linked to this mechanism and that their presence may simply act as a probe of a dense pre-existing gas.
They also highlighted how in at least a fraction of narrow-lined SNe, these may not be related to energy sources in addition to canonical recombination and radioactive decays.
A handful of SNe Ibn with well-sampled spectroscopic evolution showed evidence of high-ionisation features \citep[typically \ion{He}{II}, \ion{C}{III/IV} and \ion{N}{III}; see][]{2015MNRAS.449.1921P,2020ApJ...889..170G,2022ApJ...930..127G,2023ApJ...959L..10P,2024A&A...691A.156W} in their early spectra, but their number is likely affected by the limited number of sufficiently early spectroscopic observations.
These are frequently interpreted as signatures of ``flash-ionisation" produced by the shock breakout radiation field \citep{2014Natur.509..471G,2014A&A...572L..11G,2023ApJ...952..119B}. 
The Bowen fluorescence mechanism is likely responsible at least for the prominent \ion{N}{III} $\lambda4100$ and $\lambda\lambda4634$, 4641 features often observed in these transients \citep[see][]{2007A&A...464..715S}, providing a more physically rigorous framework for the excitation of these specific lines.
\begin{figure}
\resizebox{0.95\hsize}{!}{\includegraphics{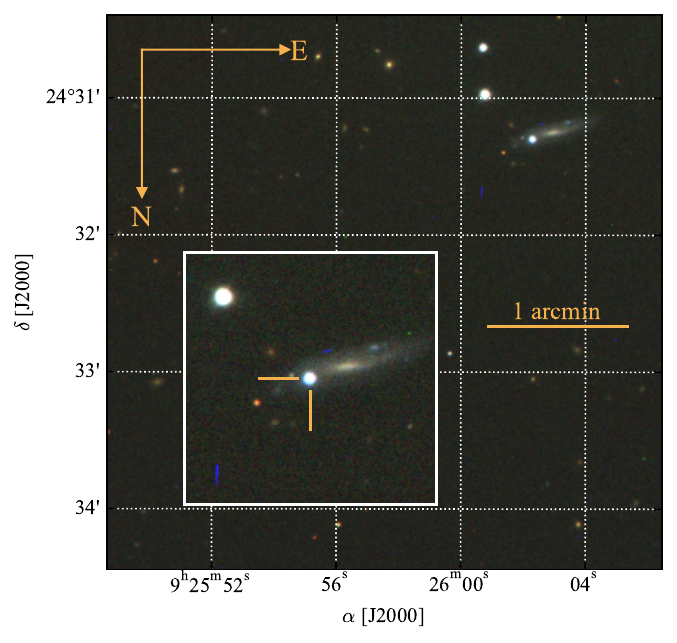}}
\caption{Colour image of SN~2020able and its host galaxy, obtained combining $g-$, $r-$ and $i-$band data obtained on 2021 January 21 with the $2.0\,\rm{m}$ Liverpool Telescope. The transient is the bright source in the middle of the inset.}
\label{fig:fchart}
\end{figure}

The physical origin and progenitor nature of SNe Ibn have been further explored by recent radiation-hydrodynamics and radiative-transfer simulations. 
\citet{2022A&A...658A.130D} showed that persistent narrow lines and modest peak luminosities in some SNe Ibn are inconsistent with the standard explosion of a massive WR star into a dense wind, suggesting that these events may also arise from low-energy explosions of lower-mass helium stars ($\le5\rm{M_{\odot}}$) in interacting binary systems. 
In this scenario, the ejecta collide with a massive, He-rich CSM shell previously ejected through binary interaction or a non-terminal nuclear flash. Notably, their simulations show that the late-time spectra of prototypical SNe Ibn like SN 2006jc and SN 2011hw are dominated by a forest of \ion{Fe}{II} emission lines below 5500~\AA, producing the hot pseudo-continuum frequently observed in interacting supernovae.

In this context, early high-cadence monitoring is crucial to constrain the CSM geometry and test progenitor models. Thanks to a dense observational cadence starting shortly after the explosion, the Type Ibn SN~2020able represents one of the most comprehensively monitored Type Ibn SNe to date, providing the opportunity to clearly trace the transition from an early, geometrically confined ejecta-CSM interaction to the later ejecta-dominated expansion within a highly stratified environment.
The discovery of SN~2020able was first reported \citep{2020TNSTR3630....1T} by the Asteroid Terrestrial-impact Last Alert System \citep[ATLAS;]{2020PASP..132h5002S} on 2020 December 2.54~UT ($\rm{MJD=59185.54}$) with an \textit{orange} magnitude of approximately $18.9\,\rm{mag}$, who assigned the internal name ATLAS20bgkn to the transient. 
The SN was also observed by the Panoramic Survey Telescope \& Rapid Response System \citep[Pan-STARRS][internal name PS20mdd]{2016arXiv161205560C} and the Zwicky Transient Facility \citep[ZTF;][]{2019PASP..131g8001G,2019PASP..131a8002B,2019PASP..131a8003M} as ZTF20aayxldg, and classified as a Type Ibn SN by \citet{2020TNSCR3728....1H}, based on its spectroscopic similarities with SN~2019uo \citep{2020ApJ...889..170G}.
In Sect.~\ref{sec:analysis}, we analyse the local environment within the host galaxy SDSS~J092602.93+243115.1 (Sect.~\ref{sec:host}). 
We then discuss the UV+optical light curves, modelling their evolution with the Modular Open-Source Fitter for Transients ({\sc MOSFiT}) package (Sect.~\ref{sec:photometry}) and examine the evolution of main features observed in the optical spectra (Sect.~\ref{sec:spectroscopy}).
Section~\ref{sec:conclusions} summarises the main results of our analysis.
Data reduction techniques are described in the Appendix~\ref{sec:obsredu}, with a log of spectroscopic observations reported in Table~\ref{tab:speclog}.

\section{Analysis and discussion} \label{sec:analysis}
\subsection{The local environment} \label{sec:host}
SN~2020able exploded at $\rm{R.A.}=$141\fdg50943, $\rm{Decl.}=+$24\fdg52169 [J2000], $-9\farcs31\,\rm{E}$ $+3\farcs14\rm{N}$ offset from the centre of its host, SDSS~J092602.93+243115.1\footnote{as reported at \url{https://ned.ipac.caltech.edu}} (also known as PGC~1709306 and WISEA~J092602.94+243114.9; see Figure~\ref{fig:fchart}).
Following \citet{2009A&A...508.1259H}, we computed the galactocentric de-projected distance of SN~2020able adopting the host geometrical parameters reported in the literature.
For an inclination $i=90^{\circ}$, position angle $\rm{P.A.}=104\fdg9$ and the coordinates of the SN and the host centre, we infer a de-projected galactocentric distance 10\farcs70. 
\citet{1996ApJ...473..576F} report a recessional velocity with respect to the 3K cosmic microwave background of $7695\pm20\,\rm{km}\,\rm{s^{-1}}$ and we therefore used this value to estimate a luminosity distance $D_L=107.6\pm7.4\,\rm{Mpc}$ ($\mu=35.16\pm0.15\,\rm{mag}$) assuming a standard Cosmology with $\Omega_\Lambda=0.73$, $\Omega_M=0.27$ and $H_0=73\,\rm{km}\,\rm{s^{-1}}\,\rm{Mpc^{-1}}$.
This choice was made on the basis of cosmologies adopted in the literature and to facilitate comparisons with other SNe Ibn; \citep[see e.g.][]{2016MNRAS.456..853P,2019ApJ...871L...9H}.

An optical spectrum of the host nucleus was obtained as part of the ``legacy" programme of the Sloan Digital Sky Survey \citep[SDSS; $\rm{Plate=2291}$, $\rm{FiberID=39}$, $\rm{MJD}=53714$;][]{2000AJ....120.1579Y}. 
We retrieved the flux-calibrated data from SDSS Data Release 12 \citep[DR12;][]{2015ApJS..219...12A} and estimated the host metallicity using standard diagnostic line ratios.
Specifically, we used the re-calibration of the O3N2 \citep{1979A&A....78..200A} and N2 \citep{2002MNRAS.330...69D} abundance indicators provided by \citet{2013A&A...559A.114M}.
The O3N2 indicator, in particular, is based on the measured line ratios $\ion{[O}{III]} \lambda5007^{}/\rm{H}\beta$ and $\rm{H}\alpha^{}/\ion{[N}{II]} \lambda6583$ and is therefore less sensitive to extinction given the small wavelength differences between the emission lines involved.
Both indicators already give oxygen abundances for the host centre significantly below the solar value \citep[$8.35\pm0.02\,\rm{dex}$ and $8.40\pm0.04\,\rm{dex}$ for O3N2 and N2, respectively, against a typical solar value $\simeq8.69\,\rm{dex}$; see][]{2009ARA&A..47..481A}.
The oxygen abundance at the galactocentric distance of SN~2020able is expected to be even lower, following the metallicity gradient of $-0.078\pm 0.040\,\rm{dex}\,r_e^{-1}$ found by \citet{2016A&A...591A..48G} for SNe~Ib/c hosts.
The disk effective radius, $r_e=9\farcs4\pm2\farcs4$, was derived directly from the definitions of the mean $B$-band surface brightness and total $B$-band magnitude \citep{1997A&AS..124..109P,2000A&AS..146...19P}, using the values $\mu_{B,e}=23.99\pm0.22\,\rm{mag\,arcsec^{-2}}$ and $b_t=17.14\pm0.50\,\rm{mag}$ retrieved from HyperLEDA\footnote{\url{http://atlas.obs-hp.fr/}}.
This yields an abundance of $12+\log(\rm{O/H})=8.28\pm0.03\,\rm{dex}$, calculated as the weighted average of the values derived using the O3N2 ($8.35\pm0.02$) and N2 ($8.40\pm0.04\,\rm{dex}$) diagnostics.
\begin{figure}
\resizebox{0.95\hsize}{!}{\includegraphics{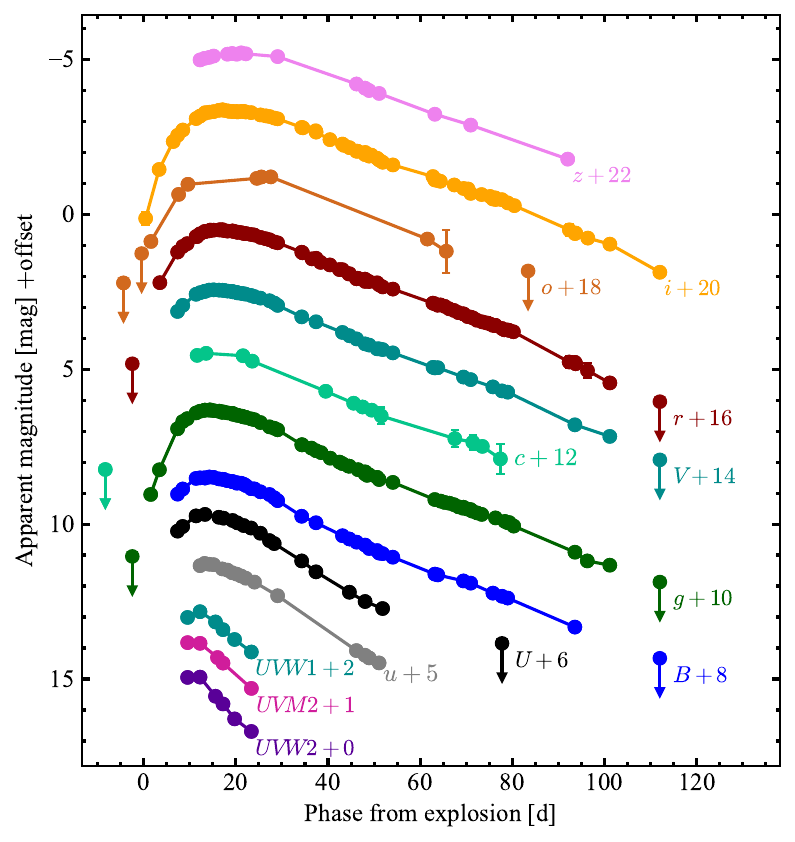}}
\caption{UV/optical light curves of SN~2020able. Rest-frame phases refer to the explosion epoch estimated through the last non-detection limit. Magnitudes were not corrected for reddening along the line of sight.}
\label{fig:lightcurves}
\end{figure}

The foreground Galactic reddening towards SN~2020able is $E(B-V)=0.028\,\rm{mag}$ \citep{2011ApJ...737..103S}, while the contribution of the local environment was derived using empirical relations between the colour excess and equivalent widths (EWs) of \ion{Na}{ID} lines, as detailed in \citet{2012MNRAS.426.1465P}.
\ion{Na}{ID} features were resolved in our spectra obtained with the $6.5\,\rm{m}$ MMT on 2020 December 13, 2021 January 5 and 2021 February 4 ($\simeq13$, $35$ and $64\,\rm{days}$ after the estimated explosion epoch; see Sect.~\ref{sec:photometry}).
We took the EWs measured from the first spectrum - the one with the highest signal-to-noise ratio (SNR) - resulting in 0.26 and 0.18\AA~for the D2 and D1 lines, respectively.
Using a weighted average of Eq.~7 and 8 of \citet{2012MNRAS.426.1465P}, these correspond to an additional contribution of $E(B-V)=0.046\pm0.012\,\rm{mag}$ to the total reddening.
The colour excess derived from the \ion{Na}{ID} Galactic features visible in the same spectrum is in agreement with that reported by \citet{2011ApJ...737..103S} for the Milky Way reddening contribution and also suggests $E(B-V)_{host}\gtrsim0.03\,\rm{mag}$ comparing the strength of the two sets of features.
In the following, we will therefore use $E(B-V)=0.074\pm0.012\,\rm{mag}$ for the total reddening along the line of sight of SN~2020able.
\begin{figure}
\resizebox{0.95\hsize}{!}{\includegraphics{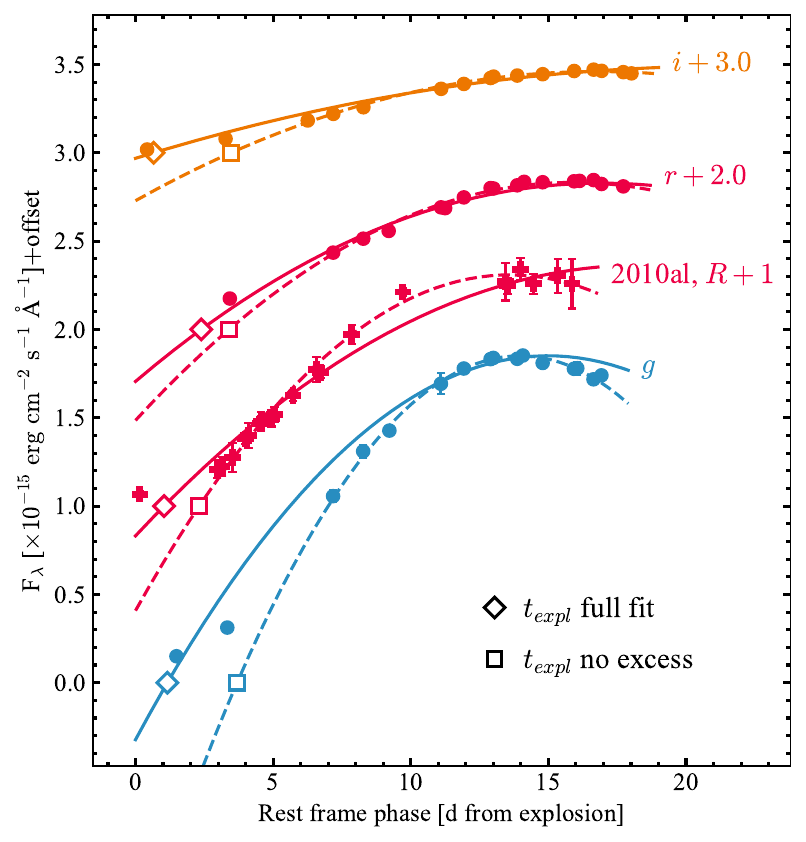}}
\caption{Early evolution of the $gri$ light curves of SN~2020able and the $R-$band light curve of SN~2010al, along with a $t^2$ fit to the derived fluxes. To highlight the early excess, we also show the best fit obtained removing the first one/two points. 
When extrapolating the flux to zero, both yield explosion dates that occur after the SN discovery (reported using open squares and diamonds). 
\label{fig:griRise}}
\end{figure}

\subsection{Photometry} \label{sec:photometry}
The ultraviolet (UV) to optical light curves of SN~2020able are shown in Figure~\ref{fig:lightcurves}. 
Details on the instrumentation used and data reduction are provided in Appendix~\ref{sec:obsredu}.

From the analysis of public ZTF data, we identified $gri$ pre-discovery detections as early as 2020 December 1.39~UT ($\rm{JD}=2459184.89$), approximately $1.15\,\rm{days}$ before the reported discovery.
The last non-detection, on the other hand, was obtained just $0.86\,\rm{days}$ earlier by the ATLAS survey ($\rm{JD}=2459184.03$, $o>19.3\,\rm{mag}$) and provides a more stringent constraint on the explosion epoch of SN~2020able. 
In the following, we will thus adopt $\rm{JD}=2459184.46\pm0.43$ (the midpoint between the ATLAS non-detection and the first ZTF detection) as the explosion epoch and refer all subsequent temporal phases to this date.
Forced photometry on ZTF and ATLAS archival frames did not reveal additional detections, ruling out pre-explosion outbursts down to an average $g-$ and $r-$band absolute magnitude $-13.7\,\rm{mag}$ in approximately the $2\,\rm{years}$ before the SN explosion. 

The early photometric evolution is relatively slow compared to that of other SNe Ibn.
A polynomial fit to the UV bands resulted in rise times of $10.6$, $10.4$ and $11.7\,\rm{days}$ (in the rest-frame) for the $UVW2-$, $UVM2-$ and $UVW1-$bands, respectively, although the lack of coverage at $t\lesssim+9\,\rm{days}$ suggests a more rapid rise to maximum cannot be definitively ruled out.
More robust estimates were derived for the optical bands, where data were available soon after discovery, confirming the relatively slow early evolution, with rise times ranging from approximately $13.3$ to $17.0\,\rm{days}$ in the $u/U-$ and $z-$bands, respectively.
We obtained similar rise times by fitting a $t^2$ power-law \citep[see e.g.][and references therein]{1982ApJ...253..785A,1999AJ....118.2675R} to the rising parts of the $gri$ light curves (the best sampled around the peaks, see Fig.~\ref{fig:griRise}), with $t_{rise}\simeq14.0$, 15.9 and $16.8\,\rm{days}$, in the $g$, $r$ and $i$ bands, respectively.
Interestingly, we note a deviation from the $t^2$ evolution at $t\lesssim+5\,\rm{days}$, with an excess in the derived flux relative to that expected by the ``fireball" model.

To investigate potential similarities among SN~2020able and other Type Ibn/Icn SNe, we selected a comparison sample (see Appendix~\ref{sec:fitRise}) based on the quality of their early light curve coverage, prioritising events with the best-sampled $r/R-$band light curves during the rising phase. 
In the few cases where these bands were unavailable or lacked sufficient early-time data, other optical bands were considered.
\begin{figure}
\resizebox{0.95\hsize}{!}{\includegraphics{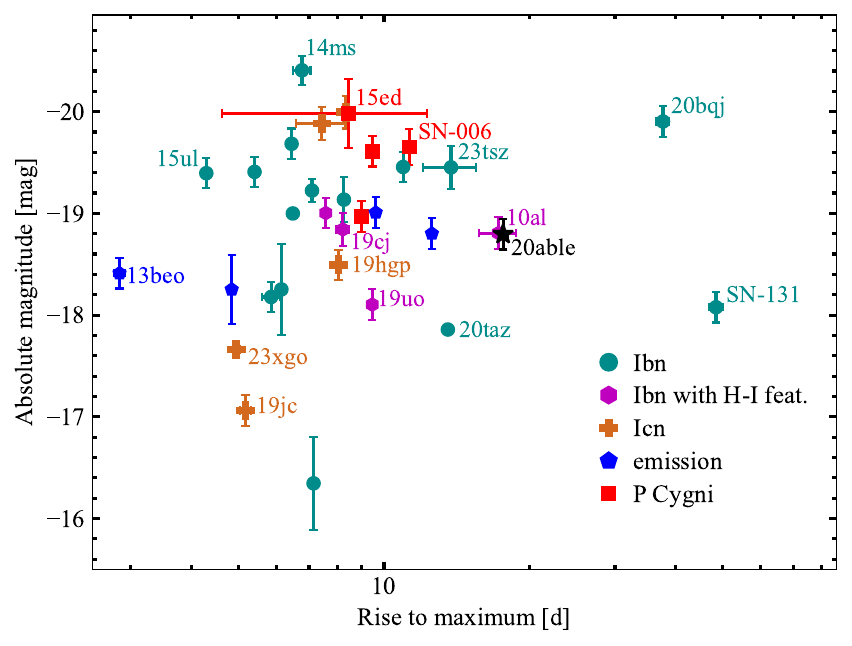}}
\caption{Absolute peak magnitudes vs. rise times for a sample of Type Ibn/Icn SNe with well-sampled early light curves. Ibn subtypes are those reported by \citet{2017ApJ...836..158H}. Data points and the corresponding literature references for the light curves used to generate this figure are provided in Table~\ref{tab:riseMax}.\label{fig:riseMax}}
\end{figure}
In Fig.~\ref{fig:riseMax}, we plot their peak magnitudes against their respective rise times, computed by fitting a second-order polynomial to the light curve. 
For each transient, we estimated the explosion and maximum light epochs with the corresponding absolute peak magnitude, by performing $10^4$ Monte Carlo simulations randomly shifting each data point within its uncertainty.
$t_{expl}$, $t_{max}$ and $m_{abs}$ and their respective errors (see Appendix~\ref{sec:fitRise} and Table~\ref{tab:riseMax}), are the mean values and standard deviations of the resulting distributions.
Absolute magnitudes were computed using the most recent redshift-independent distance to each transient or the host redshift, assuming a standard cosmology with $\Omega_{\Lambda}=0.73$, $\Omega_{M}=0.27$ and $H_0=73\,\rm{km}\,\rm{s^{-1}}\,\rm{Mpc^{-1}}$.
For SNe~2010al and 2020able, we considered the explosion epochs constrained by the pre-discovery non-detection limits \citep[see][and above]{2015MNRAS.449.1921P}, based on their peculiar early photometric evolution (see Figure~\ref{fig:griRise}).
The resulting rise times for SNe~2010al and 2020able are, on average, approximately 54\% longer than the mean value of the full sample, 92\% without considering the exceptionally long ones, SNe~2020bqj and OGLE-2014-SN-131 \citep[][]{2021A&A...652A.136K,2017A&A...602A..93K}.
In particular, Ibn SNe showing high-ionisation features in their early spectra have a systematically faster evolution with respect to SN~2020able, with the exception of SN~2010al.
Following the same procedure, we measured the post-peak decline rates of the entire sample up to $20\,\rm{days}$ from the maximum light. 
This yielded a remarkably slow decline rate of approximately $0.05\,\rm{mag}\,\rm{days^{-1}}$ for SN~2020able, close to the slow-evolving behaviour inferred for iPTF13beo, ASASSN-14ms, 2020taz, OGLE-2014-SN-131, OGLE-2012-SN-006, and 2020bqj (see Figure~\ref{fig:declines}). Specifically, SN~2020able exhibits a post-peak decay slower than 84.4\% of the full sample (81.5\% limiting the comparison to SN Ibn).

As already suggested by the comparison in Fig.~\ref{fig:riseMax}, SN~2010al is the only transient showing a similar early evolution to that of SN~2020able, with a comparable rise to the maximum and early spectroscopic evolution.
SN~2010al also showed an early excess in its early $R-$band light curve similar to the one observed in SN~2020able (see Fig.~\ref{fig:griRise}), suggesting a common mechanism responsible for the extra luminosity output.
To further explore these similarities and their physical meaning, we modelled the UV to optical light curves of SNe~2010al and 2020able with the {\sc MOSFiT}\footnote{\url{https://github.com/guillochon/MOSFiT}} \citep{2018ApJS..236....6G} tool.
\begin{figure}
\resizebox{0.95\hsize}{!}{\includegraphics{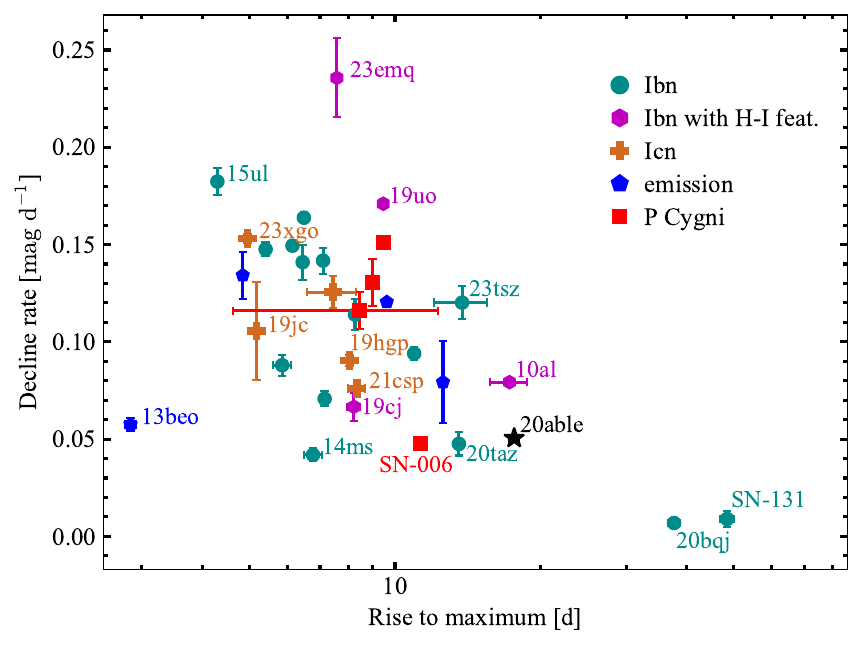}}
\caption{Decline rates vs. rise times for the sample of Fig.~\ref{fig:riseMax}. Data points to generate this figure are provided in Table~\ref{tab:riseMax}. \label{fig:declines}}
\end{figure}
Given the rapid photometric evolution of both transients, we initially adopted a non-interacting model, which accounts for the evolution solely through the $^{56}\rm{Ni}$ to $^{56}\rm{Co}$ radioactive decay (the {\sc default} model in {\sc MOSFiT}; \citealt{1994ApJS...92..527N}).
This model fails to reproduce the light curves of both SNe~2010al and 2020able without invoking unphysical nickel masses (specifically, $\rm{M_{^{56}Ni}}\gtrsim\rm{M_{\rm{ej}}}$; see Table~\ref{tab:mosfitPardefault}).
The most significant discrepancy between the models and the observed light curves occurs around the peak, particularly within the UV bands (see Appendix~\ref{sec:mosfit}, Table~\ref{tab:mosfitPardefault} and Figure~\ref{fig:lcmodeldefault}).
We therefore used the {\sc csmni} analytical model, which combines the contribution of interaction between the SN ejecta and a dense CSM \citep[e.g.][]{2013ApJ...773...76C,2017ApJ...849...70V,2020RNAAS...4...16J} and the luminosity produced by radioactive decays \citep[see e.g.][]{1980ApJ...237..541A,1994ApJS...92..527N}.
The parameter space was explored using the Dynamic Nested Sampling algorithm {\sc dynesty} \citep{2020MNRAS.493.3132S}, to ensure robust convergence of the posterior distributions, allowing {\sc MOSFiT} to run until the default convergence criterion\footnote{as discussed in \url{https://mosfit.readthedocs.io/en/latest/fitting.html}}. Flat priors were used for all parameters.
\begin{figure*}
\centering
\resizebox{0.95\hsize}{!}{\includegraphics{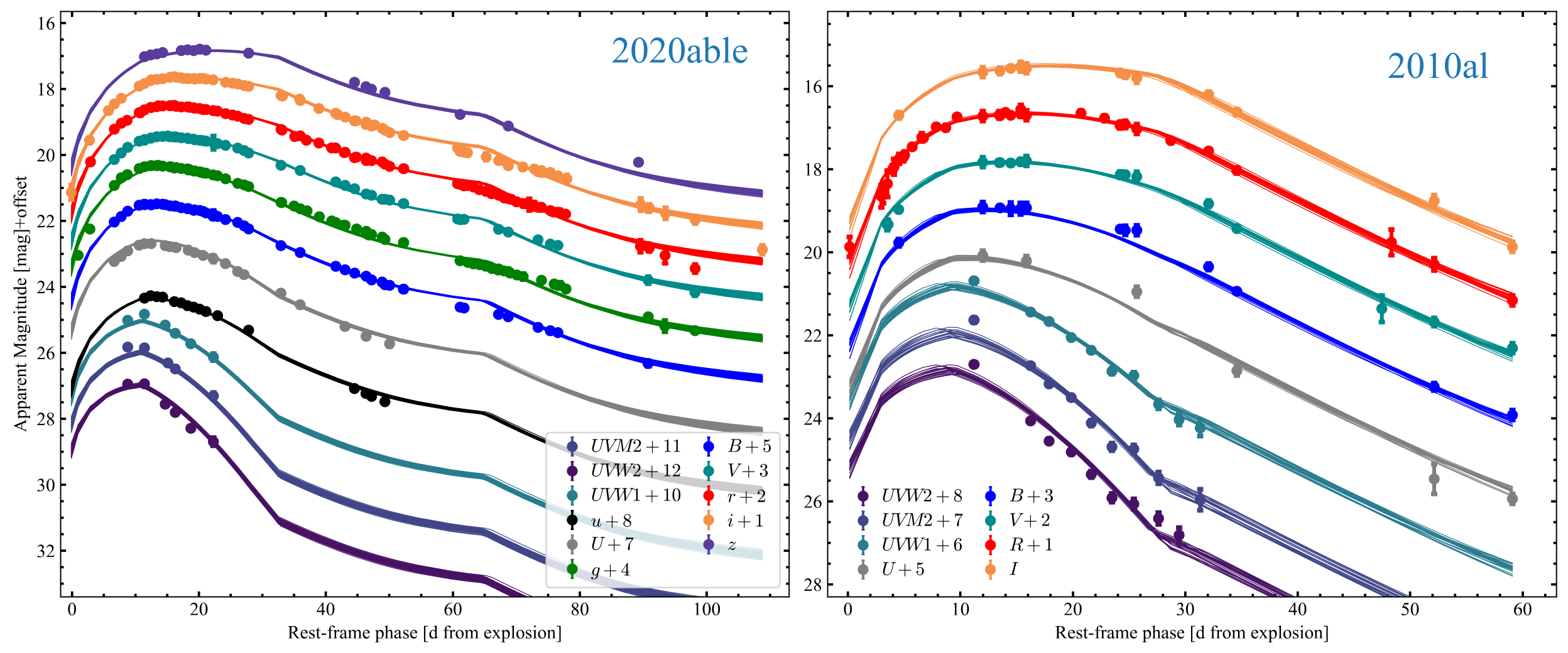}}
\caption{Multi-band light curves of SNe~2020able and 2010al along with an ensemble of models generated by {\sc MOSFiT}. Light curves have been vertically displaced by the indicated offsets. Rest-frame phases refer to the estimated explosion epoch.}
\label{fig:lcsmosfit}
\end{figure*}
To minimise the number of free parameters, we considered a gamma-ray opacity $\kappa_{\gamma}=0.03\,\rm{cm^2\,\rm{g^{-1}}}$ \citep[typical for a He-rich medium;][]{2015ApJ...814...63M} and SN ejecta with a constant inner density profile ($\delta=0$).
Although a standard radiative envelope for a pure WR progenitor suggests an outer ejecta profile index of $n\simeq10$, adopting this value in our modelling severely underestimates the host galaxy extinction derived from the \ion{Na}{ID} features. 
As discussed by \citet{1999ApJ...510..379M}, deviations towards shallower ejecta profiles ($n\simeq7$) are expected for stripped-envelope progenitors that experienced violent pre-explosion mass loss \citep[see e.g. the case of SN~1993J;][]{1995PhR...256..173N}. We therefore let $n$ vary within reasonable values for stars with radiative envelopes (i.e. 6--10).
The radius at which the transition from $\rho\propto r^{-\delta}$ to $r^{-n}$ occurs in the density profile of the ejecta is defined by the dimensionless radius $x_0=r_0(t)/R_{SN}(t)$, where the density evolves as $\rho_{\rm{SN}}=g^n\,t^{n-3}\,r^{-n}$. 
The normalisation constant $g^n=1/(4\pi(\delta-n))[2(5-\delta)(n-5)E_{k}]^{(n-3)/2}/[(3-\delta)(n-3)M_{\rm{ej}}]^{(n-5)/2}$ depends on the total ejected mass $M_{\rm{ej}}$, the total SN kinetic energy $E_{k}$ and the assumed power-law indices $n$ and $\delta$ \citep[see][]{1994ApJ...420..268C}.
Within this framework, {\sc MOSFiT} determines the transition radius $r_0$ to ensure that the ejecta density profile remains physically consistent with the main explosion parameters.
We also assumed a CSM with a density profile $\rho\propto r^{-2}$ (the ``wind scenario").
We note that although \citet{2022ApJ...927...25M} invoked a steeper density profile ($\propto r^{-3}$) to reproduce the rapid post-peak decay typically observed in SNe Ibn, such a steep density gradient is observationally incompatible with the slow post-peak evolution of SN~2020able.
Consequently, whilst the standard wind-like profile ($s=2$) is dictated by the native limitations of MOSFiT (which requires $s\le2$), it remains observationally well-justified by the slow evolution of the transient.
To further test the efficiency of interaction, we explored a pure {\sc csm} model for both transients.
For SN~2020able, we also used the spectroscopic velocity and host extinction estimates derived for SN~2020able, fixing $v_{\rm{ej}}=5.5\times10^3\,\rm{km\,s^{-1}}$, matching the late-time broad \ion{He}{I} lines at $+98\,\rm{days}$, and $N_{\rm{H}}=3.1\times 10^{20}\,\rm{cm^{-2}}$, based on the \ion{Na}{ID} equivalent widths.
All models converged successfully for both objects (see Figs.~\ref{fig:lcsmosfit}, \ref{fig:corner20able} and \ref{fig:corner10al}), but revealed distinct physical configurations despite their photometric similarities (see Tables~\ref{tab:mosfitParcsmni} and \ref{tab:mosfitParcsm}). 
Although both events exhibit sub-canonical kinetic energies ($E_{\rm{k}}\simeq2.9\times10^{50}\,\rm{erg}$ for SN~2020able and $3.8\times10^{50}\,\rm{erg}$ for SN~2010al) and high temperature floors ($T_{\rm{min}}\simeq7500-7700\,\rm{K}$), consistent with the helium recombination temperature \citep[e.g.][]{2020A&A...642A.106D}, their power sources and ejecta properties diverge significantly.
For SN~2020able, the best-fit {\sc csmni} model yields an ejecta mass of $M_{\rm{ej}}\simeq1.3\,\rm{M_{\sun}}$, a circumstellar shell mass of $M_{\rm{CSM}}\simeq0.6\,\rm{M_{\sun}}$ located at an inner radius $R_0\simeq0.3\,\rm{AU}$ ($\simeq64\,\rm{R_{\sun}}$), and a synthesised nickel mass of $M_{\rm{^{56}Ni}}\simeq3.2\times10^{-2}\,\rm{M_{\sun}}$. 
We note that both $R_0$ and $\rho_{\rm{csm}}$ exhibit broad posterior distributions spanning over an order of magnitude (see Figs.~\ref{fig:corner20able} and \ref{fig:corner10al}), a known manifestation of parameter degeneracies between shell location, density, and interaction efficiency in light curve modelling. 
However, even considering the $16\text{th}-84\text{th}$ percentile confidence intervals ($R_0\simeq0.1-0.6\,\rm{AU}$), the inner boundary remains strictly constrained within $\lesssim1\,\rm{AU}$, preserving a compact circumstellar environment.
In contrast, for SN~2010al, the {\sc csmni} fit indicates a negligible nickel contribution ($M_{\rm{^{56}Ni}}\simeq6.5\times10^{-3}\,\rm{M_{\sun}}$), suggesting that its light curve is largely dominated by ejecta-CSM interaction.
Indeed, a pure {\sc csm} model for SN~2010al provides a viable alternative, but requires a substantially more massive shell ($M_{\rm{CSM}}\simeq2.64\,\rm{M_{\sun}}$) extending up to $R_0\simeq6.9\,\rm{AU}$.
For SN~2020able, the {\sc csm} model can mathematically fit the light curve, but forces the inner boundary of the shell to an unrealistically large radius of $R_0 \simeq 35.5\,\rm{AU}$ with an extremely low density ($\rho_{\rm{csm}}\simeq5.6\times10^{-13}\,\rm{g\,cm^{-3}}$). 

Differences in the preferred model can be physically linked to their different post-peak light curve decline rates and the intrinsic limitations of the {\sc MOSFiT} code. 
As highlighted by \citet{2022ApJ...927...25M}, fast-evolving Type Ibn SNe like SN~2010al require a steeper CSM density profile ($\rho\propto r^{-s}$ with $s\simeq3$) to account for their rapid post-peak declines while maintaining physically realistic explosion and CSM parameters. 
Since {\sc MOSFiT} natively requires $s\le2$, it cannot easily fit rapidly declining transients without artificially boosting $M_{\rm{CSM}}$ or reducing $M_{^{56}\rm{Ni}}$.
Conversely, SN~2020able displays a remarkably slower post-peak evolution, making the standard wind-like profile ($s=2$) both mathematically suitable and observationally well-justified.
Interestingly, for SN~2010al, all models require a non-negligible host reddening, in contrast to the value reported in the literature, with the {\sc csm} model suggesting a neutral hydrogen column density $N_{\rm{H}}=(9.1\pm0.65)\times10^{20}\,\rm{cm^{-2}}$ \citep[corresponding to $A_V=0.412\pm0.034\,\rm{mag}$, according to][]{2009MNRAS.400.2050G}.
To further explore this discrepancy, we checked the mid-resolution X-shooter spectra of SN~2010al published by \citet{2011AN....332..266P}.
Specifically, we estimated the host contribution to the total reddening by measuring \ion{Na}{ID} EWs in spectra obtained around maximum light and approximately $2\,\rm{months}$ after explosion (see Sect.~\ref{sec:host}), and the derived values were roughly in agreement with that predicted by {\sc MOSFiT} at both phases \citep[$0.60\pm0.16\,\rm{mag}$, averaging Eq.~7 and 8 in][]{2012MNRAS.426.1465P}.

Despite the modest CSM masses derived from the modelling, both transients reached high peak luminosities, indicating efficient conversion of kinetic energy into radiation. 
The high recombination temperatures confirm stripped-envelope progenitors for both events, but the significantly larger synthesised $^{56}\rm{Ni}$ mass in SN~2020able points toward a more massive and compact progenitor star. 
Combined with the CSM expansion velocities inferred from the spectral lines (Section~\ref{sec:spectroscopy}), this evidence supports a Wolf-Rayet progenitor for SN~2020able \citep[e.g.][]{2007ARA&A..45..177C}.
 
\begin{figure}
\centering
\resizebox{0.95\hsize}{!}{\includegraphics{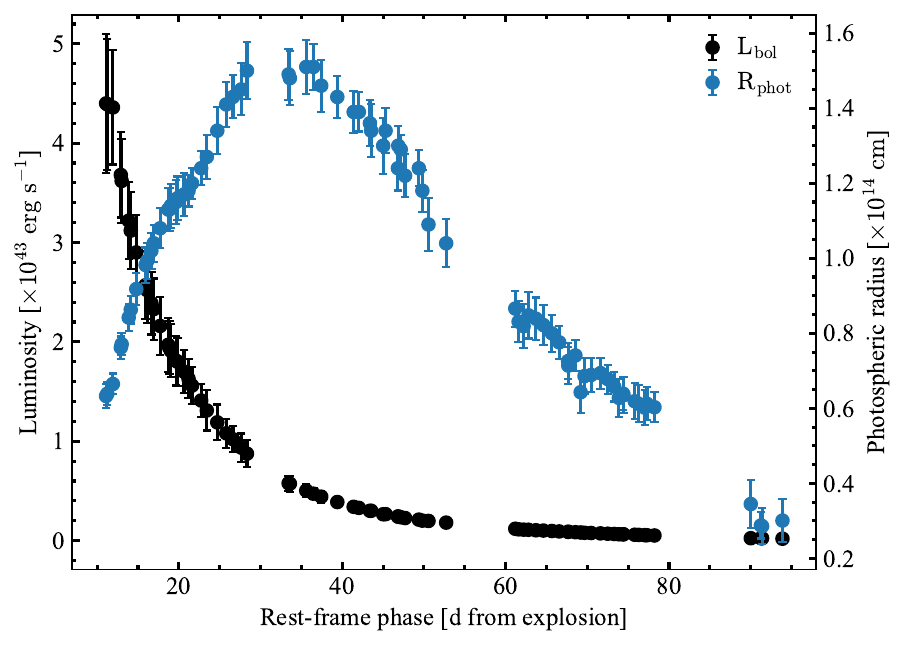}}
\caption{Evolution of the photospheric radius and the bolometric luminosity as derived by fitting a blackbody to the SED of SN~2020able. Rest-frame phases refer to the estimated explosion epoch.}
\label{fig:bolometric}
\end{figure}

The temporal evolution of the photospheric properties of SN~2020able was estimated by constructing the spectral energy distributions (SEDs) from multi-band photometry. 
To ensure consistent sampling, light curves were linearly interpolated to the epochs of the $r-$band observations, with a baseline tied to the earliest Swift/UV data to capture the ultraviolet flux that dominates the early SED of core-collapse SNe \citep[e.g.][]{2013MNRAS.434.1636T}. 
Sequential blackbody fits yielded a peak bolometric luminosity $L_{\rm{peak}} \simeq 4.4 \times 10^{43}\,\rm{erg\,s^{-1}}$ (see Fig.~\ref{fig:bolometric} and Table~\ref{tab:bolometric}). 
At $+11\,\rm{days}$, the photosphere remains hot ($T_{\rm{ph}}\simeq2.0\times10^4\,\rm{K}$), approximately 35\% higher than estimated from the optical continuum alone, underscoring the need for UV coverage in interacting transients.
Conversely, the derived temperatures and photospheric radii vary rapidly, arguing against interaction significantly shaping the overall evolution.
The rapid decline of the light curves at later phases, along with the observed spectroscopic evolution (see Sect.~\ref{sec:spectroscopy}), also highlight a ``mild" interaction scenario. 
In this framework, interaction contributes to the transient luminosity, but remains secondary to the dynamics of the underlying ejecta at later phases, consistent with the evolution of fast-evolving SNe Ibn \citep[e.g.][]{2015MNRAS.449.1921P,2016MNRAS.456..853P}. 
This interpretation is also supported by the observed spectroscopic evolution (see Sect.~\ref{sec:spectroscopy}) as well as the agreement between the photospheric radius obtained from SED fitting and the physical parameters estimated with {\sc MOSFiT}, which also explain the early $gri$-band excess (see Figure~\ref{fig:griRise}). 
Given the best-fit {\sc csmni} average density of $\rho\simeq9.3\times10^{-9}\,\rm{g}\,\rm{cm^{-3}}$ and $M_{\rm{CSM}}\simeq0.6\,\rm{M_{\sun}}$, we infer an outer radius for the interacting shell of approximately $3.1\times10^{13}\,\rm{cm}$ (assuming an inner radius $R_0\simeq3.7\times10^{12}\,\rm{cm}$; see Table~\ref{tab:mosfitParcsmni}).  
This indicates a remarkably compact interaction region ($\Delta R\simeq2.7\times10^{13}\,\rm{cm}$) that is rapidly swept up by the expanding SN ejecta. 
In this picture, the interaction is confined both spatially ($\lesssim3\times10^{13}\,\rm{cm}$) and temporally (within the first few days post-explosion), reinforcing that the late-time behaviour of SN~2020able is likely dominated by the expanding ejecta rather than ongoing interaction.

\subsection{Spectroscopy} \label{sec:spectroscopy}
\begin{figure*}
\centering
\resizebox{0.85\hsize}{!}{\includegraphics{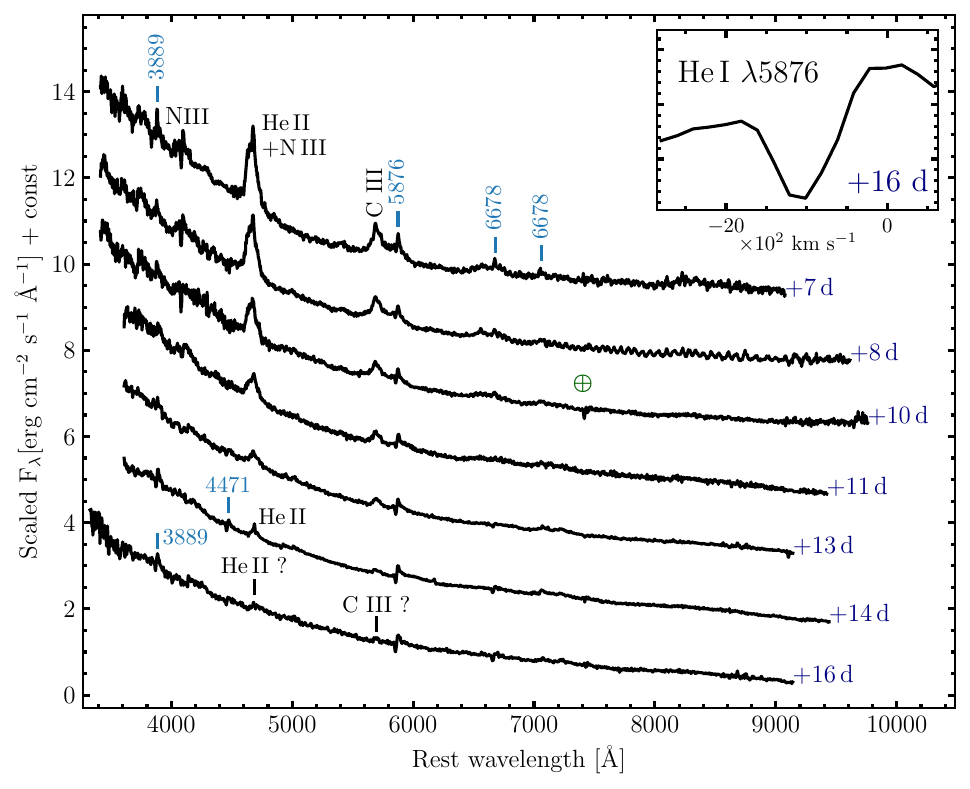}}
\caption{Early spectroscopic evolution of SN~2020able at $t<+16\,\rm{days}$ along with a line identification of the main emission features. \ion{He}{I} lines are reported with their corresponding rest wavelengths. The $\oplus$ symbol marks the position of the $O_2$ telluric feature (Band A) around 7500\AA. The inset shows a zoom-in of \ion{He}{I} $\lambda5876$ at $+16\,\rm{days}$ in the velocity plane. Rest-frame phases refer to the estimated explosion epoch.}
\label{fig:earlySpec}
\end{figure*}

\begin{figure*}
\resizebox{\hsize}{!}{\includegraphics{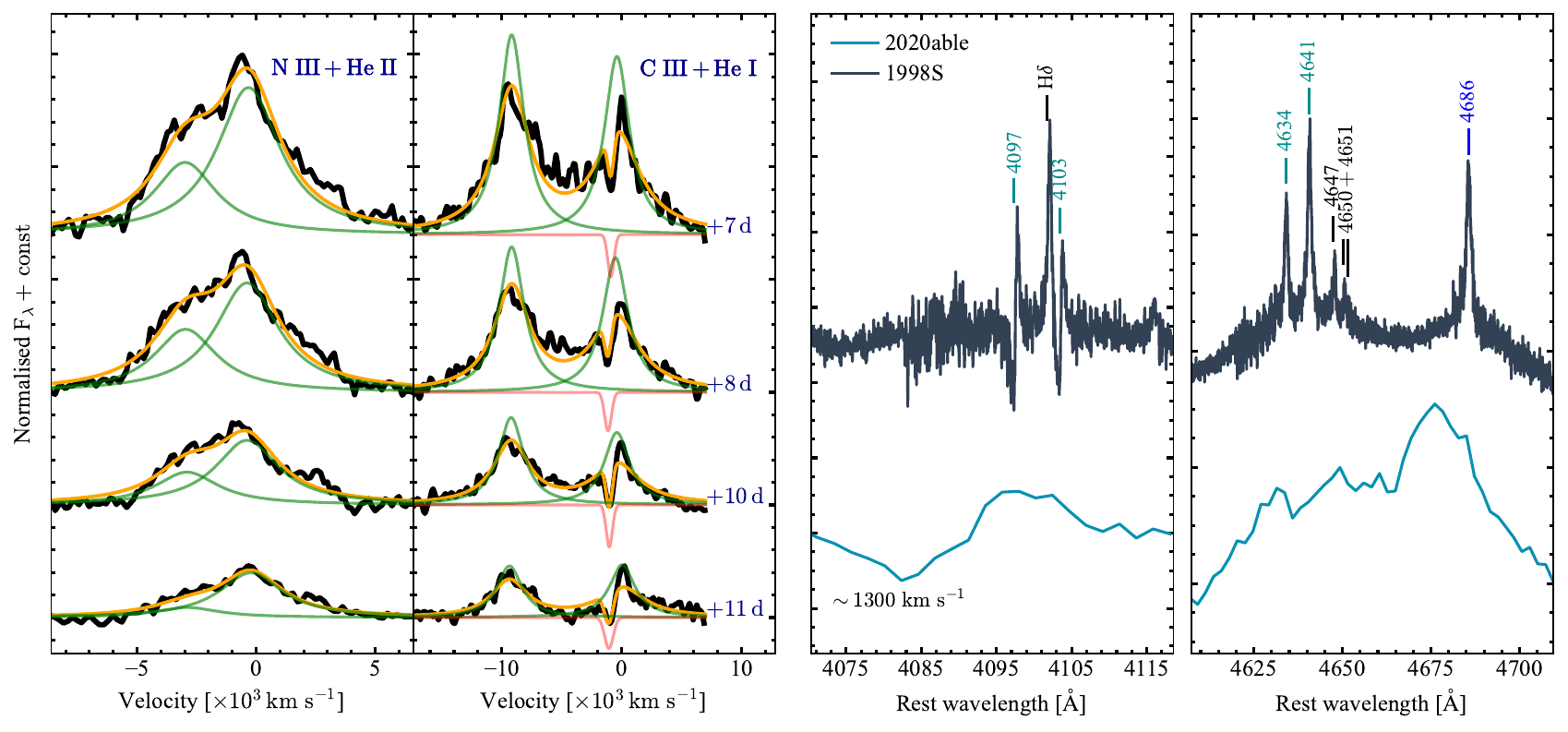}}
\caption{{\bf Left:} Multi-component fit to the \ion{N}{III}+\ion{He}{II} and \ion{C}{III}+\ion{He}{I} regions. A Gaussian absorption was used to reproduce the overall profile of \ion{He}{I} $\lambda5876$. The widths and peaks of the features remain constant throughout the evolution of the high-ionisation feature, with a $\rm{FWHM_{vel}}\simeq3.0\times10^3\,\rm{km}\,\rm{s^{-1}}$. {\bf Right:} Comparison of the spectrum obtained at $+7\,\rm{days}$ with the one of SN~1998S, with a much higher resolution, obtained approximately $5\,\rm{days}$ after explosion \citep[see][]{2015ApJ...806..213S}.}
\label{fig:98SComp}
\end{figure*}

\begin{figure*}
\resizebox{\hsize}{!}{\includegraphics{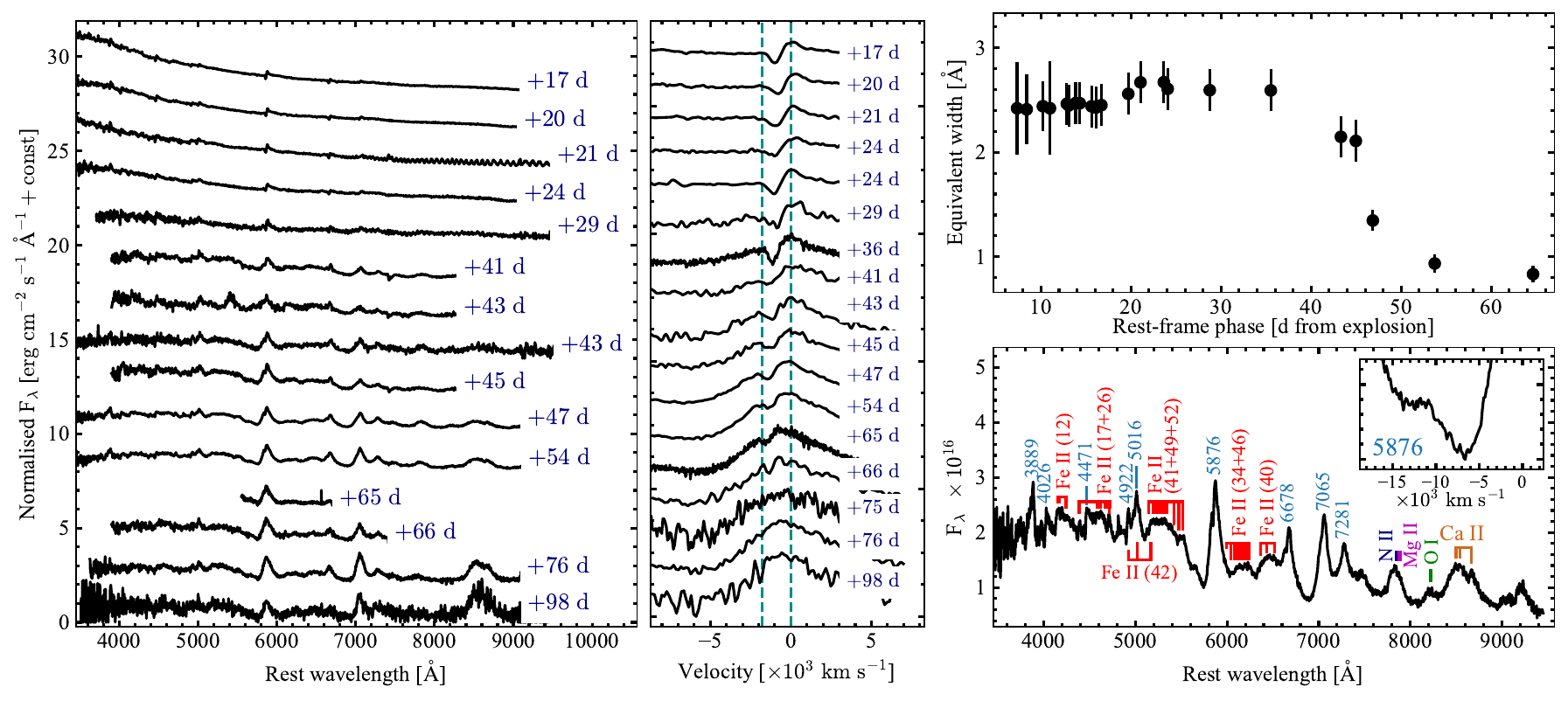}}
\caption{{\bf Left:} Spectroscopic evolution of SN~2020able at $t>+17\,\rm{days}$. {\bf Middle:} Zoom-in around \ion{He}{I} $\lambda5876$ showing the evolution of its narrow absorption feature. {\bf Right (top):} Evolution of the EW of the \ion{He}{I} $\lambda5876$ narrow absorption showing a sudden drop at $t\gtrsim+47\,\rm{days}$. {\bf Right (bottom):} Line identification in the $+47\,\rm{days}$ spectrum, with a zoom-in over the complex broad absorption feature observed in the \ion{He}{I} $\lambda5876$ line. Rest-frame phases refer to the estimated epoch of the explosion.}
\label{fig:lateSpec}
\end{figure*}

Optical spectra are shown in Fig.~\ref{fig:earlySpec}, \ref{fig:98SComp} and \ref{fig:lateSpec}, while details on instruments used to carry out the follow-up campaign, reduction steps, and a log of the spectroscopic observations are reported in Appendix~\ref{sec:obsredu} and Table~\ref{tab:speclog}.

Early spectra showed a hot continuum rapidly decreasing from $\simeq2.0\times10^4$ to $\simeq1.4\times10^4\,\rm{K}$ within the first $16\,\rm{days}$ after the explosion, as estimated by fitting a blackbody to the spectral continuum after excluding regions dominated by strong emission features.
Prominent \ion{C}{III}, \ion{N}{III}, \ion{He}{II} and \ion{He}{I} lines, often visible during the early phases of CC SNe, were clearly detected within the same period.
In Type Ibn SNe with similar early features (e.g. SNe~2010al, 2019cj, 2019uo, 2019wep and 2023emq; see Fig.~\ref{fig:riseMax} and Table~\ref{tab:riseMax}), these typically last longer than in hydrogen-rich SNe before disappearing below the continuum level.
We identified relatively narrow (FWHM velocity $\simeq3\times10^3\,\rm{km}\,\rm{s^{-1}}$) \ion{He}{I} $\lambda3889$, 5876 and 6678 lines, along with prominent \ion{C}{III} $\lambda5696$, \ion{N}{III} $\lambda4634-4641$ \citep[multiplet 2; see][]{1945CoPri..20....1M} and $\lambda4100$ and \ion{He}{II} $\lambda4686$.
\ion{N}{III} $\lambda4100$ is likely formed by the blend of the $\lambda\lambda4097$, 4103 doublet \citep[see][and Figure~\ref{fig:98SComp}]{1972AJ.....77..312W}.
This is the only high-ionisation line with an absorption component, corresponding to an expansion velocity $\simeq1.3\times10^3\,\rm{km}\,\rm{s^{-1}}$, comparable to those measured from the P Cygni absorption minima of the \ion{He}{I} lines (see the inset in Figure~\ref{fig:earlySpec}).
High-ionisation features were clearly visible up to $+14\,\rm{days}$, while we only tentatively identified a shallow emission corresponding to \ion{C}{III} $\lambda5696$ at $+14$ and $+16\,\rm{days}$.
\citet{2025A&A...703A.177T} showed that the \ion{N}{III} lines in the early spectra of SN~2024bch were likely produced by the Bowen fluorescence mechanism. 
This is triggered by the near-coincidence of \ion{He}{II} Ly-$\alpha$ $\lambda303.783$ and \ion{O}{III} $\lambda303.799$, followed by a second resonance between \ion{O}{III} $\lambda374.436$ and \ion{N}{III} $\lambda\lambda374.434, 374.441$ \citep[see e.g.][]{2007A&A...464..715S}. 
In this framework, \ion{He}{II} $\lambda4686$ and \ion{C}{IV} $\lambda5696$ emerge from the subsequent recombination cascades of He and C.
This interpretation may also apply to SN~2020able, given the high temperatures inferred from its spectral continuum up to $+16\,\rm{days}$. 
Indeed, these estimates should be regarded as conservative lower limits, since the SED peak falls at wavelengths significantly shorter than those covered by our optical spectra (see Sect.~\ref{sec:photometry} where we estimated $T_{\rm{ph}}\simeq2.0\times10^4\,\rm{K}$ at $+11\,\rm{days}$); the actual temperatures are likely higher, further supporting the presence of a high ionisation regime.
\begin{table}
\caption{Photospheric parameters obtained fitting a blackbody to the SED of SN~2020able using UV and optical data.}
\label{tab:bolometric}
\centering
\begin{tabular}{cccc}
\toprule
Phase & $\rm{T_{ph}}$ (err) & $\rm{R_{phot}}$ (err)   & $\rm{L_{bol}}$ (err) \\
\midrule
 (d)             & (K)                & $\times10^{14}\,\rm{cm}$ & $\times10^{43}\,\rm{erg}\,\rm{s^{-1}}$ \\  
\midrule
$+11$ & 19817(610) & 6.33(0.32) & 4.40(0.70) \\   
$+11$ & 19710(573) & 6.39(0.31) & 4.39(0.66) \\   
$+12$ & 19292(491) & 6.65(0.28) & 4.36(0.58) \\   
$+13$ & 17265(375) & 7.62(0.30) & 3.68(0.43) \\   
$+13$ & 17101(370) & 7.71(0.30) & 3.62(0.42) \\   
$+14$ & 15889(350) & 8.42(0.35) & 3.22(0.39) \\ 
\bottomrule
\end{tabular}
\tablefoot{Rest-frame phases are relative to the estimated explosion epoch. The table is published in its entirety in a machine-readable format, available at the CDS. A portion is shown here for guidance regarding its shape and content.}
\end{table}
Assuming that these features are driven by the Bowen fluorescence mechanism, their prolonged persistence in Type Ibn SNe - compared to hydrogen-rich events - is likely due to the hydrogen-deficient nature of their circumstellar envelopes.
This reduces the competition for ionising photons and prevents the absorption of the \ion{He}{II} resonance line by neutral hydrogen, thus sustaining the emission more efficiently.

Whereas the main \ion{He}{I} lines ($\lambda3889$, 5876, and 6678) were unambiguously identified due to the absence of nearby features, our early spectra lacked the resolution to distinguish the individual components of the complex emission around 4670~\AA.
For the same reason, we cannot rule out the presence of \ion{C}{IV} $\lambda\lambda5801$, 5812 \citep[typically observed along with \ion{C}{III} $\lambda5696$ and \ion{C}{IV} $\lambda4658$ in the spectra of WR stars;][]{1998MNRAS.296..367C}, which may be blended with \ion{C}{III} $\lambda5696$ and \ion{He}{I} $\lambda5876$.
We therefore compared our $+7\,\rm{days}$ spectrum with the much higher resolution one of SN~1998S obtained approximately $5\,\rm{days}$ after explosion in the same spectral region (see Figure~\ref{fig:98SComp}).
Beyond the Balmer lines, consistent with its classification as a hydrogen-rich CC SN, SN~1998S exhibited both \ion{N}{III} $\lambda4097$ and $\lambda4103$. 
These features displayed narrow P Cygni profiles, indicating expansion velocities of $\simeq20\,\rm{km}\,\rm{s^{-1}}$, notably lower than the $\simeq1.3\times10^3\,\rm{km}\,\rm{s^{-1}}$ observed in SN~2020able, highlighting the remarkable disparity in velocities of the absorbing shells.
As in SN~2020able, the \ion{N}{III} $\lambda4100$ doublet is the only high-ionisation line clearly showing an absorption component, while \ion{N}{III} $\lambda\lambda4634$, 4641, \ion{C}{IV} $\lambda4647-4651$ and \ion{He}{II} $\lambda4686$ are purely in emission.
Although the resolution of our $+7\,\rm{days}$ spectrum does not allow us to identify all these features, we cannot rule out their presence in the early spectra of SN~2020able.
A resolution of $R\gtrsim4000$ is therefore crucial in order to fully identify spectroscopic features in the early spectra of interacting transients such as Type Ibn supernovae.

In Fig.~\ref{fig:98SComp}, we also show a tentative fit to the main features within the first $11\,\rm{days}$ from explosion, a period during which all components remain clearly discernible above the spectral continuum.
In our fit, we considered the contributions of \ion{N}{III} $\lambda4634$, \ion{He}{II} $\lambda4686$, \ion{C}{III} $\lambda5696$, and \ion{He}{I} $\lambda5876$ and adopted Lorentzian profiles to follow the shape of the electron-scattering wings. 
Although we note that such profiles may not be fully captured by a Lorentzian shape, particularly at higher resolutions \citep{2018MNRAS.475.1261H,2020A&A...635A..39T}, the overall profiles are well-reproduced assuming a constant FWHM of approximately $\simeq3\times10^3\,\rm{km}\,\rm{s^{-1}}$ at all times, leaving only the peak wavelengths and line fluxes as free parameters.
Interestingly, all emission features have centroids roughly at their corresponding rest wavelengths, with no evidence of efficient radiative acceleration \citep[see e.g.][]{2014ApJ...797..118F}.
This is consistent with the evolution of high-ionisation features observed in SN~2024bch \citep{2025A&A...703A.177T} and potentially places the emitting shell at a distance where radiative acceleration becomes negligible.
This scenario is further supported by the constant FWHM of the emission lines. 
A compact, nearby shell being swept up by the rapidly expanding SN ejecta would have introduced a shocked component, thus modifying the FWHM within the first $16\,\rm{days}$, which is ruled out by our fitting procedure.
This points towards a stratified circumstellar environment, likely composed of multiple distinct shells: an inner one contributing to the early interaction and luminous peak, and outer regions responsible for the high-ionisation lines and the persistent narrow \ion{He}{I} features observed up to approximately $+70\,\rm{days}$ (see below).

At $t\gtrsim+16\,\rm{days}$, high-ionisation features fade and narrow \ion{He}{I} lines with prominent P Cygni profiles are the only features visible above a hot continuum.
We identified \ion{He}{I} $\lambda3889$, $\lambda4471$ (appearing already at $+14\,\rm{days}$), $\lambda4686$ and $\lambda5876$.
All helium lines showed an absorption minimum corresponding to an expansion velocity of approximately $1.2\times10^{3}\,\rm{km}\,\rm{s^{-1}}$ and a blue-velocity-at-zero-intensity \citep[BVZI, see e.g.][]{2020A&A...638A..92T} of approximately $1.8\times10^3\,\rm{km}\,\rm{s^{-1}}$, which remained constant up to $+75\,\rm{days}$, after which the absorption component is no longer visible.
Within the same period, its EW, measured by a simple Gaussian fit, remained roughly constant until approximately $+47\,\rm{days}$, when it showed a sudden drop (see Figure~\ref{fig:lateSpec}).
At $+43\,\rm{days}$, the \ion{He}{I} $\lambda5876$ profile underwent a significant metamorphosis, displaying a much broader emission accompanied by a blue-shifted absorption component. 
From its minimum, we derived a photospheric velocity of approximately $6.7\times10^3\,\rm{km}\,\rm{s^{-1}}$, although the nature of this absorption may be considered ambiguous due to the proximity of \ion{Fe}{II} lines \citep[e.g. multiplets 41, 42, 48, 49, and 52;][]{1945CoPri..20....1M}.
However, its minimum decreased monotonically to $5.5\times10^3\,\rm{km}\,\rm{s^{-1}}$, as observed at $+98\,\rm{days}$, consistent with a receding photosphere mapping progressively deeper layers of the SN ejecta.
At $+43\,\rm{days}$, we also tentatively identified a distinct absorption consistent with \ion{Fe}{II} $\lambda5169$, resulting in a velocity of approximately $4.0\times10^3\,\rm{km}\,\rm{s^{-1}}$, consistent with the value inferred from the SED fitting (see Section~\ref{sec:photometry}). 
These results, along with the rapid photometric decline and the monotonically decreasing blackbody radius, collectively suggest that the ejecta-CSM interaction does not contribute to the energetic output of SN~2020able at these phases.

The most straightforward interpretation for the sudden drop in the EW of the narrow \ion{He}{I} absorptions at $t\gtrsim+47\,\rm{days}$ and their disappearance at $t\gtrsim+70\,\rm{days}$ is that the SN ejecta have finally overtaken the circumstellar shell. 
In this scenario, we can estimate the geometric extent of this He-rich material by adopting the ejecta expansion velocity derived above, which provides a direct constraint on its inner and outer radii.
Specifically, if the drop in the EW of the narrow absorption corresponds to the time the expanding ejecta reach the inner boundary of the shell, assuming $v_{ej}=6.7\times10^3\,\rm{km}\,\rm{s^{-1}}$ and a CSM expansion velocity of $1.2\times10^3\,\rm{km}\,\rm{s^{-1}}$ would result in an inner radius $R_{in}\simeq2.2\times10^{15}\,\rm{cm}$.
Furthermore, the disappearance of the narrow absorption at approximately $+70\,\rm{days}$ suggests an outer radius $R_{out}\simeq3.3\times10^{15}\,\rm{cm}$.
Alternatively, the absorption components may be diluted in the blue wing of the emerging \ion{He}{I} broad emission lines, which dominates the spectra since $+43\,\rm{days}$ (see Figure~\ref{fig:lateSpec}).

The geometric extent of the shell, combined with the expansion velocity of the bulk material estimated from the minimum of the narrow P Cygni profiles, can be used to constrain the physical properties of the CSM and the mass-loss episode that generated it. 
Assuming a constant expansion velocity of $1.2\times10^3\,\rm{km\,s^{-1}}$ and the calculated inner and outer radii, we estimate that the mass-loss episode began approximately $321\,\rm{days}$ and ended roughly $216\,\rm{days}$ before the SN explosion.
Furthermore, \ion{He}{II} $\lambda4686$ visible in the early spectra of SN~2020able implies a recombination timescale $t_{\rm{rec}}<7\,\rm{days}$ for the He-rich shell, providing a minimum electron density $n_e > (t_{rec}\alpha_{rec})^{-1}$. 
Adopting $\alpha_{\rm{rec}}=9.08\times10^{-13}\,\rm{cm^3\,s^{-1}}$ \citep[typical for He-rich gas at $T_{\rm{e}}=2.0\times10^4\,\rm{K}$;][]{2006agna.book.....O}, we obtain $n_{\rm{e}}>1.8\times10^6\,\rm{cm^{-3}}$.
A more stringent constraint for the local density can be derived from the narrow emission lines visible until $+16\,\rm{days}$. 
As derived above, their FWHM ($\simeq3\times10^3\,\rm{km\,s^{-1}}$) remained constant throughout the entire evolution of the high-ionisation features (see Figure~\ref{fig:98SComp}). 
For line profiles dominated by electron-scattering, this width is not indicative of the expansion velocity of the shell \citep[see][]{2018MNRAS.475.1261H}, which we directly estimated at roughly a third of this value from the absorption minima of the \ion{He}{I} and \ion{N}{III} lines.
The discrepancy is even more relevant for relatively compact shells, where the FWHM rather probes the Thomson optical depth ($\tau_{\rm{es}}$) for a given $T_{\rm{e}}$. 
For $\eta=R_{\rm{out}}/R_{\rm{in}}\simeq1.5$, a FWHM of $3\times10^3\,\rm{km\,s^{-1}}$ corresponds to $\tau_{es} \simeq6.0-7.0$ for an electron temperature $T_{\rm{e}}=2.0\times10^4\,\rm{K}$ and reasonable density profiles \citep[$\rho\propto r^{-2}$ with $s=0$, 1 and 2; see Fig.~5 in][]{2018MNRAS.475.1261H}.
This range of $\tau_{\rm{es}}$ is also supported by the overall shape of the emission lines: while the expansion velocity of the shell is expected to produce asymmetric electron-scattering profiles (i.e. with an excess in their red wings), \citet{2018MNRAS.475.1261H} showed that the combined effects of a thin shell and a high optical depth can still produce the symmetric features observed in SN~2020able even in shells with relatively high expansion velocities.
Taking $\tau_{\rm{es}}=6.5$, we derive $n_e=\tau_{\rm{es}}/(\sigma_T\Delta R)\simeq1.4\times 10^9\,\rm{cm^{-3}}$, which implies a mass-loss rate of $8.7\times10^{-2}\,\rm{M_{\sun}\,yr^{-1}}$ assuming a typical filling factor $f=0.1$ for a WR wind \citep[see e.g.][]{2023MNRAS.518.5001F}, a fully ionised helium composition and an event lasting approximately $105\,\rm{days}$.
This value is inconsistent with the wind velocity measured from the \ion{He}{I} P Cygni minima - which are comparable to those of steady winds in late WN (WN7-8) or very late WC stars \citep[e.g. WC9; see][]{2007ARA&A..45..177C} - but rather suggests that the layer with $\tau_{\rm{es}}=6.5$ was generated by a shorter eruptive event similar to the one observed two years prior to the explosion of SN~2006jc \citep{2007Natur.447..829P,2007ApJ...657L.105F}.
We then suggest that at least a fraction of the outer shell was produced by an eruptive episode, whereas a more extended CSM was generated by the wind of a WR star.

The evolution of the high-ionisation lines indicates that this inner denser shell is not reached by the fast ejecta ($v_{\rm{ej}}\simeq6.7\times10^3\,\rm{km}\,\rm{s^{-1}}$) within the first $16\,\rm{days}$ after the explosion. 
This allows us to estimate a lower limit for its radius at $R_{\rm{in,dense}}\simeq9.3\times10^{14}\,\rm{cm}$, compatible with the photospheric radius estimated by the SED fitting (see Section~\ref{sec:photometry}).
On the other hand, the evolution of the EW of the \ion{He}{I} P Cygni profiles indicates that the outer shell has an inner radius of $R_{\rm{in}}\simeq2.2\times10^{15}\,\rm{cm}$. 
This implies a thickness of $\Delta R\lesssim1.3\times10^{15}\,\rm{cm}$ for the denser shell.
With $\tau_{\rm{es}}=6.5$ and a filling factor $f=0.1$, this results in a density of $n_{\rm{e}}\simeq7.5\times10^8\,\rm{cm^{-3}}$ and a mass of approximately $0.02\,\rm{M_{\sun}}$.
Consequently, an eruptive episode of similar duration to that observed in SN~2006jc ($10\,\rm{days}$) with these physical properties must have had an average mass-loss rate of approximately $0.7\,\rm{M_{\sun}\,\rm{yr^{-1}}}$.
Given that eruptive events of evolved massive stars are not expected to largely exceed the expansion velocity of their steady winds \citep{2014ARA&A..52..487S}, we can estimate the energetics of the outburst by assuming an expansion velocity of approximately $1.5\times10^3\,\rm{km}\,\rm{s^{-1}}$.
This would yield a substantial kinetic luminosity $L_{\rm{mech}}=1^{}/2\,\dot{M}\,v^2_{\rm{out}}\simeq4.7\times10^{41}\,\rm{erg}\,\rm{s}^{-1}$ assuming the mass-loss rate $0.7\,\rm{M_{\sun}\,\rm{yr^{-1}}}$ derived above.
To reconcile this with the archival non-detection limits ($M_{\rm{bol}}\geq-14.0\,\rm{mag}$ assuming a $g-$band bolometric correction of $-0.3\,\rm{mag}$, consistent with an outburst with a photospheric temperature between $10^4$ and $1.2\times10^4\,\rm{K}$), a radiative efficiency $\epsilon=L_{\rm{bol}}/L_{\rm{mech}}\lesssim26\%$ is required.
Although the relatively high optical depth suggests efficient thermalisation of the kinetic energy, this radiative efficiency remains physically consistent with the value of $\tau_{\rm{es}}=6.5$ derived from the electron-scattering profiles. 
In addition, a filling factor $f=0.1$ reflecting the clumpy nature of the medium can limit the overall conversion efficiency, as a significant fraction of the kinetic energy can flow through the inter-clump medium without being thermalised.

At $t\gtrsim+46\,\rm{days}$, we also identified broad lines ($\rm{FWHM}\simeq5.0\times10^3\,\rm{km}\,\rm{s^{-1}}$) of \ion{Ca}{II} (near infrared triplet) and \ion{O}{I} $\lambda8222$, along with numerous relatively narrow \ion{Fe}{II} lines. 
The latter form the blends responsible for the blue pseudo-continuum at $\lambda\lesssim5600$~\AA, frequently observed in the spectra of interacting SNe \citep[see][]{2016MNRAS.458.2094D}. 
At approximately 7800~\AA, we also observed a strong emission that was not easily identifiable. 
Its peak is inconsistent with \ion{Mg}{II} $\lambda\lambda7877$, 7896, which would require a blueshift of approximately $2.0\times10^3\,\rm{km}\,\rm{s^{-1}}$. 
Similarly, the \ion{O}{I} $\lambda7774$ multiplet would be redshifted by about $2.4\times10^3\,\rm{km}\,\rm{s^{-1}}$. 
Such shifts were not observed in any other broad line within the spectrum.
Furthermore, we can exclude \ion{Fe}{II} $\lambda7841$ - typically one of the most intense lines of multiplet 73 - as we do not detect other lines from the same multiplet that should have comparable intensity (e.g. $\lambda7712$ and $\lambda7320$). 
A possible alternative identification is \ion{N}{II} multiplet 11, which has a relatively intense line at 7841.27\AA~(see Figure~\ref{fig:lateSpec}). 
The presence of nitrogen could be linked to the nature of the progenitor: in WN stars, nitrogen is indeed the second most abundant element after helium, being a primary product of the CNO cycle. 
Wind velocities measured from the P Cygni minima of the \ion{He}{I} lines would also be consistent with a WR progenitor in the WN/C transition phase.

In Fig.~\ref{fig:specModels}, we compare the spectral evolution of SN~2020able at $t>43\,\rm{days}$ with non-local thermodynamic equilibrium (LTE) radiative-transfer models.
For each spectrum, we considered the ``best-fit" model through a $\chi^2$ minimisation algorithm against the full grid of models computed by \citet{2022A&A...658A.130D}.
These represent low-mass helium stars with initial helium core masses - after removal of the hydrogen envelope - of $3.3-5.0\,\rm{M_{\sun}}$ evolved in interacting binary systems. 
Such progenitors are characterised by a high fractional helium abundance (typically $\ge 50\%$), which is essential to reproduce the persistence of He I emission lines observed in the data.
In these simulations, parameters such as the radius of the cold dense shell (CDS), the mass accumulated within the CDS, and the power deposition from the ejecta/CSM interaction ($2.0-1.0\times10^{42}\,\rm{erg}\,\rm{s^{-1}}$) are variables dependent on time. 
Consequently, the shifting agreement between each observed spectrum and specific model (i.e. from {\sc he3p3} to {\sc he4p5}) does not imply a change in the underlying progenitor, but rather captures the physical evolution of the spectrum-formation region as the CDS expands and the ionisation state of the gas evolves.
Although the $\chi^2$ procedure suggests a good agreement with the {\sc he3p3}-{\sc he5p0} models - pointing toward a progenitor with a total ejecta mass between $1.15$ and $1.90\,\rm{M_{\sun}}$ and a helium mass of $0.92-1.10\,\rm{M_{\sun}}$ - a significant discrepancy remains in the observed line profiles.
Specifically, the models fail to reproduce the broad emission and absorption components clearly visible in the spectra of SN~2020able. 
This mismatch likely arises from the fundamental assumptions of the \citet{2022A&A...658A.130D} simulations, which are specifically designed to explore the spectral signatures of a CDS formed through strong ejecta-CSM interaction. 
In this regime, the compression of material into a narrow velocity space naturally produces the narrow lines characteristic of Type Ibn supernovae. 
Since the spectrophotometric evolution of SN~2020able suggests that at these phases the interaction power is no longer the dominant energy source, the observed broad features likely originate from the inner supernova ejecta now fully exposed.
Consequently, while the models provide a fundamental benchmark for the chemical signature of Type Ibn SNe, the dynamical evidence in SN~2020able supports a higher-mass progenitor channel, where the ejecta maintain a complex velocity stratification typical of massive stripped-envelope stars.

\section{Summary and conclusions} \label{sec:conclusions}
In this paper, we discussed the UV/optical spectrophotometric evolution of the Type Ibn SN~2020able.
The transient exploded in an environment with significantly sub-solar metallicity, with an estimated $12+\log(\rm{O/H})=8.28\pm0.03\,\rm{dex}$.

The early photospheric evolution highlights the need for a source of energy in addition to canonical recombination and radioactive decays.
The combined information obtained from the early SED fitting, the observed deviation from a $t^{-2}$ power-law evolution of $gri-$band light curves and the fit with {\sc MOSFiT} indicates that the interaction must be geometrically and temporally confined to the very early post-explosion phases.
In addition, the analysis of observed data indicates a stratified environment with three distinct regions at different distances from the progenitor star.
For the innermost shocked shell, the best-fit {\sc MOSFiT} parameters ($M_{\rm{CSM}}\simeq0.6\,\rm{M_{\sun}}$, $\rho\simeq9.3\times10^{-9}\,\rm{g}\,\rm{cm^{-3}}$, and $R_0\simeq3.7\times10^{12}\,\rm{cm}$) imply a remarkably compact outer radius of $R_{\rm{out,shocked}}\simeq3.1\times10^{13}\,\rm{cm}$.
This shell is likely fully ionised and does not contribute to the early spectroscopic evolution of SN~2020able, which is dominated by relatively narrow high-ionisation lines originating from an outer dense shell at $R_{\rm{in,dense}}\gtrsim9.3\times10^{14}\,\rm{cm}$ with $\tau_{es}\simeq6.5$.
The subsequent conversion of kinetic energy into radiation contributes significantly to the early excess observed in the $gri-$band light curves, as well as the relatively high peak luminosity ($4.4\times10^{43}\,\rm{erg}\,\rm{s^{-1}}$), while at later phases it does not seem to contribute significantly to the energy output and spectroscopic evolution of the transient.

At a distance of $R_{in}\gtrsim9.3\times10^{14}\,\rm{cm}$, the CSM exhibits a significant Thomson optical depth ($\tau_e\simeq6.5$), as indicated by the FWHM of the electron scattering profiles, resulting in an average density $n_{\rm{e}}\simeq1.4\times10^9\,\rm{cm^{-3}}$.
However, the expansion velocity of $v\simeq1.2\times10^3\,\rm{km}\,\rm{s^{-1}}$, derived from the P Cygni minima of \ion{He}{I} and \ion{N}{III} $\lambda4100$, is typical of steady winds of late WN to WC-type WR stars. 
This kinematic evidence is in stark contrast with the high mass-loss rate inferred from our density constraints. 
To maintain such density and optical depth, a steady-state wind would require a mass-loss rate ($\dot{M}\simeq8.7\times10^{-2}\,\rm{M}_{\sun}\,\rm{yr^{-1}}$) that is physically inconsistent with radiation-driven wind theory for WR stars \citep{2006ApJ...645L..45S}. 
We therefore conclude that at least the inner part of the un-shocked CSM was produced by a discrete pre-SN eruptive event, similar to that observed for SN~2006jc.
\begin{figure}
\centering
\resizebox{0.95\hsize}{!}{\includegraphics{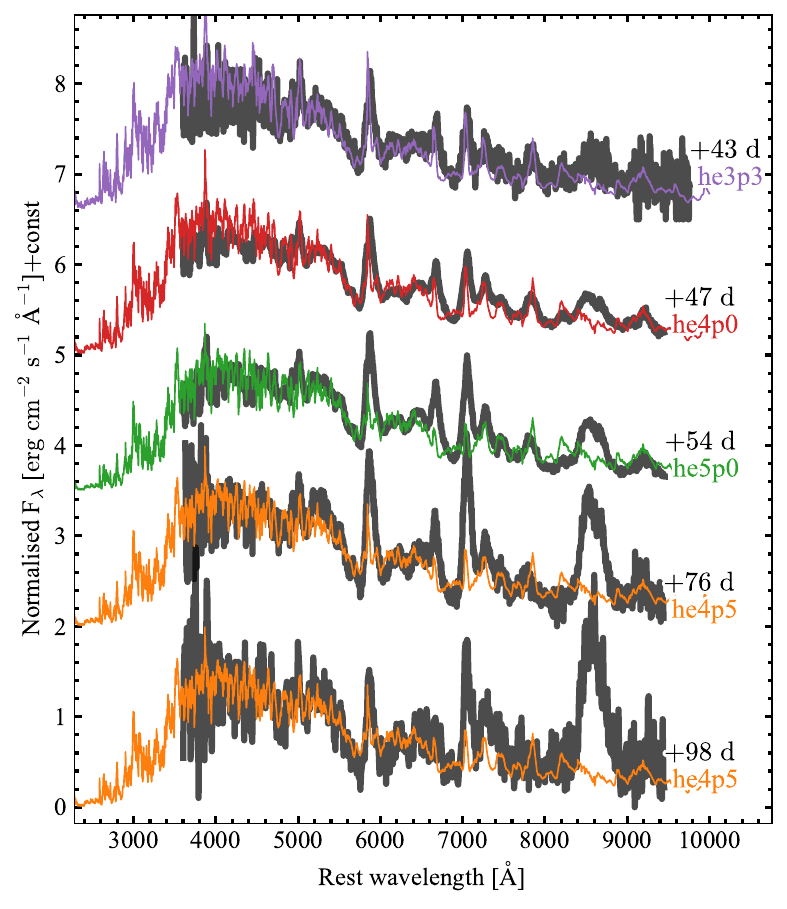}}
\caption{Comparison of optical spectra of SN~2020able with non-LTE radiative-transfer models from \citet{2022A&A...658A.130D}. Spectra were normalised to their flux at 5500~\AA.}
\label{fig:specModels}
\end{figure}

The discrepancy between the ``standard" velocity and the ``extreme" density can be reconciled by a steep density gradient within the un-shocked shell, where the bulk of the scattering mass ($M_{\rm{dense}}\simeq1.8\times10^{-2}\,\rm{M_{\sun}}$) is concentrated near the inner radius (at $R_{\rm{in,dense}}\gtrsim9.3\times10^{14}\,\rm{cm}$), while the P Cygni profiles arise from the lower-density outer layers produced by the steady wind of a WR star.
The evolution of the EW of narrow absorptions further suggests that the inner and outer radii of the outermost shell must be approximately $2.2\times10^{15}$ and $3.3\times10^{15}\,\rm{cm}$, respectively, highlighting a complex stratified environment that mirrors the erratic terminal mass-loss history of its progenitor.

Assuming a 30\% radiative efficiency (consistent with $\tau=6.5$ estimated from electron-scattering profiles) and a filling factor $f=0.1$ \citep[see][]{2023MNRAS.518.5001F}, the conversion of kinetic energy into radiation produced by the interaction of the ejecta with the denser shell would yield a luminosity $L_{\rm{int}}=1^{}/2\,\epsilon\,f\,\dot{M}\,v^3_{\rm{rel}}/{v_{\rm{shell}}}\simeq5.8\times10^{41}\,\rm{erg}\,\rm{s^{-1}}$ \citep[where we assumed $v_{shell}=1.5\times10^3\,\rm{km}\,\rm{s^{-1}}\gtrsim v_{wind}$; e.g.][]{2014ARA&A..52..487S}.
With these assumptions, the interaction power would represent only a minor fraction ($2-3\%$) of the bolometric luminosity at $+16\,\rm{days}$ ($L_{\rm{bol}}=2.45\times10^{43}\,\rm{erg}\,\rm{s^{-1}}$) calculated in Section~\ref{sec:photometry}. 
Even assuming a higher radiative efficiency, a dense medium confined between $9.3\times10^{14}$ and $2.2\times10^{15}\,\rm{cm}$ would be traversed by a shock at a relative velocity of $5.2\times10^3\,\rm{km}\,\rm{s^{-1}}$ in approximately $29\,\rm{days}$. 
The ejecta-shell interaction would therefore be temporally confined within the first $45\,\rm{days}$ post-explosion, assuming the shell is reached at $t\gtrsim+16\,\rm{days}$. 
This is in agreement with the spectroscopic evolution of the transient at $t\gtrsim+16\,\rm{days}$ and $t\gtrsim+43\,\rm{days}$ (see Section~\ref{sec:spectroscopy}). 
A compact shell would increase the contribution of the interaction to the total luminosity of the transient (in any case $\lesssim24$\% assuming $f=1$ and a 30\% efficiency), but this scenario is inconsistent with the observations.
Specifically, the fast photometric evolution, the lack of shocked components in the line profiles (e.g. intermediate-width boxy profiles) and the high ejecta velocities measured at $+43\,\rm{days}$ (see Fig.~\ref{fig:lateSpec}) suggest a highly fragmented dense shell and a marginal impact of the interaction on the total bolometric luminosity at these phases.
This is also highlighted by the comparison with the non-LTE radiative-transfer models of \citet{2022A&A...658A.130D} (see Fig.~\ref{fig:specModels}), which fail to reproduce the broad spectral features observed in SN~2020able at $t\gtrsim+43\,\rm{days}$. 
Although these models assume a spectrum dominated by a narrow-line-forming CDS, the mismatch suggests that ejecta-CSM interaction is not the primary driver of the spectroscopic evolution of the transient at these phases. 
The presence of broad features suggests that the inner, high-velocity ejecta have become the dominant source of emission, also supporting a more massive progenitor scenario characterised by the steep velocity gradients typical of stripped-envelope stars.

A consistent and detailed study of interaction processes in optical transients is of paramount importance, providing key parameters for deciphering the elusive final stages of massive star evolution, whilst also serving the field of multi-messenger Astronomy.
Such analyses are instrumental in mapping the diverse mass-loss histories of these progenitors, allowing us to bridge the gap between their late-stage instabilities and the diversity of observed cosmic explosions.
Furthermore, providing tight constraints on the physical conditions within these environments is essential for identifying which transients possess the characteristics required to accelerate particles to relativistic velocities and efficiently produce high-energy neutrinos.

\section*{Data Availability}
Data shown in Fig.~\ref{fig:lightcurves} are only available in electronic form at the CDS via anonymous ftp to \url{cdsarc.u-strasbg.fr} (130.79.128.5) or via \url{http://cdsweb.u-strasbg.fr/cgi-bin/qcat?J/A+A/}. Light curves for the {\sc csm} and corner plots for the {\sc default} and {\sc csm} {\sc MOSFiT} models are available at \url{https://zenodo.org/records/21678623}.

\begin{acknowledgements}
L.~T., E.~C., A.~P., S.~B., A.~R., G.~V., N.~E.-R. and P.~O. acknowledge support from the PRIN-INAF 2022 ``Shedding light on the nature of gap transients: from the observations to the models". The Las Cumbres Observatory team is supported by NSF grants AST-2308113 and AST-1911151.
Y.-Z.~C is supported by the National Natural Science Foundation of China (No. 12303054), the Yunnan Fundamental Research Projects (Grant Nos. 202401AU070063, 202501AS070078), the National Key Research and Development Program of China (Grant No. 2024YFA1611603) and the International Centre of Supernovae, Yunnan Key Laboratory (No. 202302AN360001). G.~V., A.~R. and Y.-Z.~C. also acknowledge financial support from the SOXS project (PI S.~Campana).
N.E.R. also acknowledges the Spanish Ministerio de Ciencia e Innovaci\'on (MCIN) and the Agencia Estatal de Investigaci\'on (AEI) 10.13039/501100011033 under the program Unidad de Excelencia Mar\'ia de Maeztu CEX2020-001058-M.
C.~P.~G. acknowledges financial support from grant RYC2024-050959-I, funded by MICIU/AEI/10.13039/501100011033 and the FSE+, as well as from projects PID2023-151307NB-I00, PIE 20215AT016, and CEX2020-001058-M, and the MaX-CSIC Excellence Award MaX4-SOMMA-ICE.
B.~K. is supported by the ``Special Project for High-End Foreign Experts", Xingdian Funding from Yunnan Province and the National Key Research and Development Program of China (2024YFA1611603).
D. acknowledges support from STFC grant No. ST/Y002253/1.
Time-domain research by the University of Arizona team and D.~J.~S. is supported by National Science Foundation (NSF) grants 2308181, 2407566, and 2432036. 
{\sc ecsnoopy} is a package for SN photometry using PSF fitting developed by E.~Cappellaro. A package description can be found at \url{https://sngroup.oapd.inaf.it/ecsnoopy.html}. 
Funding for the Sloan Digital Sky Survey IV has been provided by the Alfred P. Sloan Foundation, the U.S. Department of Energy Office of Science, and the Participating Institutions. SDSS-IV acknowledges support and resources from the Center for High Performance Computing at the University of Utah. The SDSS website is \url{www.sdss.org}. 
This work makes use of observations from the Las Cumbres Observatory network.
ZTF is a fully-automated, wide-field survey aimed at a systematic exploration of the optical transient sky, supported by the National Science Foundation under Grants No. AST-1440341 and AST-2034437 and a collaboration including current partners Caltech, IPAC, the Oskar Klein Centre at Stockholm University, the University of Maryland, University of California, Berkeley, University of Wisconsin at Milwaukee, University of Warwick, Ruhr University, Cornell University, Northwestern University and Drexel University. Operations are conducted by COO, IPAC, and UW. 
This research made use of the Spanish Virtual Observatory project, the NASA/IPAC Extragalactic Database (NED) and the HyperLeda database. 
Based on observations collected with the Copernico and Schmidt telescopes (Asiago, Italy) of the INAF - Osservatorio Astronomico di Padova. 
Data were in part obtained with ALFOSC, provided by the Instituto de Astrofisica de Andalucia (IAA) under a joint agreement with the University of Copenhagen and NOT.
The Liverpool Telescope is operated on the island of La Palma by Liverpool John Moores University in the Spanish Observatorio del Roque de los Muchachos of the Instituto de Astrofisica de Canarias with financial support from the UK Science and Technology Facilities Council.
Some observations reported here were obtained at the MMT Observatory, a joint facility of the University of Arizona and the Smithsonian Institution.
Based on observations obtained at the $3.6\,\rm{m}$ Devasthal Optical Telescope (DOT), a National Facility run and managed by Aryabhatta Research Institute of Observational Sciences (ARIES), an autonomous Institute under the Department of Science and Technology, Government of India.
The IAO facilities are operated by the Indian Institute of Astrophysics, Bangalore.
\end{acknowledgements}

\bibliographystyle{aa}
\bibliography{mybib}

@ARTICLE{2018ApJS..236....6G,
       author = {{Guillochon}, James and {Nicholl}, Matt and {Villar}, V. Ashley and {Mockler}, Brenna and {Narayan}, Gautham and {Mandel}, Kaisey S. and {Berger}, Edo and {Williams}, Peter K.~G.},
        title = "{MOSFiT: Modular Open Source Fitter for Transients}",
      journal = {\apjs},
         year = 2018,
        month = may,
       volume = {236},
       number = {1},
          eid = {6},
        pages = {6},
          doi = {10.3847/1538-4365/aab761},
archivePrefix = {arXiv},
       eprint = {1710.02145},
 primaryClass = {astro-ph.IM},
       adsurl = {https://ui.adsabs.harvard.edu/abs/2018ApJS..236....6G}
}

@ARTICLE{2013ApJ...773...76C,
       author = {{Chatzopoulos}, E. and {Wheeler}, J. Craig and {Vinko}, J. and {Horvath}, Z.~L. and {Nagy}, A.},
        title = "{Analytical Light Curve Models of Superluminous Supernovae: {\ensuremath{\chi}}$^{2}$-minimization of Parameter Fits}",
      journal = {\apj},
         year = 2013,
        month = aug,
       volume = {773},
       number = {1},
          eid = {76},
        pages = {76},
          doi = {10.1088/0004-637X/773/1/76},
archivePrefix = {arXiv},
       eprint = {1306.3447},
 primaryClass = {astro-ph.HE},
       adsurl = {https://ui.adsabs.harvard.edu/abs/2013ApJ...773...76C}
}

@ARTICLE{2017ApJ...849...70V,
       author = {{Villar}, V. Ashley and {Berger}, Edo and {Metzger}, Brian D. and {Guillochon}, James},
        title = "{Theoretical Models of Optical Transients. I. A Broad Exploration of the Duration-Luminosity Phase Space}",
      journal = {\apj},
         year = 2017,
        month = nov,
       volume = {849},
       number = {1},
          eid = {70},
        pages = {70},
          doi = {10.3847/1538-4357/aa8fcb},
archivePrefix = {arXiv},
       eprint = {1707.08132},
 primaryClass = {astro-ph.HE},
       adsurl = {https://ui.adsabs.harvard.edu/abs/2017ApJ...849...70V}
}

@ARTICLE{2020RNAAS...4...16J,
       author = {{Jiang}, Brighten and {Jiang}, Shuai and {Ashley Villar}, V.},
        title = "{Extended Self-similar Solution for Circumstellar Material-supernova Ejecta Interaction}",
      journal = {RNAAS},
         year = 2020,
        month = jan,
       volume = {4},
       number = {1},
          eid = {16},
        pages = {16},
          doi = {10.3847/2515-5172/ab7128},
archivePrefix = {arXiv},
       eprint = {2008.10397},
 primaryClass = {astro-ph.HE},
       adsurl = {https://ui.adsabs.harvard.edu/abs/2020RNAAS...4...16J}
}

@ARTICLE{1994ApJS...92..527N,
       author = {{Nadyozhin}, D.~K.},
        title = "{The Properties of NI CO Fe Decay}",
      journal = {\apjs},
         year = 1994,
        month = jun,
       volume = {92},
        pages = {527},
          doi = {10.1086/192008},
       adsurl = {https://ui.adsabs.harvard.edu/abs/1994ApJS...92..527N}
}

@ARTICLE{2009MNRAS.400.2050G,
       author = {{G{\"u}ver}, Tolga and {{\"O}zel}, Feryal},
        title = "{The relation between optical extinction and hydrogen column density in the Galaxy}",
      journal = {\mnras},
         year = 2009,
        month = dec,
       volume = {400},
       number = {4},
        pages = {2050-2053},
          doi = {10.1111/j.1365-2966.2009.15598.x},
archivePrefix = {arXiv},
       eprint = {0903.2057},
 primaryClass = {astro-ph.GA},
       adsurl = {https://ui.adsabs.harvard.edu/abs/2009MNRAS.400.2050G}
}

@ARTICLE{2020A&A...642A.106D,
       author = {{Dessart}, Luc and {Yoon}, Sung-Chul and {Aguilera-Dena}, David R. and {Langer}, Norbert},
        title = "{Supernovae Ib and Ic from the explosion of helium stars}",
      journal = {\aap},
         year = 2020,
        month = oct,
       volume = {642},
          eid = {A106},
        pages = {A106},
          doi = {10.1051/0004-6361/202038763},
archivePrefix = {arXiv},
       eprint = {2008.07601},
 primaryClass = {astro-ph.SR},
       adsurl = {https://ui.adsabs.harvard.edu/abs/2020A&A...642A.106D}
}

@ARTICLE{2020MNRAS.493.3132S,
       author = {{Speagle}, Joshua S.},
        title = "{DYNESTY: a dynamic nested sampling package for estimating Bayesian posteriors and evidences}",
      journal = {\mnras},
         year = 2020,
        month = apr,
       volume = {493},
       number = {3},
        pages = {3132-3158},
          doi = {10.1093/mnras/staa278},
archivePrefix = {arXiv},
       eprint = {1904.02180},
 primaryClass = {astro-ph.IM},
       adsurl = {https://ui.adsabs.harvard.edu/abs/2020MNRAS.493.3132S}
}

@ARTICLE{2016MNRAS.456..853P,
       author = {{Pastorello}, A. and {Wang}, X. -F. and {Ciabattari}, F. and {Bersier}, D. and {Mazzali}, P.~A. and {Gao}, X. and {Xu}, Z. and {Zhang}, J. -J. and {Tokuoka}, S. and {Benetti}, S. and {Cappellaro}, E. and {Elias-Rosa}, N. and {Harutyunyan}, A. and {Huang}, F. and {Miluzio}, M. and {Mo}, J. and {Ochner}, P. and {Tartaglia}, L. and {Terreran}, G. and {Tomasella}, L. and {Turatto}, M.},
        title = "{Massive stars exploding in a He-rich circumstellar medium - IX. SN 2014av, and characterization of Type Ibn SNe}",
      journal = {\mnras},
         year = 2016,
        month = feb,
       volume = {456},
       number = {1},
        pages = {853-869},
          doi = {10.1093/mnras/stv2634},
archivePrefix = {arXiv},
       eprint = {1509.09069},
 primaryClass = {astro-ph.SR},
       adsurl = {https://ui.adsabs.harvard.edu/abs/2016MNRAS.456..853P}
}

@ARTICLE{2019ApJ...871L...9H,
       author = {{Hosseinzadeh}, Griffin and {McCully}, Curtis and {Zabludoff}, Ann I. and {Arcavi}, Iair and {French}, K. Decker and {Howell}, D. Andrew and {Berger}, Edo and {Hiramatsu}, Daichi},
        title = "{Type Ibn Supernovae May not all Come from Massive Stars}",
      journal = {\apjl},
         year = 2019,
        month = jan,
       volume = {871},
       number = {1},
          eid = {L9},
        pages = {L9},
          doi = {10.3847/2041-8213/aafc61},
archivePrefix = {arXiv},
       eprint = {1901.03332},
 primaryClass = {astro-ph.HE},
       adsurl = {https://ui.adsabs.harvard.edu/abs/2019ApJ...871L...9H}
}

@ARTICLE{2011ApJ...737..103S,
       author = {{Schlafly}, Edward F. and {Finkbeiner}, Douglas P.},
        title = "{Measuring Reddening with Sloan Digital Sky Survey Stellar Spectra and Recalibrating SFD}",
      journal = {\apj},
         year = 2011,
        month = aug,
       volume = {737},
       number = {2},
          eid = {103},
        pages = {103},
          doi = {10.1088/0004-637X/737/2/103},
archivePrefix = {arXiv},
       eprint = {1012.4804},
 primaryClass = {astro-ph.GA},
       adsurl = {https://ui.adsabs.harvard.edu/abs/2011ApJ...737..103S}
}

@ARTICLE{2012MNRAS.426.1465P,
       author = {{Poznanski}, Dovi and {Prochaska}, J. Xavier and {Bloom}, Joshua S.},
        title = "{An empirical relation between sodium absorption and dust extinction}",
      journal = {\mnras},
         year = 2012,
        month = oct,
       volume = {426},
       number = {2},
        pages = {1465-1474},
          doi = {10.1111/j.1365-2966.2012.21796.x},
archivePrefix = {arXiv},
       eprint = {1206.6107},
 primaryClass = {astro-ph.IM},
       adsurl = {https://ui.adsabs.harvard.edu/abs/2012MNRAS.426.1465P}
}

@ARTICLE{1996ApJ...473..576F,
       author = {{Fixsen}, D.~J. and {Cheng}, E.~S. and {Gales}, J.~M. and {Mather}, J.~C. and {Shafer}, R.~A. and {Wright}, E.~L.},
        title = "{The Cosmic Microwave Background Spectrum from the Full COBE FIRAS Data Set}",
      journal = {\apj},
         year = 1996,
        month = dec,
       volume = {473},
        pages = {576},
          doi = {10.1086/178173},
archivePrefix = {arXiv},
       eprint = {astro-ph/9605054},
 primaryClass = {astro-ph},
       adsurl = {https://ui.adsabs.harvard.edu/abs/1996ApJ...473..576F}
}

@ARTICLE{2020ApJ...889..170G,
       author = {{Gangopadhyay}, Anjasha and {Misra}, Kuntal and {Hiramatsu}, Daichi and {Wang}, Shan-Qin and {Hosseinzadeh}, Griffin and {Wang}, Xiaofeng and {Valenti}, Stefano and {Zhang}, Jujia and {Howell}, D. Andrew and {Arcavi}, Iair and {Anupama}, G.~C. and {Burke}, Jamison and {Dastidar}, Raya and {Itagaki}, Koichi and {Kumar}, Brajesh and {Kumar}, Brijesh and {Li}, Long and {McCully}, Curtis and {Mo}, Jun and {Pandey}, Shashi Bhushan and {Pellegrino}, Craig and {Sai}, Hanna and {Sahu}, D.~K. and {Sanwal}, Pankaj and {Singh}, Avinash and {Singh}, Mridweeka and {Zhang}, Jicheng and {Zhang}, Tianmeng and {Zhang}, Xinhan},
        title = "{Flash Ionization Signatures in the Type Ibn Supernova SN 2019uo}",
      journal = {\apj},
         year = 2020,
        month = feb,
       volume = {889},
       number = {2},
          eid = {170},
        pages = {170},
          doi = {10.3847/1538-4357/ab6328},
archivePrefix = {arXiv},
       eprint = {1912.07878},
 primaryClass = {astro-ph.HE},
       adsurl = {https://ui.adsabs.harvard.edu/abs/2020ApJ...889..170G}
}

@ARTICLE{2022ApJ...930..127G,
       author = {{Gangopadhyay}, Anjasha and {Misra}, Kuntal and {Hosseinzadeh}, Griffin and {Arcavi}, Iair and {Pellegrino}, Craig and {Wang}, Xiaofeng and {Andrew Howell}, D. and {Burke}, Jamison and {Zhang}, Jujia and {Kawabata}, Koji and {Singh}, Mridweeka and {Dastidar}, Raya and {Hiramatsu}, Daichi and {McCully}, Curtis and {Mo}, Jun and {Chen}, Zhihao and {Xiang}, Danfeng},
        title = "{Evolution of a Peculiar Type Ibn Supernova SN 2019wep}",
      journal = {\apj},
         year = 2022,
        month = may,
       volume = {930},
       number = {2},
          eid = {127},
        pages = {127},
          doi = {10.3847/1538-4357/ac6187},
archivePrefix = {arXiv},
       eprint = {2203.15194},
 primaryClass = {astro-ph.HE},
       adsurl = {https://ui.adsabs.harvard.edu/abs/2022ApJ...930..127G}
}

@ARTICLE{2015MNRAS.449.1921P,
       author = {{Pastorello}, A. and {Benetti}, S. and {Brown}, P.~J. and {Tsvetkov}, D.~Y. and {Inserra}, C. and {Taubenberger}, S. and {Tomasella}, L. and {Fraser}, M. and {Rich}, D.~J. and {Botticella}, M.~T. and {Bufano}, F. and {Cappellaro}, E. and {Ergon}, M. and {Gorbovskoy}, E.~S. and {Harutyunyan}, A. and {Huang}, F. and {Kotak}, R. and {Lipunov}, V.~M. and {Magill}, L. and {Miluzio}, M. and {Morrell}, N. and {Ochner}, P. and {Smartt}, S.~J. and {Sollerman}, J. and {Spiro}, S. and {Stritzinger}, M.~D. and {Turatto}, M. and {Valenti}, S. and {Wang}, X. and {Wright}, D.~E. and {Yurkov}, V.~V. and {Zampieri}, L. and {Zhang}, T.},
        title = "{Massive stars exploding in a He-rich circumstellar medium - IV. Transitional Type Ibn supernovae}",
      journal = {\mnras},
         year = 2015,
        month = may,
       volume = {449},
       number = {2},
        pages = {1921-1940},
          doi = {10.1093/mnras/stu2745},
archivePrefix = {arXiv},
       eprint = {1502.04946},
 primaryClass = {astro-ph.SR},
       adsurl = {https://ui.adsabs.harvard.edu/abs/2015MNRAS.449.1921P}
}

@ARTICLE{2024A&A...691A.156W,
       author = {{Wang}, Z. -Y. and {Pastorello}, A. and {Maeda}, K. and {Reguitti}, A. and {Cai}, Y. -Z. and {Andrew Howell}, D. and {Benetti}, S. and {Buckley}, D.~A.~H. and {Cappellaro}, E. and {Carini}, R. and {Cartier}, R. and {Chen}, T. -W. and {Elias-Rosa}, N. and {Fang}, Q. -L. and {Gal-Yam}, A. and {Gangopadhyay}, A. and {Gromadzki}, M. and {Gan}, W. -P. and {Hiramatsu}, D. and {Hu}, M. -K. and {Inserra}, C. and {McCully}, C. and {Nicholl}, M. and {Olivares E.}, F. and {Pignata}, G. and {Pineda-Garc{\'\i}a}, J. and {Pursiainen}, M. and {Ragosta}, F. and {Rau}, A. and {Roy}, R. and {Sollerman}, J. and {Tartaglia}, L. and {Terreran}, G. and {Valerin}, G. and {Wang}, Q. and {Wang}, S. -Q. and {Young}, D.~R. and {Aryan}, A. and {Bronikowski}, M. and {Concepcion}, E. and {Galbany}, L. and {Lin}, H. and {Melandri}, A. and {Petrushevska}, T. and {Ramirez}, M. and {Shi}, D. -D. and {Warwick}, B. and {Zhang}, J. -J. and {Wang}, B. and {Wang}, X. -F. and {Zhu}, X. -J.},
        title = "{Massive stars exploding in a He-rich circumstellar medium: X. Flash spectral features in the Type Ibn SN 2019cj and observations of SN 2018jmt}",
      journal = {\aap},
         year = 2024,
        month = nov,
       volume = {691},
          eid = {A156},
        pages = {A156},
          doi = {10.1051/0004-6361/202451131},
archivePrefix = {arXiv},
       eprint = {2408.12393},
 primaryClass = {astro-ph.HE},
       adsurl = {https://ui.adsabs.harvard.edu/abs/2024A&A...691A.156W}
}

@ARTICLE{2023ApJ...959L..10P,
       author = {{Pursiainen}, M. and {Leloudas}, G. and {Schulze}, S. and {Charalampopoulos}, P. and {Angus}, C.~R. and {Anderson}, J.~P. and {Bauer}, F. and {Chen}, T. -W. and {Galbany}, L. and {Gromadzki}, M. and {Guti{\'e}rrez}, C.~P. and {Inserra}, C. and {Lyman}, J. and {M{\"u}ller-Bravo}, T.~E. and {Nicholl}, M. and {Smartt}, S.~J. and {Tartaglia}, L. and {Wiseman}, P. and {Young}, D.~R.},
        title = "{SN 2023emq: A Flash-ionized Ibn Supernova with Possible C III Emission}",
      journal = {\apjl},
         year = 2023,
        month = dec,
       volume = {959},
       number = {1},
          eid = {L10},
        pages = {L10},
          doi = {10.3847/2041-8213/ad103d},
archivePrefix = {arXiv},
       eprint = {2306.09804},
 primaryClass = {astro-ph.HE},
       adsurl = {https://ui.adsabs.harvard.edu/abs/2023ApJ...959L..10P}
}

@BOOK{2006agna.book.....O,
       author = {{Osterbrock}, Donald E. and {Ferland}, Gary J.},
        title = "{Astrophysics of gaseous nebulae and active galactic nuclei}",
    publisher = "{University Science Books}",
         year = 2006,
       adsurl = {https://ui.adsabs.harvard.edu/abs/2006agna.book.....O}
}

@ARTICLE{2007A&A...464..715S,
       author = {{Selvelli}, P. and {Danziger}, J. and {Bonifacio}, P.},
        title = "{The He$_{II}$ Fowler lines and the O$_{III}$ and N$_{III}$ Bowen fluorescence lines in the symbiotic nova RR Telescopii}",
      journal = {\aap},
         year = 2007,
        month = mar,
       volume = {464},
       number = {2},
        pages = {715-734},
          doi = {10.1051/0004-6361:20066175},
archivePrefix = {arXiv},
       eprint = {astro-ph/0611943},
 primaryClass = {astro-ph},
       adsurl = {https://ui.adsabs.harvard.edu/abs/2007A&A...464..715S}
}

@ARTICLE{2009A&A...508.1259H,
       author = {{Hakobyan}, A.~A. and {Mamon}, G.~A. and {Petrosian}, A.~R. and {Kunth}, D. and {Turatto}, M.},
        title = "{The radial distribution of core-collapse supernovae in spiral host galaxies}",
      journal = {\aap},
         year = 2009,
        month = dec,
       volume = {508},
       number = {3},
        pages = {1259-1268},
          doi = {10.1051/0004-6361/200912795},
archivePrefix = {arXiv},
       eprint = {0910.1801},
 primaryClass = {astro-ph.CO},
       adsurl = {https://ui.adsabs.harvard.edu/abs/2009A&A...508.1259H}
}

@ARTICLE{2013A&A...559A.114M,
       author = {{Marino}, R.~A. and {Rosales-Ortega}, F.~F. and {S{\'a}nchez}, S.~F. and {Gil de Paz}, A. and {V{\'\i}lchez}, J. and {Miralles-Caballero}, D. and {Kehrig}, C. and {P{\'e}rez-Montero}, E. and {Stanishev}, V. and {Iglesias-P{\'a}ramo}, J. and {D{\'\i}az}, A.~I. and {Castillo-Morales}, A. and {Kennicutt}, R. and {L{\'o}pez-S{\'a}nchez}, A.~R. and {Galbany}, L. and {Garc{\'\i}a-Benito}, R. and {Mast}, D. and {Mendez-Abreu}, J. and {Monreal-Ibero}, A. and {Husemann}, B. and {Walcher}, C.~J. and {Garc{\'\i}a-Lorenzo}, B. and {Masegosa}, J. and {Del Olmo Orozco}, A. and {Mour{\~a}o}, A.~M. and {Ziegler}, B. and {Moll{\'a}}, M. and {Papaderos}, P. and {S{\'a}nchez-Bl{\'a}zquez}, P. and {Gonz{\'a}lez Delgado}, R.~M. and {Falc{\'o}n-Barroso}, J. and {Roth}, M.~M. and {van de Ven}, G. and {CALIFA Team}},
        title = "{The O3N2 and N2 abundance indicators revisited: improved calibrations based on CALIFA and T$_{e}$-based literature data}",
      journal = {\aap},
         year = 2013,
        month = nov,
       volume = {559},
          eid = {A114},
        pages = {A114},
          doi = {10.1051/0004-6361/201321956},
archivePrefix = {arXiv},
       eprint = {1307.5316},
 primaryClass = {astro-ph.CO},
       adsurl = {https://ui.adsabs.harvard.edu/abs/2013A&A...559A.114M}
}

@ARTICLE{1979A&A....78..200A,
       author = {{Alloin}, D. and {Collin-Souffrin}, S. and {Joly}, M. and {Vigroux}, L.},
        title = "{Nitrogen and oxygen abundances in galaxies.}",
      journal = {\aap},
         year = 1979,
        month = sep,
       volume = {78},
        pages = {200-216},
       adsurl = {https://ui.adsabs.harvard.edu/abs/1979A&A....78..200A}
}

@ARTICLE{2002MNRAS.330...69D,
       author = {{Denicol{\'o}}, Glenda and {Terlevich}, Roberto and {Terlevich}, Elena},
        title = "{New light on the search for low-metallicity galaxies - I. The N2 calibrator}",
      journal = {\mnras},
         year = 2002,
        month = feb,
       volume = {330},
       number = {1},
        pages = {69-74},
          doi = {10.1046/j.1365-8711.2002.05041.x},
archivePrefix = {arXiv},
       eprint = {astro-ph/0110356},
 primaryClass = {astro-ph},
       adsurl = {https://ui.adsabs.harvard.edu/abs/2002MNRAS.330...69D}
}

@ARTICLE{2016A&A...591A..48G,
       author = {{Galbany}, L. and {Stanishev}, V. and {Mour{\~a}o}, A.~M. and {Rodrigues}, M. and {Flores}, H. and {Walcher}, C.~J. and {S{\'a}nchez}, S.~F. and {Garc{\'\i}a-Benito}, R. and {Mast}, D. and {Badenes}, C. and {Gonz{\'a}lez Delgado}, R.~M. and {Kehrig}, C. and {Lyubenova}, M. and {Marino}, R.~A. and {Moll{\'a}}, M. and {Meidt}, S. and {P{\'e}rez}, E. and {van de Ven}, G. and {V{\'\i}lchez}, J.~M.},
        title = "{Nearby supernova host galaxies from the CALIFA survey. II. Supernova environmental metallicity}",
      journal = {\aap},
         year = 2016,
        month = jun,
       volume = {591},
          eid = {A48},
        pages = {A48},
          doi = {10.1051/0004-6361/201528045},
archivePrefix = {arXiv},
       eprint = {1603.07808},
 primaryClass = {astro-ph.GA},
       adsurl = {https://ui.adsabs.harvard.edu/abs/2016A&A...591A..48G}
}

@ARTICLE{2000AJ....120.1579Y,
       author = {{York}, Donald G. and {Adelman}, J. and {Anderson}, Jr., John E. and {Anderson}, Scott F. and {Annis}, James and {Bahcall}, Neta A. and {Bakken}, J.~A. and {Barkhouser}, Robert and {Bastian}, Steven and {Berman}, Eileen and {Boroski}, William N. and {Bracker}, Steve and {Briegel}, Charlie and {Briggs}, John W. and {Brinkmann}, J. and {Brunner}, Robert and {Burles}, Scott and {Carey}, Larry and {Carr}, Michael A. and {Castander}, Francisco J. and {Chen}, Bing and {Colestock}, Patrick L. and {Connolly}, A.~J. and {Crocker}, J.~H. and {Csabai}, Istv{\'a}n and {Czarapata}, Paul C. and {Davis}, John Eric and {Doi}, Mamoru and {Dombeck}, Tom and {Eisenstein}, Daniel and {Ellman}, Nancy and {Elms}, Brian R. and {Evans}, Michael L. and {Fan}, Xiaohui and {Federwitz}, Glenn R. and {Fiscelli}, Larry and {Friedman}, Scott and {Frieman}, Joshua A. and {Fukugita}, Masataka and {Gillespie}, Bruce and {Gunn}, James E. and {Gurbani}, Vijay K. and {de Haas}, Ernst and {Haldeman}, Merle and {Harris}, Frederick H. and {Hayes}, J. and {Heckman}, Timothy M. and {Hennessy}, G.~S. and {Hindsley}, Robert B. and {Holm}, Scott and {Holmgren}, Donald J. and {Huang}, Chi-hao and {Hull}, Charles and {Husby}, Don and {Ichikawa}, Shin-Ichi and {Ichikawa}, Takashi and {Ivezi{\'c}}, {\v{Z}}eljko and {Kent}, Stephen and {Kim}, Rita S.~J. and {Kinney}, E. and {Klaene}, Mark and {Kleinman}, A.~N. and {Kleinman}, S. and {Knapp}, G.~R. and {Korienek}, John and {Kron}, Richard G. and {Kunszt}, Peter Z. and {Lamb}, D.~Q. and {Lee}, B. and {Leger}, R. French and {Limmongkol}, Siriluk and {Lindenmeyer}, Carl and {Long}, Daniel C. and {Loomis}, Craig and {Loveday}, Jon and {Lucinio}, Rich and {Lupton}, Robert H. and {MacKinnon}, Bryan and {Mannery}, Edward J. and {Mantsch}, P.~M. and {Margon}, Bruce and {McGehee}, Peregrine and {McKay}, Timothy A. and {Meiksin}, Avery and {Merelli}, Aronne and {Monet}, David G. and {Munn}, Jeffrey A. and {Narayanan}, Vijay K. and {Nash}, Thomas and {Neilsen}, Eric and {Neswold}, Rich and {Newberg}, Heidi Jo and {Nichol}, R.~C. and {Nicinski}, Tom and {Nonino}, Mario and {Okada}, Norio and {Okamura}, Sadanori and {Ostriker}, Jeremiah P. and {Owen}, Russell and {Pauls}, A. George and {Peoples}, John and {Peterson}, R.~L. and {Petravick}, Donald and {Pier}, Jeffrey R. and {Pope}, Adrian and {Pordes}, Ruth and {Prosapio}, Angela and {Rechenmacher}, Ron and {Quinn}, Thomas R. and {Richards}, Gordon T. and {Richmond}, Michael W. and {Rivetta}, Claudio H. and {Rockosi}, Constance M. and {Ruthmansdorfer}, Kurt and {Sandford}, Dale and {Schlegel}, David J. and {Schneider}, Donald P. and {Sekiguchi}, Maki and {Sergey}, Gary and {Shimasaku}, Kazuhiro and {Siegmund}, Walter A. and {Smee}, Stephen and {Smith}, J. Allyn and {Snedden}, S. and {Stone}, R. and {Stoughton}, Chris and {Strauss}, Michael A. and {Stubbs}, Christopher and {SubbaRao}, Mark and {Szalay}, Alexander S. and {Szapudi}, Istvan and {Szokoly}, Gyula P. and {Thakar}, Anirudda R. and {Tremonti}, Christy and {Tucker}, Douglas L. and {Uomoto}, Alan and {Vanden Berk}, Dan and {Vogeley}, Michael S. and {Waddell}, Patrick and {Wang}, Shu-i. and {Watanabe}, Masaru and {Weinberg}, David H. and {Yanny}, Brian and {Yasuda}, Naoki and {SDSS Collaboration}},
        title = "{The Sloan Digital Sky Survey: Technical Summary}",
      journal = {\aj},
         year = 2000,
        month = sep,
       volume = {120},
       number = {3},
        pages = {1579-1587},
          doi = {10.1086/301513},
archivePrefix = {arXiv},
       eprint = {astro-ph/0006396},
 primaryClass = {astro-ph},
       adsurl = {https://ui.adsabs.harvard.edu/abs/2000AJ....120.1579Y}
}

@ARTICLE{2015ApJS..219...12A,
       author = {{Alam}, Shadab and {Albareti}, Franco D. and {Allende Prieto}, Carlos and {Anders}, F. and {Anderson}, Scott F. and {Anderton}, Timothy and {Andrews}, Brett H. and {Armengaud}, Eric and {Aubourg}, {\'E}ric and {Bailey}, Stephen and {Basu}, Sarbani and {Bautista}, Julian E. and {Beaton}, Rachael L. and {Beers}, Timothy C. and {Bender}, Chad F. and {Berlind}, Andreas A. and {Beutler}, Florian and {Bhardwaj}, Vaishali and {Bird}, Jonathan C. and {Bizyaev}, Dmitry and {Blake}, Cullen H. and {Blanton}, Michael R. and {Blomqvist}, Michael and {Bochanski}, John J. and {Bolton}, Adam S. and {Bovy}, Jo and {Shelden Bradley}, A. and {Brandt}, W.~N. and {Brauer}, D.~E. and {Brinkmann}, J. and {Brown}, Peter J. and {Brownstein}, Joel R. and {Burden}, Angela and {Burtin}, Etienne and {Busca}, Nicol{\'a}s G. and {Cai}, Zheng and {Capozzi}, Diego and {Carnero Rosell}, Aurelio and {Carr}, Michael A. and {Carrera}, Ricardo and {Chambers}, K.~C. and {Chaplin}, William James and {Chen}, Yen-Chi and {Chiappini}, Cristina and {Chojnowski}, S. Drew and {Chuang}, Chia-Hsun and {Clerc}, Nicolas and {Comparat}, Johan and {Covey}, Kevin and {Croft}, Rupert A.~C. and {Cuesta}, Antonio J. and {Cunha}, Katia and {da Costa}, Luiz N. and {Da Rio}, Nicola and {Davenport}, James R.~A. and {Dawson}, Kyle S. and {De Lee}, Nathan and {Delubac}, Timoth{\'e}e and {Deshpande}, Rohit and {Dhital}, Saurav and {Dutra-Ferreira}, Let{\'\i}cia and {Dwelly}, Tom and {Ealet}, Anne and {Ebelke}, Garrett L. and {Edmondson}, Edward M. and {Eisenstein}, Daniel J. and {Ellsworth}, Tristan and {Elsworth}, Yvonne and {Epstein}, Courtney R. and {Eracleous}, Michael and {Escoffier}, Stephanie and {Esposito}, Massimiliano and {Evans}, Michael L. and {Fan}, Xiaohui and {Fern{\'a}ndez-Alvar}, Emma and {Feuillet}, Diane and {Filiz Ak}, Nurten and {Finley}, Hayley and {Finoguenov}, Alexis and {Flaherty}, Kevin and {Fleming}, Scott W. and {Font-Ribera}, Andreu and {Foster}, Jonathan and {Frinchaboy}, Peter M. and {Galbraith-Frew}, J.~G. and {Garc{\'\i}a}, Rafael A. and {Garc{\'\i}a-Hern{\'a}ndez}, D.~A. and {Garc{\'\i}a P{\'e}rez}, Ana E. and {Gaulme}, Patrick and {Ge}, Jian and {G{\'e}nova-Santos}, R. and {Georgakakis}, A. and {Ghezzi}, Luan and {Gillespie}, Bruce A. and {Girardi}, L{\'e}o and {Goddard}, Daniel and {Gontcho}, Satya Gontcho A. and {Gonz{\'a}lez Hern{\'a}ndez}, Jonay I. and {Grebel}, Eva K. and {Green}, Paul J. and {Grieb}, Jan Niklas and {Grieves}, Nolan and {Gunn}, James E. and {Guo}, Hong and {Harding}, Paul and {Hasselquist}, Sten and {Hawley}, Suzanne L. and {Hayden}, Michael and {Hearty}, Fred R. and {Hekker}, Saskia and {Ho}, Shirley and {Hogg}, David W. and {Holley-Bockelmann}, Kelly and {Holtzman}, Jon A. and {Honscheid}, Klaus and {Huber}, Daniel and {Huehnerhoff}, Joseph and {Ivans}, Inese I. and {Jiang}, Linhua and {Johnson}, Jennifer A. and {Kinemuchi}, Karen and {Kirkby}, David and {Kitaura}, Francisco and {Klaene}, Mark A. and {Knapp}, Gillian R. and {Kneib}, Jean-Paul and {Koenig}, Xavier P. and {Lam}, Charles R. and {Lan}, Ting-Wen and {Lang}, Dustin and {Laurent}, Pierre and {Le Goff}, Jean-Marc and {Leauthaud}, Alexie and {Lee}, Khee-Gan and {Lee}, Young Sun and {Licquia}, Timothy C. and {Liu}, Jian and {Long}, Daniel C. and {L{\'o}pez-Corredoira}, Mart{\'\i}n and {Lorenzo-Oliveira}, Diego and {Lucatello}, Sara and {Lundgren}, Britt and {Lupton}, Robert H. and {Mack}, III, Claude E. and {Mahadevan}, Suvrath and {Maia}, Marcio A.~G. and {Majewski}, Steven R. and {Malanushenko}, Elena and {Malanushenko}, Viktor and {Manchado}, A. and {Manera}, Marc and {Mao}, Qingqing and {Maraston}, Claudia and {Marchwinski}, Robert C. and {Margala}, Daniel and {Martell}, Sarah L. and {Martig}, Marie and {Masters}, Karen L. and {Mathur}, Savita and {McBride}, Cameron K. and {McGehee}, Peregrine M. and {McGreer}, Ian D. and {McMahon}, Richard G. and {M{\'e}nard}, Brice and {Menzel}, Marie-Luise and {Merloni}, Andrea and {M{\'e}sz{\'a}ros}, Szabolcs and {Miller}, Adam A. and {Miralda-Escud{\'e}}, Jordi and {Miyatake}, Hironao and {Montero-Dorta}, Antonio D. and {More}, Surhud and {Morganson}, Eric and {Morice-Atkinson}, Xan and {Morrison}, Heather L. and {Mosser}, Ben{\^o}it and {Muna}, Demitri and {Myers}, Adam D. and {Nandra}, Kirpal and {Newman}, Jeffrey A. and {Neyrinck}, Mark and {Nguyen}, Duy Cuong and {Nichol}, Robert C. and {Nidever}, David L. and {Noterdaeme}, Pasquier and {Nuza}, Sebasti{\'a}n E. and {O'Connell}, Julia E. and {O'Connell}, Robert W. and {O'Connell}, Ross and {Ogando}, Ricardo L.~C. and {Olmstead}, Matthew D. and {Oravetz}, Audrey E. and {Oravetz}, Daniel J. and {Osumi}, Keisuke and {Owen}, Russell and {Padgett}, Deborah L. and {Padmanabhan}, Nikhil and {Paegert}, Martin and {Palanque-Delabrouille}, Nathalie and {Pan}, Kaike},
        title = "{The Eleventh and Twelfth Data Releases of the Sloan Digital Sky Survey: Final Data from SDSS-III}",
      journal = {\apjs},
         year = 2015,
        month = jul,
       volume = {219},
       number = {1},
          eid = {12},
        pages = {12},
          doi = {10.1088/0067-0049/219/1/12},
archivePrefix = {arXiv},
       eprint = {1501.00963},
 primaryClass = {astro-ph.IM},
       adsurl = {https://ui.adsabs.harvard.edu/abs/2015ApJS..219...12A}
}

@ARTICLE{2000A&AS..146...19P,
       author = {{Paturel}, G. and {Fang}, Y. and {Petit}, C. and {Garnier}, R. and {Rousseau}, J.},
        title = "{An image database. III. Automatic extraction for millions of galaxies}",
      journal = {\aaps},
         year = 2000,
        month = oct,
       volume = {146},
        pages = {19-29},
          doi = {10.1051/aas:2000358},
       adsurl = {https://ui.adsabs.harvard.edu/abs/2000A&AS..146...19P}
}

@ARTICLE{1997A&AS..124..109P,
       author = {{Paturel}, G. and {Andernach}, H. and {Bottinelli}, L. and {di Nella}, H. and {Durand}, N. and {Garnier}, R. and {Gouguenheim}, L. and {Lanoix}, P. and {Marthinet}, M.~C. and {Petit}, C. and {Rousseau}, J. and {Theureau}, G. and {Vauglin}, I.},
        title = "{Extragalactic database. VII. Reduction of astrophysical parameters}",
      journal = {\aaps},
         year = 1997,
        month = jul,
       volume = {124},
        pages = {109-122},
          doi = {10.1051/aas:1997354},
archivePrefix = {arXiv},
       eprint = {astro-ph/9806140},
 primaryClass = {astro-ph},
       adsurl = {https://ui.adsabs.harvard.edu/abs/1997A&AS..124..109P}
}

@ARTICLE{2020TNSTR3630....1T,
       author = {{Tonry}, J. and {Denneau}, L. and {Heinze}, A. and {Weiland}, H. and {Stalder}, B. and {Rest}, A. and {Stubbs}, C. and {Smith}, K.~W. and {Smartt}, S.~J. and {Young}, D.~R. and {Srivastav}, S. and {McBrien}, O. and {Fulton}, M. and {Gillanders}, J. and {Chen}, T.~W. and {Wright}, D.~E. and {Anderson}, J.},
        title = "{ATLAS Transient Discovery Report for 2020-12-02}",
      journal = {TNS Discovery Report},
         year = 2020,
        month = dec,
       volume = {2020-3630},
        pages = {1},
       adsurl = {https://ui.adsabs.harvard.edu/abs/2020TNSTR3630....1T}
}

@ARTICLE{2020PASP..132h5002S,
       author = {{Smith}, K.~W. and {Smartt}, S.~J. and {Young}, D.~R. and {Tonry}, J.~L. and {Denneau}, L. and {Flewelling}, H. and {Heinze}, A.~N. and {Weiland}, H.~J. and {Stalder}, B. and {Rest}, A. and {Stubbs}, C.~W. and {Anderson}, J.~P. and {Chen}, T. -W. and {Clark}, P. and {Do}, A. and {F{\"o}rster}, F. and {Fulton}, M. and {Gillanders}, J. and {McBrien}, O.~R. and {O'Neill}, D. and {Srivastav}, S. and {Wright}, D.~E.},
        title = "{Design and Operation of the ATLAS Transient Science Server}",
      journal = {\pasp},
         year = 2020,
        month = aug,
       volume = {132},
       number = {1014},
          eid = {085002},
        pages = {085002},
          doi = {10.1088/1538-3873/ab936e},
archivePrefix = {arXiv},
       eprint = {2003.09052},
 primaryClass = {astro-ph.IM},
       adsurl = {https://ui.adsabs.harvard.edu/abs/2020PASP..132h5002S}
}

@ARTICLE{2016arXiv161205560C,
       author = {{Chambers}, K.~C. and {Magnier}, E.~A. and {Metcalfe}, N. and {Flewelling}, H.~A. and {Huber}, M.~E. and {Waters}, C.~Z. and {Denneau}, L. and {Draper}, P.~W. and {Farrow}, D. and {Finkbeiner}, D.~P. and {Holmberg}, C. and {Koppenhoefer}, J. and {Price}, P.~A. and {Rest}, A. and {Saglia}, R.~P. and {Schlafly}, E.~F. and {Smartt}, S.~J. and {Sweeney}, W. and {Wainscoat}, R.~J. and {Burgett}, W.~S. and {Chastel}, S. and {Grav}, T. and {Heasley}, J.~N. and {Hodapp}, K.~W. and {Jedicke}, R. and {Kaiser}, N. and {Kudritzki}, R. -P. and {Luppino}, G.~A. and {Lupton}, R.~H. and {Monet}, D.~G. and {Morgan}, J.~S. and {Onaka}, P.~M. and {Shiao}, B. and {Stubbs}, C.~W. and {Tonry}, J.~L. and {White}, R. and {Ba{\~n}ados}, E. and {Bell}, E.~F. and {Bender}, R. and {Bernard}, E.~J. and {Boegner}, M. and {Boffi}, F. and {Botticella}, M.~T. and {Calamida}, A. and {Casertano}, S. and {Chen}, W. -P. and {Chen}, X. and {Cole}, S. and {Deacon}, N. and {Frenk}, C. and {Fitzsimmons}, A. and {Gezari}, S. and {Gibbs}, V. and {Goessl}, C. and {Goggia}, T. and {Gourgue}, R. and {Goldman}, B. and {Grant}, P. and {Grebel}, E.~K. and {Hambly}, N.~C. and {Hasinger}, G. and {Heavens}, A.~F. and {Heckman}, T.~M. and {Henderson}, R. and {Henning}, T. and {Holman}, M. and {Hopp}, U. and {Ip}, W. -H. and {Isani}, S. and {Jackson}, M. and {Keyes}, C.~D. and {Koekemoer}, A.~M. and {Kotak}, R. and {Le}, D. and {Liska}, D. and {Long}, K.~S. and {Lucey}, J.~R. and {Liu}, M. and {Martin}, N.~F. and {Masci}, G. and {McLean}, B. and {Mindel}, E. and {Misra}, P. and {Morganson}, E. and {Murphy}, D.~N.~A. and {Obaika}, A. and {Narayan}, G. and {Nieto-Santisteban}, M.~A. and {Norberg}, P. and {Peacock}, J.~A. and {Pier}, E.~A. and {Postman}, M. and {Primak}, N. and {Rae}, C. and {Rai}, A. and {Riess}, A. and {Riffeser}, A. and {Rix}, H.~W. and {R{\"o}ser}, S. and {Russel}, R. and {Rutz}, L. and {Schilbach}, E. and {Schultz}, A.~S.~B. and {Scolnic}, D. and {Strolger}, L. and {Szalay}, A. and {Seitz}, S. and {Small}, E. and {Smith}, K.~W. and {Soderblom}, D.~R. and {Taylor}, P. and {Thomson}, R. and {Taylor}, A.~N. and {Thakar}, A.~R. and {Thiel}, J. and {Thilker}, D. and {Unger}, D. and {Urata}, Y. and {Valenti}, J. and {Wagner}, J. and {Walder}, T. and {Walter}, F. and {Watters}, S.~P. and {Werner}, S. and {Wood-Vasey}, W.~M. and {Wyse}, R.},
        title = "{The Pan-STARRS1 Surveys}",
      journal = {arXiv e-prints},
         year = 2016,
        month = dec,
          eid = {arXiv:1612.05560},
        pages = {arXiv:1612.05560},
          doi = {10.48550/arXiv.1612.05560},
archivePrefix = {arXiv},
       eprint = {1612.05560},
 primaryClass = {astro-ph.IM},
       adsurl = {https://ui.adsabs.harvard.edu/abs/2016arXiv161205560C}
}

@ARTICLE{2019PASP..131g8001G,
       author = {{Graham}, Matthew J. and {Kulkarni}, S.~R. and {Bellm}, Eric C. and {Adams}, Scott M. and {Barbarino}, Cristina and {Blagorodnova}, Nadejda and {Bodewits}, Dennis and {Bolin}, Bryce and {Brady}, Patrick R. and {Cenko}, S. Bradley and {Chang}, Chan-Kao and {Coughlin}, Michael W. and {De}, Kishalay and {Eadie}, Gwendolyn and {Farnham}, Tony L. and {Feindt}, Ulrich and {Franckowiak}, Anna and {Fremling}, Christoffer and {Gezari}, Suvi and {Ghosh}, Shaon and {Goldstein}, Daniel A. and {Golkhou}, V. Zach and {Goobar}, Ariel and {Ho}, Anna Y.~Q. and {Huppenkothen}, Daniela and {Ivezi{\'c}}, {\v{Z}}eljko and {Jones}, R. Lynne and {Juric}, Mario and {Kaplan}, David L. and {Kasliwal}, Mansi M. and {Kelley}, Michael S.~P. and {Kupfer}, Thomas and {Lee}, Chien-De and {Lin}, Hsing Wen and {Lunnan}, Ragnhild and {Mahabal}, Ashish A. and {Miller}, Adam A. and {Ngeow}, Chow-Choong and {Nugent}, Peter and {Ofek}, Eran O. and {Prince}, Thomas A. and {Rauch}, Ludwig and {van Roestel}, Jan and {Schulze}, Steve and {Singer}, Leo P. and {Sollerman}, Jesper and {Taddia}, Francesco and {Yan}, Lin and {Ye}, Quan-Zhi and {Yu}, Po-Chieh and {Barlow}, Tom and {Bauer}, James and {Beck}, Ron and {Belicki}, Justin and {Biswas}, Rahul and {Brinnel}, Valery and {Brooke}, Tim and {Bue}, Brian and {Bulla}, Mattia and {Burruss}, Rick and {Connolly}, Andrew and {Cromer}, John and {Cunningham}, Virginia and {Dekany}, Richard and {Delacroix}, Alex and {Desai}, Vandana and {Duev}, Dmitry A. and {Feeney}, Michael and {Flynn}, David and {Frederick}, Sara and {Gal-Yam}, Avishay and {Giomi}, Matteo and {Groom}, Steven and {Hacopians}, Eugean and {Hale}, David and {Helou}, George and {Henning}, John and {Hover}, David and {Hillenbrand}, Lynne A. and {Howell}, Justin and {Hung}, Tiara and {Imel}, David and {Ip}, Wing-Huen and {Jackson}, Edward and {Kaspi}, Shai and {Kaye}, Stephen and {Kowalski}, Marek and {Kramer}, Emily and {Kuhn}, Michael and {Landry}, Walter and {Laher}, Russ R. and {Mao}, Peter and {Masci}, Frank J. and {Monkewitz}, Serge and {Murphy}, Patrick and {Nordin}, Jakob and {Patterson}, Maria T. and {Penprase}, Bryan and {Porter}, Michael and {Rebbapragada}, Umaa and {Reiley}, Dan and {Riddle}, Reed and {Rigault}, Mickael and {Rodriguez}, Hector and {Rusholme}, Ben and {van Santen}, Jakob and {Shupe}, David L. and {Smith}, Roger M. and {Soumagnac}, Maayane T. and {Stein}, Robert and {Surace}, Jason and {Szkody}, Paula and {Terek}, Scott and {Van Sistine}, Angela and {van Velzen}, Sjoert and {Vestrand}, W. Thomas and {Walters}, Richard and {Ward}, Charlotte and {Zhang}, Chaoran and {Zolkower}, Jeffry},
        title = "{The Zwicky Transient Facility: Science Objectives}",
      journal = {\pasp},
         year = 2019,
        month = jul,
       volume = {131},
       number = {1001},
        pages = {078001},
          doi = {10.1088/1538-3873/ab006c},
archivePrefix = {arXiv},
       eprint = {1902.01945},
 primaryClass = {astro-ph.IM},
       adsurl = {https://ui.adsabs.harvard.edu/abs/2019PASP..131g8001G}
}

@ARTICLE{2019PASP..131a8002B,
       author = {{Bellm}, Eric C. and {Kulkarni}, Shrinivas R. and {Graham}, Matthew J. and {Dekany}, Richard and {Smith}, Roger M. and {Riddle}, Reed and {Masci}, Frank J. and {Helou}, George and {Prince}, Thomas A. and {Adams}, Scott M. and {Barbarino}, C. and {Barlow}, Tom and {Bauer}, James and {Beck}, Ron and {Belicki}, Justin and {Biswas}, Rahul and {Blagorodnova}, Nadejda and {Bodewits}, Dennis and {Bolin}, Bryce and {Brinnel}, Valery and {Brooke}, Tim and {Bue}, Brian and {Bulla}, Mattia and {Burruss}, Rick and {Cenko}, S. Bradley and {Chang}, Chan-Kao and {Connolly}, Andrew and {Coughlin}, Michael and {Cromer}, John and {Cunningham}, Virginia and {De}, Kishalay and {Delacroix}, Alex and {Desai}, Vandana and {Duev}, Dmitry A. and {Eadie}, Gwendolyn and {Farnham}, Tony L. and {Feeney}, Michael and {Feindt}, Ulrich and {Flynn}, David and {Franckowiak}, Anna and {Frederick}, S. and {Fremling}, C. and {Gal-Yam}, Avishay and {Gezari}, Suvi and {Giomi}, Matteo and {Goldstein}, Daniel A. and {Golkhou}, V. Zach and {Goobar}, Ariel and {Groom}, Steven and {Hacopians}, Eugean and {Hale}, David and {Henning}, John and {Ho}, Anna Y.~Q. and {Hover}, David and {Howell}, Justin and {Hung}, Tiara and {Huppenkothen}, Daniela and {Imel}, David and {Ip}, Wing-Huen and {Ivezi{\'c}}, {\v{Z}}eljko and {Jackson}, Edward and {Jones}, Lynne and {Juric}, Mario and {Kasliwal}, Mansi M. and {Kaspi}, S. and {Kaye}, Stephen and {Kelley}, Michael S.~P. and {Kowalski}, Marek and {Kramer}, Emily and {Kupfer}, Thomas and {Landry}, Walter and {Laher}, Russ R. and {Lee}, Chien-De and {Lin}, Hsing Wen and {Lin}, Zhong-Yi and {Lunnan}, Ragnhild and {Giomi}, Matteo and {Mahabal}, Ashish and {Mao}, Peter and {Miller}, Adam A. and {Monkewitz}, Serge and {Murphy}, Patrick and {Ngeow}, Chow-Choong and {Nordin}, Jakob and {Nugent}, Peter and {Ofek}, Eran and {Patterson}, Maria T. and {Penprase}, Bryan and {Porter}, Michael and {Rauch}, Ludwig and {Rebbapragada}, Umaa and {Reiley}, Dan and {Rigault}, Mickael and {Rodriguez}, Hector and {van Roestel}, Jan and {Rusholme}, Ben and {van Santen}, Jakob and {Schulze}, S. and {Shupe}, David L. and {Singer}, Leo P. and {Soumagnac}, Maayane T. and {Stein}, Robert and {Surace}, Jason and {Sollerman}, Jesper and {Szkody}, Paula and {Taddia}, F. and {Terek}, Scott and {Van Sistine}, Angela and {van Velzen}, Sjoert and {Vestrand}, W. Thomas and {Walters}, Richard and {Ward}, Charlotte and {Ye}, Quan-Zhi and {Yu}, Po-Chieh and {Yan}, Lin and {Zolkower}, Jeffry},
        title = "{The Zwicky Transient Facility: System Overview, Performance, and First Results}",
      journal = {\pasp},
         year = 2019,
        month = jan,
       volume = {131},
       number = {995},
        pages = {018002},
          doi = {10.1088/1538-3873/aaecbe},
archivePrefix = {arXiv},
       eprint = {1902.01932},
 primaryClass = {astro-ph.IM},
       adsurl = {https://ui.adsabs.harvard.edu/abs/2019PASP..131a8002B}
}

@ARTICLE{2019PASP..131a8003M,
       author = {{Masci}, Frank J. and {Laher}, Russ R. and {Rusholme}, Ben and {Shupe}, David L. and {Groom}, Steven and {Surace}, Jason and {Jackson}, Edward and {Monkewitz}, Serge and {Beck}, Ron and {Flynn}, David and {Terek}, Scott and {Landry}, Walter and {Hacopians}, Eugean and {Desai}, Vandana and {Howell}, Justin and {Brooke}, Tim and {Imel}, David and {Wachter}, Stefanie and {Ye}, Quan-Zhi and {Lin}, Hsing-Wen and {Cenko}, S. Bradley and {Cunningham}, Virginia and {Rebbapragada}, Umaa and {Bue}, Brian and {Miller}, Adam A. and {Mahabal}, Ashish and {Bellm}, Eric C. and {Patterson}, Maria T. and {Juri{\'c}}, Mario and {Golkhou}, V. Zach and {Ofek}, Eran O. and {Walters}, Richard and {Graham}, Matthew and {Kasliwal}, Mansi M. and {Dekany}, Richard G. and {Kupfer}, Thomas and {Burdge}, Kevin and {Cannella}, Christopher B. and {Barlow}, Tom and {Van Sistine}, Angela and {Giomi}, Matteo and {Fremling}, Christoffer and {Blagorodnova}, Nadejda and {Levitan}, David and {Riddle}, Reed and {Smith}, Roger M. and {Helou}, George and {Prince}, Thomas A. and {Kulkarni}, Shrinivas R.},
        title = "{The Zwicky Transient Facility: Data Processing, Products, and Archive}",
      journal = {\pasp},
         year = 2019,
        month = jan,
       volume = {131},
       number = {995},
        pages = {018003},
          doi = {10.1088/1538-3873/aae8ac},
archivePrefix = {arXiv},
       eprint = {1902.01872},
 primaryClass = {astro-ph.IM},
       adsurl = {https://ui.adsabs.harvard.edu/abs/2019PASP..131a8003M}
}

@ARTICLE{2020TNSCR3728....1H,
       author = {{Hiramatsu}, D. and {Hosseinzadeh}, G. and {Burke}, J. and {Howell}, D.~A. and {McCully}, C. and {Gonzalez}, E.~P. and {Pellegrino}, C.},
        title = "{Global SN Project Transient Classification Report for 2020-12-09}",
      journal = {TNS Classification Report},
         year = 2020,
        month = dec,
       volume = {2020-3728},
        pages = {1},
       adsurl = {https://ui.adsabs.harvard.edu/abs/2020TNSCR3728....1H}
}

@INPROCEEDINGS{2004AAS...20511601G,
       author = {{Gehrels}, N. and {Swift}},
        title = "{The Swift Gamma-Ray Burst Mission}",
    booktitle = {American Astronomical Society Meeting Abstracts},
         year = 2004,
       series = {American Astronomical Society Meeting Abstracts},
       volume = {205},
        month = dec,
          eid = {116.01},
        pages = {116.01},
       adsurl = {https://ui.adsabs.harvard.edu/abs/2004AAS...20511601G}
}

@ARTICLE{2013PASP..125.1031B,
       author = {{Brown}, T.~M. and {Baliber}, N. and {Bianco}, F.~B. and {Bowman}, M. and {Burleson}, B. and {Conway}, P. and {Crellin}, M. and {Depagne}, {\'E}. and {De Vera}, J. and {Dilday}, B. and {Dragomir}, D. and {Dubberley}, M. and {Eastman}, J.~D. and {Elphick}, M. and {Falarski}, M. and {Foale}, S. and {Ford}, M. and {Fulton}, B.~J. and {Garza}, J. and {Gomez}, E.~L. and {Graham}, M. and {Greene}, R. and {Haldeman}, B. and {Hawkins}, E. and {Haworth}, B. and {Haynes}, R. and {Hidas}, M. and {Hjelstrom}, A.~E. and {Howell}, D.~A. and {Hygelund}, J. and {Lister}, T.~A. and {Lobdill}, R. and {Martinez}, J. and {Mullins}, D.~S. and {Norbury}, M. and {Parrent}, J. and {Paulson}, R. and {Petry}, D.~L. and {Pickles}, A. and {Posner}, V. and {Rosing}, W.~E. and {Ross}, R. and {Sand}, D.~J. and {Saunders}, E.~S. and {Shobbrook}, J. and {Shporer}, A. and {Street}, R.~A. and {Thomas}, D. and {Tsapras}, Y. and {Tufts}, J.~R. and {Valenti}, S. and {Vander Horst}, K. and {Walker}, Z. and {White}, G. and {Willis}, M.},
        title = "{Las Cumbres Observatory Global Telescope Network}",
      journal = {\pasp},
         year = 2013,
        month = sep,
       volume = {125},
       number = {931},
        pages = {1031},
          doi = {10.1086/673168},
archivePrefix = {arXiv},
       eprint = {1305.2437},
 primaryClass = {astro-ph.IM},
       adsurl = {https://ui.adsabs.harvard.edu/abs/2013PASP..125.1031B}
}

@ARTICLE{2009AJ....137.4517B,
       author = {{Brown}, Peter J. and {Holland}, Stephen T. and {Immler}, Stefan and {Milne}, Peter and {Roming}, Peter W.~A. and {Gehrels}, Neil and {Nousek}, John and {Panagia}, Nino and {Still}, Martin and {Vanden Berk}, Daniel},
        title = "{Ultraviolet Light Curves of Supernovae with the Swift Ultraviolet/Optical Telescope}",
      journal = {\aj},
         year = 2009,
        month = may,
       volume = {137},
       number = {5},
        pages = {4517-4525},
          doi = {10.1088/0004-6256/137/5/4517},
archivePrefix = {arXiv},
       eprint = {0803.1265},
 primaryClass = {astro-ph},
       adsurl = {https://ui.adsabs.harvard.edu/abs/2009AJ....137.4517B}
}

@ARTICLE{2022JAI....1140004J,
       author = {{Joshi}, Y.~C. and {Bangia}, T. and {Jaiswar}, M.~K. and {Pant}, J. and {Reddy}, K. and {Yadav}, S.},
        title = "{ARIES 130-cm Devasthal Fast Optical Telescope {\textemdash} Operation and Outcome}",
      journal = {JAI},
         year = 2022,
        month = jan,
       volume = {11},
       number = {4},
          eid = {2240004},
        pages = {2240004},
          doi = {10.1142/S2251171722400049},
       adsurl = {https://ui.adsabs.harvard.edu/abs/2022JAI....1140004J}
}

@ARTICLE{2018BSRSL..87...29K,
       author = {{Kumar}, Brijesh and {Omar}, Amitesh and {Maheswar}, Gopinathan and {Pandey}, Anil Kumar and {Sagar}, Ram and {Uddin}, Wahab and {Sanwal}, Basant Ballabh and {Bangia}, Tarun and {Kumar}, Tripurari Satyanarayana and {Yadav}, Shobhit and {Sahu}, Sanjit and {Pant}, Jayshreekar and {Reddy}, Bheemireddy Krishna and {Gupta}, Alok Chandra and {Chand}, Hum and {Pandey}, Jeewan Chandra and {Joshi}, Mohit Kumar and {Jaiswar}, Mukeshkuma and {Nanjappa}, Nandish and {Purushottam} and {Yadav}, Rama Kant Singh and {Sharma}, Saurabh and {Pandey}, Shashi Bhushan and {Joshi}, Santosh and {Joshi}, Yogesh Chandra and {Lata}, Sneh and {Mehdi}, Biman Jyoti and {Misra}, Kuntal and {Singh}, Mahendra},
        title = "{3.6-m Devasthal Optical Telescope Project: Completion and first results}",
      journal = {Bulletin de la Societe Royale des Sciences de Liege},
         year = 2018,
        month = apr,
       volume = {87},
        pages = {29-41},
       adsurl = {https://ui.adsabs.harvard.edu/abs/2018BSRSL..87...29K}
}

@misc{2014ascl.soft08004N,
       author = {{Nasa High Energy Astrophysics Science Archive Research Center (Heasarc)}},
        title = "{HEAsoft: Unified Release of FTOOLS and XANADU}",
 howpublished = {Astrophysics Source Code Library, record ascl:1408.004},
         year = 2014,
        month = aug,
          eid = {ascl:1408.004},
       adsurl = {https://ui.adsabs.harvard.edu/abs/2014ascl.soft08004N}
}

@ARTICLE{2021TNSAN...7....1S,
       author = {{Shingles}, L. and {Smith}, K.~W. and {Young}, D.~R. and {Smartt}, S.~J. and {Tonry}, J. and {Denneau}, L. and {Heinze}, A. and {Weiland}, H. and {Flewelling}, H. and {Stalder}, B. and {Clocchiatti}, A. and {F{\"o}rster}, F. and {Pignata}, G. and {Rest}, A. and {Anderson}, J. and {Stubbs}, C. and {Erasmus}, N.},
        title = "{Release of the ATLAS Forced Photometry server for public use}",
      journal = {TNS AstroNote},
         year = 2021,
        month = jan,
       volume = {7},
        pages = {1-7},
       adsurl = {https://ui.adsabs.harvard.edu/abs/2021TNSAN...7....1S}
}

@ARTICLE{1982ApJ...253..785A,
       author = {{Arnett}, W.~D.},
        title = "{Type I supernovae. I - Analytic solutions for the early part of the light curve}",
      journal = {\apj},
         year = 1982,
        month = feb,
       volume = {253},
        pages = {785-797},
          doi = {10.1086/159681},
       adsurl = {https://ui.adsabs.harvard.edu/abs/1982ApJ...253..785A}
}

@ARTICLE{1999AJ....118.2675R,
       author = {{Riess}, Adam G. and {Filippenko}, Alexei V. and {Li}, Weidong and {Treffers}, Richard R. and {Schmidt}, Brian P. and {Qiu}, Yulei and {Hu}, Jingyao and {Armstrong}, Mark and {Faranda}, Chuck and {Thouvenot}, Eric and {Buil}, Christian},
        title = "{The Rise Time of Nearby Type IA Supernovae}",
      journal = {\aj},
         year = 1999,
        month = dec,
       volume = {118},
       number = {6},
        pages = {2675-2688},
          doi = {10.1086/301143},
archivePrefix = {arXiv},
       eprint = {astro-ph/9907037},
 primaryClass = {astro-ph},
       adsurl = {https://ui.adsabs.harvard.edu/abs/1999AJ....118.2675R}
}

@ARTICLE{2022ApJS..259...35A,
       author = {{Abdurro'uf} and {Accetta}, Katherine and {Aerts}, Conny and {Silva Aguirre}, V{\'\i}ctor and {Ahumada}, Romina and {Ajgaonkar}, Nikhil and {Filiz Ak}, N. and {Alam}, Shadab and {Allende Prieto}, Carlos and {Almeida}, Andr{\'e}s and {Anders}, Friedrich and {Anderson}, Scott F. and {Andrews}, Brett H. and {Anguiano}, Borja and {Aquino-Ort{\'\i}z}, Erik and {Arag{\'o}n-Salamanca}, Alfonso and {Argudo-Fern{\'a}ndez}, Maria and {Ata}, Metin and {Aubert}, Marie and {Avila-Reese}, Vladimir and {Badenes}, Carles and {Barb{\'a}}, Rodolfo H. and {Barger}, Kat and {Barrera-Ballesteros}, Jorge K. and {Beaton}, Rachael L. and {Beers}, Timothy C. and {Belfiore}, Francesco and {Bender}, Chad F. and {Bernardi}, Mariangela and {Bershady}, Matthew A. and {Beutler}, Florian and {Bidin}, Christian Moni and {Bird}, Jonathan C. and {Bizyaev}, Dmitry and {Blanc}, Guillermo A. and {Blanton}, Michael R. and {Boardman}, Nicholas Fraser and {Bolton}, Adam S. and {Boquien}, M{\'e}d{\'e}ric and {Borissova}, Jura and {Bovy}, Jo and {Brandt}, W.~N. and {Brown}, Jordan and {Brownstein}, Joel R. and {Brusa}, Marcella and {Buchner}, Johannes and {Bundy}, Kevin and {Burchett}, Joseph N. and {Bureau}, Martin and {Burgasser}, Adam and {Cabang}, Tuesday K. and {Campbell}, Stephanie and {Cappellari}, Michele and {Carlberg}, Joleen K. and {Wanderley}, F{\'a}bio Carneiro and {Carrera}, Ricardo and {Cash}, Jennifer and {Chen}, Yan-Ping and {Chen}, Wei-Huai and {Cherinka}, Brian and {Chiappini}, Cristina and {Choi}, Peter Doohyun and {Chojnowski}, S. Drew and {Chung}, Haeun and {Clerc}, Nicolas and {Cohen}, Roger E. and {Comerford}, Julia M. and {Comparat}, Johan and {da Costa}, Luiz and {Covey}, Kevin and {Crane}, Jeffrey D. and {Cruz-Gonzalez}, Irene and {Culhane}, Connor and {Cunha}, Katia and {Dai}, Y. Sophia and {Damke}, Guillermo and {Darling}, Jeremy and {Davidson}, Jr., James W. and {Davies}, Roger and {Dawson}, Kyle and {De Lee}, Nathan and {Diamond-Stanic}, Aleksandar M. and {Cano-D{\'\i}az}, Mariana and {S{\'a}nchez}, Helena Dom{\'\i}nguez and {Donor}, John and {Duckworth}, Chris and {Dwelly}, Tom and {Eisenstein}, Daniel J. and {Elsworth}, Yvonne P. and {Emsellem}, Eric and {Eracleous}, Mike and {Escoffier}, Stephanie and {Fan}, Xiaohui and {Farr}, Emily and {Feng}, Shuai and {Fern{\'a}ndez-Trincado}, Jos{\'e} G. and {Feuillet}, Diane and {Filipp}, Andreas and {Fillingham}, Sean P. and {Frinchaboy}, Peter M. and {Fromenteau}, Sebastien and {Galbany}, Llu{\'\i}s and {Garc{\'\i}a}, Rafael A. and {Garc{\'\i}a-Hern{\'a}ndez}, D.~A. and {Ge}, Junqiang and {Geisler}, Doug and {Gelfand}, Joseph and {G{\'e}ron}, Tobias and {Gibson}, Benjamin J. and {Goddy}, Julian and {Godoy-Rivera}, Diego and {Grabowski}, Kathleen and {Green}, Paul J. and {Greener}, Michael and {Grier}, Catherine J. and {Griffith}, Emily and {Guo}, Hong and {Guy}, Julien and {Hadjara}, Massinissa and {Harding}, Paul and {Hasselquist}, Sten and {Hayes}, Christian R. and {Hearty}, Fred and {Hern{\'a}ndez}, Jes{\'u}s and {Hill}, Lewis and {Hogg}, David W. and {Holtzman}, Jon A. and {Horta}, Danny and {Hsieh}, Bau-Ching and {Hsu}, Chin-Hao and {Hsu}, Yun-Hsin and {Huber}, Daniel and {Huertas-Company}, Marc and {Hutchinson}, Brian and {Hwang}, Ho Seong and {Ibarra-Medel}, H{\'e}ctor J. and {Chitham}, Jacob Ider and {Ilha}, Gabriele S. and {Imig}, Julie and {Jaekle}, Will and {Jayasinghe}, Tharindu and {Ji}, Xihan and {Johnson}, Jennifer A. and {Jones}, Amy and {J{\"o}nsson}, Henrik and {Katkov}, Ivan and {Khalatyan}, Dr., Arman and {Kinemuchi}, Karen and {Kisku}, Shobhit and {Knapen}, Johan H. and {Kneib}, Jean-Paul and {Kollmeier}, Juna A. and {Kong}, Miranda and {Kounkel}, Marina and {Kreckel}, Kathryn and {Krishnarao}, Dhanesh and {Lacerna}, Ivan and {Lane}, Richard R. and {Langgin}, Rachel and {Lavender}, Ramon and {Law}, David R. and {Lazarz}, Daniel and {Leung}, Henry W. and {Leung}, Ho-Hin and {Lewis}, Hannah M. and {Li}, Cheng and {Li}, Ran and {Lian}, Jianhui and {Liang}, Fu-Heng and {Lin}, Lihwai and {Lin}, Yen-Ting and {Lin}, Sicheng and {Lintott}, Chris and {Long}, Dan and {Longa-Pe{\~n}a}, Pen{\'e}lope and {L{\'o}pez-Cob{\'a}}, Carlos and {Lu}, Shengdong and {Lundgren}, Britt F. and {Luo}, Yuanze and {Mackereth}, J. Ted and {de la Macorra}, Axel and {Mahadevan}, Suvrath and {Majewski}, Steven R. and {Manchado}, Arturo and {Mandeville}, Travis and {Maraston}, Claudia and {Margalef-Bentabol}, Berta and {Masseron}, Thomas and {Masters}, Karen L. and {Mathur}, Savita and {McDermid}, Richard M. and {Mckay}, Myles and {Merloni}, Andrea and {Merrifield}, Michael and {Meszaros}, Szabolcs and {Miglio}, Andrea and {Di Mille}, Francesco and {Minniti}, Dante and {Minsley}, Rebecca and {Monachesi}, Antonela},
        title = "{The Seventeenth Data Release of the Sloan Digital Sky Surveys: Complete Release of MaNGA, MaStar, and APOGEE-2 Data}",
      journal = {\apjs},
         year = 2022,
        month = apr,
       volume = {259},
       number = {2},
          eid = {35},
        pages = {35},
          doi = {10.3847/1538-4365/ac4414},
archivePrefix = {arXiv},
       eprint = {2112.02026},
 primaryClass = {astro-ph.GA},
       adsurl = {https://ui.adsabs.harvard.edu/abs/2022ApJS..259...35A}
}

@PHDTHESIS{2016PhDT.......149T,
       author = {{Tartaglia}, Leonardo},
        title = "{Interacting supernovae and supernova impostors}",
       school = {Astronomical Observatory of Padua; University of Padua, Department of
        Physics and Astronomy},
         year = 2016,
        month = feb,
       adsurl = {https://ui.adsabs.harvard.edu/abs/2016PhDT.......149T}
}

@ARTICLE{2008AJ....135..264C,
       author = {{Chonis}, Taylor S. and {Gaskell}, C. Martin},
        title = "{Setting UBVRI Photometric Zero-Points Using Sloan Digital Sky Survey ugriz Magnitudes}",
      journal = {\aj},
         year = 2008,
        month = jan,
       volume = {135},
       number = {1},
        pages = {264-267},
          doi = {10.1088/0004-6256/135/1/264},
archivePrefix = {arXiv},
       eprint = {0710.5801},
 primaryClass = {astro-ph},
       adsurl = {https://ui.adsabs.harvard.edu/abs/2008AJ....135..264C}
}

@ARTICLE{2017ApJ...836..158H,
       author = {{Hosseinzadeh}, Griffin and {Arcavi}, Iair and {Valenti}, Stefano and {McCully}, Curtis and {Howell}, D. Andrew and {Johansson}, Joel and {Sollerman}, Jesper and {Pastorello}, Andrea and {Benetti}, Stefano and {Cao}, Yi and {Cenko}, S. Bradley and {Clubb}, Kelsey I. and {Corsi}, Alessandra and {Duggan}, Gina and {Elias-Rosa}, Nancy and {Filippenko}, Alexei V. and {Fox}, Ori D. and {Fremling}, Christoffer and {Horesh}, Assaf and {Karamehmetoglu}, Emir and {Kasliwal}, Mansi and {Marion}, G.~H. and {Ofek}, Eran and {Sand}, David and {Taddia}, Francesco and {Zheng}, WeiKang and {Fraser}, Morgan and {Gal-Yam}, Avishay and {Inserra}, Cosimo and {Laher}, Russ and {Masci}, Frank and {Rebbapragada}, Umaa and {Smartt}, Stephen and {Smith}, Ken W. and {Sullivan}, Mark and {Surace}, Jason and {Wo{\'z}niak}, Przemek},
        title = "{Type Ibn Supernovae Show Photometric Homogeneity and Spectral Diversity at Maximum Light}",
      journal = {\apj},
         year = 2017,
        month = feb,
       volume = {836},
       number = {2},
          eid = {158},
        pages = {158},
          doi = {10.3847/1538-4357/836/2/158},
archivePrefix = {arXiv},
       eprint = {1608.01998},
 primaryClass = {astro-ph.HE},
       adsurl = {https://ui.adsabs.harvard.edu/abs/2017ApJ...836..158H}
}

@ARTICLE{2017A&A...602A..93K,
       author = {{Karamehmetoglu}, E. and {Taddia}, F. and {Sollerman}, J. and {Wyrzykowski}, {\L}. and {Schmidl}, S. and {Fraser}, M. and {Fremling}, C. and {Greiner}, J. and {Inserra}, C. and {Kostrzewa-Rutkowska}, Z. and {Maguire}, K. and {Smartt}, S. and {Sullivan}, M. and {Young}, D.~R.},
        title = "{OGLE-2014-SN-131: A long-rising Type Ibn supernova from a massive progenitor}",
      journal = {\aap},
         year = 2017,
        month = jun,
       volume = {602},
          eid = {A93},
        pages = {A93},
          doi = {10.1051/0004-6361/201629619},
archivePrefix = {arXiv},
       eprint = {1703.08222},
 primaryClass = {astro-ph.HE},
       adsurl = {https://ui.adsabs.harvard.edu/abs/2017A&A...602A..93K}
}

@ARTICLE{2015MNRAS.454.4293P,
       author = {{Pastorello}, A. and {Tartaglia}, L. and {Elias-Rosa}, N. and {Morales-Garoffolo}, A. and {Terreran}, G. and {Taubenberger}, S. and {Noebauer}, U.~M. and {Benetti}, S. and {Cappellaro}, E. and {Ciabattari}, F. and {Dennefeld}, M. and {Dimai}, A. and {Ishida}, E.~E.~O. and {Harutyunyan}, A. and {Leonini}, S. and {Ochner}, P. and {Sollerman}, J. and {Taddia}, F. and {Zaggia}, S.},
        title = "{Massive stars exploding in a He-rich circumstellar medium - VIII. PSN J07285387+3349106, a highly reddened supernova Ibn}",
      journal = {\mnras},
         year = 2015,
        month = dec,
       volume = {454},
       number = {4},
        pages = {4293-4303},
          doi = {10.1093/mnras/stv2256},
archivePrefix = {arXiv},
       eprint = {1509.09065},
 primaryClass = {astro-ph.SR},
       adsurl = {https://ui.adsabs.harvard.edu/abs/2015MNRAS.454.4293P}
}

@ARTICLE{2015MNRAS.453.3649P,
       author = {{Pastorello}, A. and {Prieto}, J.~L. and {Elias-Rosa}, N. and {Bersier}, D. and {Hosseinzadeh}, G. and {Morales-Garoffolo}, A. and {Noebauer}, U.~M. and {Taubenberger}, S. and {Tomasella}, L. and {Kochanek}, C.~S. and {Falco}, E. and {Basu}, U. and {Beacom}, J.~F. and {Benetti}, S. and {Brimacombe}, J. and {Cappellaro}, E. and {Danilet}, A.~B. and {Dong}, Subo and {Fernandez}, J.~M. and {Goss}, N. and {Granata}, V. and {Harutyunyan}, A. and {Holoien}, T.~W.-S. and {Ishida}, E.~E.~O. and {Kiyota}, S. and {Krannich}, G. and {Nicholls}, B. and {Ochner}, P. and {Pojma{\'n}ski}, G. and {Shappee}, B.~J. and {Simonian}, G.~V. and {Stanek}, K.~Z. and {Starrfield}, S. and {Szczygie{\l}}, D. and {Tartaglia}, L. and {Terreran}, G. and {Thompson}, T.~A. and {Turatto}, M. and {Wagner}, R.~M. and {Wiethoff}, W.~S. and {Wilber}, A. and {Wo{\'z}niak}, P.~R.},
        title = "{Massive stars exploding in a He-rich circumstellar medium - VII. The metamorphosis of ASASSN-15ed from a narrow line Type Ibn to a normal Type Ib Supernova}",
      journal = {\mnras},
         year = 2015,
        month = nov,
       volume = {453},
       number = {4},
        pages = {3649-3661},
          doi = {10.1093/mnras/stv1812},
archivePrefix = {arXiv},
       eprint = {1509.09062},
 primaryClass = {astro-ph.SR},
       adsurl = {https://ui.adsabs.harvard.edu/abs/2015MNRAS.453.3649P}
}

@ARTICLE{2015MNRAS.449.1941P,
       author = {{Pastorello}, A. and {Wyrzykowski}, {\L}. and {Valenti}, S. and {Prieto}, J.~L. and {Koz{\l}owski}, S. and {Udalski}, A. and {Elias-Rosa}, N. and {Morales-Garoffolo}, A. and {Anderson}, J.~P. and {Benetti}, S. and {Bersten}, M. and {Botticella}, M.~T. and {Cappellaro}, E. and {Fasano}, G. and {Fraser}, M. and {Gal-Yam}, A. and {Gillone}, M. and {Graham}, M.~L. and {Greiner}, J. and {Hachinger}, S. and {Howell}, D.~A. and {Inserra}, C. and {Parrent}, J. and {Rau}, A. and {Schulze}, S. and {Smartt}, S.~J. and {Smith}, K.~W. and {Turatto}, M. and {Yaron}, O. and {Young}, D.~R. and {Kubiak}, M. and {Szyma{\'n}ski}, M.~K. and {Pietrzy{\'n}ski}, G. and {Soszy{\'n}ski}, I. and {Ulaczyk}, K. and {Poleski}, R. and {Pietrukowicz}, P. and {Skowron}, J. and {Mr{\'o}z}, P.},
        title = "{Massive stars exploding in a He-rich circumstellar medium - V. Observations of the slow-evolving SN Ibn OGLE-2012-SN-006}",
      journal = {\mnras},
         year = 2015,
        month = may,
       volume = {449},
       number = {2},
        pages = {1941-1953},
          doi = {10.1093/mnras/stu2621},
archivePrefix = {arXiv},
       eprint = {1502.04945},
 primaryClass = {astro-ph.SR},
       adsurl = {https://ui.adsabs.harvard.edu/abs/2015MNRAS.449.1941P}
}

@ARTICLE{2021A&A...649A.163K,
       author = {{Karamehmetoglu}, E. and {Fransson}, C. and {Sollerman}, J. and {Tartaglia}, L. and {Taddia}, F. and {De}, K. and {Fremling}, C. and {Bagdasaryan}, A. and {Barbarino}, C. and {Bellm}, E.~C. and {Dekany}, R. and {Dugas}, A.~M. and {Giomi}, M. and {Goobar}, A. and {Graham}, M. and {Ho}, A. and {Laher}, R.~R. and {Masci}, F.~J. and {Neill}, J.~D. and {Perley}, D. and {Riddle}, R. and {Rusholme}, B. and {Soumagnac}, M.~T.},
        title = "{The luminous and rapidly evolving SN 2018bcc. Clues toward the origin of Type Ibn SNe from the Zwicky Transient Facility}",
      journal = {\aap},
         year = 2021,
        month = may,
       volume = {649},
          eid = {A163},
        pages = {A163},
          doi = {10.1051/0004-6361/201936308},
archivePrefix = {arXiv},
       eprint = {1910.06016},
 primaryClass = {astro-ph.HE},
       adsurl = {https://ui.adsabs.harvard.edu/abs/2021A&A...649A.163K}
}

@ARTICLE{2021ApJ...917...97W,
       author = {{Wang}, Xiaofeng and {Lin}, Weili and {Zhang}, Jujia and {Zhang}, Tianmeng and {Cai}, Yongzhi and {Zhang}, Kaicheng and {Filippenko}, Alexei V. and {Graham}, Melissa and {Maeda}, Keiichi and {Mo}, Jun and {Xiang}, Danfeng and {Xi}, Gaobo and {Yan}, Shengyu and {Wang}, Lifan and {Wang}, Lingjun and {Kawabata}, Koji and {Zhai}, Qian},
        title = "{ASASSN-14ms: The Most Energetic Known Explosion of a Type Ibn Supernova and Its Physical Origin}",
      journal = {\apj},
         year = 2021,
        month = aug,
       volume = {917},
       number = {2},
          eid = {97},
        pages = {97},
          doi = {10.3847/1538-4357/ac0c17},
archivePrefix = {arXiv},
       eprint = {2106.06690},
 primaryClass = {astro-ph.HE},
       adsurl = {https://ui.adsabs.harvard.edu/abs/2021ApJ...917...97W}
}

@ARTICLE{2021A&A...652A.136K,
       author = {{Kool}, E.~C. and {Karamehmetoglu}, E. and {Sollerman}, J. and {Schulze}, S. and {Lunnan}, R. and {Reynolds}, T.~M. and {Barbarino}, C. and {Bellm}, E.~C. and {De}, K. and {Duev}, D.~A. and {Fremling}, C. and {Golkhou}, V.~Z. and {Graham}, M.~L. and {Green}, D.~A. and {Horesh}, A. and {Kaye}, S. and {Kim}, Y.-L. and {Laher}, R.~R. and {Masci}, F.~J. and {Nordin}, J. and {Perley}, D.~A. and {Phinney}, E.~S. and {Porter}, M. and {Reiley}, D. and {Rodriguez}, H. and {van Roestel}, J. and {Rusholme}, B. and {Sharma}, Y. and {Sfaradi}, I. and {Soumagnac}, M.~T. and {Taggart}, K. and {Tartaglia}, L. and {Williams}, D.~R.~A. and {Yan}, L.},
        title = "{SN 2020bqj: A Type Ibn supernova with a long-lasting peak plateau}",
      journal = {\aap},
         year = 2021,
        month = aug,
       volume = {652},
          eid = {A136},
        pages = {A136},
          doi = {10.1051/0004-6361/202039137},
archivePrefix = {arXiv},
       eprint = {2008.04056},
 primaryClass = {astro-ph.HE},
       adsurl = {https://ui.adsabs.harvard.edu/abs/2021A&A...652A.136K}
}

@ARTICLE{2015MNRAS.449.1954P,
       author = {{Pastorello}, A. and {Hadjiyska}, E. and {Rabinowitz}, D. and {Valenti}, S. and {Turatto}, M. and {Fasano}, G. and {Benitez-Herrera}, S. and {Baltay}, C. and {Benetti}, S. and {Botticella}, M.~T. and {Cappellaro}, E. and {Elias-Rosa}, N. and {Ellman}, N. and {Feindt}, U. and {Filippenko}, A.~V. and {Fraser}, M. and {Gal-Yam}, A. and {Graham}, M.~L. and {Howell}, D.~A. and {Inserra}, C. and {Kelly}, P.~L. and {Kotak}, R. and {Kowalski}, M. and {McKinnon}, R. and {Morales-Garoffolo}, A. and {Nugent}, P.~E. and {Smartt}, S.~J. and {Smith}, K.~W. and {Stritzinger}, M.~D. and {Sullivan}, M. and {Taubenberger}, S. and {Walker}, E.~S. and {Yaron}, O. and {Young}, D.~R.},
        title = "{Massive stars exploding in a He-rich circumstellar medium - VI. Observations of two distant Type Ibn supernova candidates discovered by La Silla-QUEST}",
      journal = {\mnras},
         year = 2015,
        month = may,
       volume = {449},
       number = {2},
        pages = {1954-1966},
          doi = {10.1093/mnras/stv335},
archivePrefix = {arXiv},
       eprint = {1502.04949},
 primaryClass = {astro-ph.SR},
       adsurl = {https://ui.adsabs.harvard.edu/abs/2015MNRAS.449.1954P}
}

@ARTICLE{2025A&A...700A.156W,
       author = {{Wang}, Z.-Y. and {Pastorello}, A. and {Cai}, Y.-Z. and {Fraser}, M. and {Reguitti}, A. and {Lin}, W.-L. and {Tartaglia}, L. and {Andrew Howell}, D. and {Benetti}, S. and {Cappellaro}, E. and {Chen}, Z.-H. and {Elias-Rosa}, N. and {Farah}, J. and {Fiore}, A. and {Hiramatsu}, D. and {Kankare}, E. and {Li}, Z.-T. and {Lundqvist}, P. and {Mazzali}, P.~A. and {McCully}, C. and {Mo}, J. and {Moran}, S. and {Newsome}, M. and {Padilla Gonzalez}, E. and {Pellegrino}, C. and {Peng}, Z.-H. and {Smartt}, S.~J. and {Srivastav}, S. and {Stritzinger}, M.~D. and {Terreran}, G. and {Tomasella}, L. and {Valerin}, G. and {Wang}, G.-J. and {Wang}, X.-F. and {de Boer}, T. and {Chambers}, K.~C. and {Gao}, H. and {Guo}, F.-Z. and {Guti{\'e}rrez}, C.~P. and {Kangas}, T. and {Karamehmetoglu}, E. and {Li}, G.-C. and {Lin}, C.-C. and {Lowe}, T.~B. and {Ma}, X.-R. and {Magnier}, E.~A. and {Minguez}, P. and {Pei}, S.-P. and {Reynolds}, T.~M. and {Wainscoat}, R.~J. and {Wang}, B. and {Williams}, S. and {Wu}, C.-Y. and {Yan}, S.-Y. and {Zhang}, J.-J. and {Zhang}, X.-H. and {Zhu}, X.-J.},
        title = "{Massive stars exploding in a He-rich circumstellar medium: XI. Diverse evolution of five Ibn SNe 2020nxt, 2020taz, 2021bbv, 2023utc, and 2024aej}",
      journal = {\aap},
         year = 2025,
        month = aug,
       volume = {700},
          eid = {A156},
        pages = {A156},
          doi = {10.1051/0004-6361/202554768},
archivePrefix = {arXiv},
       eprint = {2506.15139},
 primaryClass = {astro-ph.HE},
       adsurl = {https://ui.adsabs.harvard.edu/abs/2025A&A...700A.156W}
}

@ARTICLE{2023A&A...673A..27N,
       author = {{Nagao}, T. and {Kuncarayakti}, H. and {Maeda}, K. and {Moore}, T. and {Pastorello}, A. and {Mattila}, S. and {Uno}, K. and {Smartt}, S.~J. and {Sim}, S.~A. and {Ferrari}, L. and {Tomasella}, L. and {Anderson}, J.~P. and {Chen}, T.-W. and {Galbany}, L. and {Gao}, H. and {Gromadzki}, M. and {Guti{\'e}rrez}, C.~P. and {Inserra}, C. and {Kankare}, E. and {Magnier}, E.~A. and {M{\"u}ller-Bravo}, T.~E. and {Reguitti}, A. and {Young}, D.~R.},
        title = "{Photometry and spectroscopy of the Type Icn supernova 2021ckj. The diverse properties of the ejecta and circumstellar matter of Type Icn supernovae}",
      journal = {\aap},
         year = 2023,
        month = may,
       volume = {673},
          eid = {A27},
        pages = {A27},
          doi = {10.1051/0004-6361/202346084},
archivePrefix = {arXiv},
       eprint = {2303.07721},
 primaryClass = {astro-ph.HE},
       adsurl = {https://ui.adsabs.harvard.edu/abs/2023A&A...673A..27N}
}

@ARTICLE{2022ApJ...938...73P,
       author = {{Pellegrino}, C. and {Howell}, D.~A. and {Terreran}, G. and {Arcavi}, I. and {Bostroem}, K.~A. and {Brown}, P.~J. and {Burke}, J. and {Dong}, Y. and {Gilkis}, A. and {Hiramatsu}, D. and {Hosseinzadeh}, G. and {McCully}, C. and {Modjaz}, M. and {Newsome}, M. and {Gonzalez}, E. Padilla and {Pritchard}, T.~A. and {Sand}, D.~J. and {Valenti}, S. and {Williamson}, M.},
        title = "{The Diverse Properties of Type Icn Supernovae Point to Multiple Progenitor Channels}",
      journal = {\apj},
         year = 2022,
        month = oct,
       volume = {938},
       number = {1},
          eid = {73},
        pages = {73},
          doi = {10.3847/1538-4357/ac8ff6},
archivePrefix = {arXiv},
       eprint = {2205.07894},
 primaryClass = {astro-ph.HE},
       adsurl = {https://ui.adsabs.harvard.edu/abs/2022ApJ...938...73P}
}

@ARTICLE{2022ApJ...927..180P,
       author = {{Perley}, Daniel A. and {Sollerman}, Jesper and {Schulze}, Steve and {Yao}, Yuhan and {Fremling}, Christoffer and {Gal-Yam}, Avishay and {Ho}, Anna Y.~Q. and {Yang}, Yi and {Kool}, Erik C. and {Irani}, Ido and {Yan}, Lin and {Andreoni}, Igor and {Baade}, Dietrich and {Bellm}, Eric C. and {Brink}, Thomas G. and {Chen}, Ting-Wan and {Cikota}, Aleksandar and {Coughlin}, Michael W. and {Dahiwale}, Aishwarya and {Dekany}, Richard and {Duev}, Dmitry A. and {Filippenko}, Alexei V. and {Hoeflich}, Peter and {Kasliwal}, Mansi M. and {Kulkarni}, S.~R. and {Lunnan}, Ragnhild and {Masci}, Frank J. and {Maund}, Justyn R. and {Medford}, Michael S. and {Riddle}, Reed and {Rosnet}, Philippe and {Shupe}, David L. and {Strotjohann}, Nora Linn and {Tzanidakis}, Anastasios and {Zheng}, WeiKang},
        title = "{The Type Icn SN 2021csp: Implications for the Origins of the Fastest Supernovae and the Fates of Wolf-Rayet Stars}",
      journal = {\apj},
         year = 2022,
        month = mar,
       volume = {927},
       number = {2},
          eid = {180},
        pages = {180},
          doi = {10.3847/1538-4357/ac478e},
archivePrefix = {arXiv},
       eprint = {2111.12110},
 primaryClass = {astro-ph.HE},
       adsurl = {https://ui.adsabs.harvard.edu/abs/2022ApJ...927..180P}
}

@ARTICLE{2024ApJ...977....2P,
       author = {{Pellegrino}, C. and {Modjaz}, M. and {Takei}, Y. and {Tsuna}, D. and {Newsome}, M. and {Pritchard}, T. and {Baer-Way}, R. and {Bostroem}, K.~A. and {Chandra}, P. and {Charalampopoulos}, P. and {Dong}, Y. and {Farah}, J. and {Howell}, D.~A. and {McCully}, C. and {Mohamed}, S. and {Padilla Gonzalez}, E. and {Terreran}, G.},
        title = "{The X-Ray Luminous Type Ibn SN 2022ablq: Estimates of Preexplosion Mass Loss and Constraints on Precursor Emission}",
      journal = {\apj},
         year = 2024,
        month = dec,
       volume = {977},
       number = {1},
          eid = {2},
        pages = {2},
          doi = {10.3847/1538-4357/ad8bc5},
archivePrefix = {arXiv},
       eprint = {2407.18291},
 primaryClass = {astro-ph.HE},
       adsurl = {https://ui.adsabs.harvard.edu/abs/2024ApJ...977....2P}
}

@ARTICLE{2026MNRAS.547f1517G,
       author = {{Gangopadhyay}, Anjasha and {Sollerman}, Jesper and {Tsalapatas}, Konstantinos and {Maeda}, Keiichi and {Dukiya}, Naveen and {Schulze}, Steve and {Fransson}, Claes and {Sarin}, Nikhil and {Pessi}, Priscila J. and {Singh}, Mridweeka and {Wise}, Jacob and {Nakaoka}, Tatsuya and {Singh}, Avinash and {Dastidar}, Raya and {Kawabata}, Miho and {Qin}, Yu-Jing and {Das}, Kaustav K. and {Perley}, Daniel and {Fremling}, Christoffer and {Taguchi}, Kenta and {Hinds}, K.-Ryan and {Lunnan}, Ragnhild and {Teja}, Rishabh Singh and {Dubey}, Monalisa and {Ailawadhi}, Bhavya and {Banerjee}, Smaranika and {Kawabata}, Koji S. and {Misra}, Kuntal and {Sahu}, Devendra K. and {Brennan}, Sea'n. J. and {Kasliwal}, Mansi M. and {Ho}, Anna Y.~C.~Q. and {Bochenek}, Aleksandra and {Rusholme}, Ben and {Laher}, Russ R. and {Smith}, Roger and {Purdum}, Josiah and {Sravan}, Niharika},
        title = "{SN 2023xgo: Helium-rich Type Icn or Carbon-Flash Type Ibn supernova?}",
      journal = {\mnras},
         year = 2026,
        month = apr,
       volume = {547},
       number = {3},
          eid = {staf1517},
        pages = {staf1517},
          doi = {10.1093/mnras/staf1517},
archivePrefix = {arXiv},
       eprint = {2506.10700},
 primaryClass = {astro-ph.HE},
       adsurl = {https://ui.adsabs.harvard.edu/abs/2026MNRAS.547f1517G}
}

@ARTICLE{2026A&A...707A.157C,
      author = {{Cai}, Y.-Z. and {Pastorello}, A. and {Maeda}, K. and {Zhao}, J.-W. and {Wang}, Z.-Y. and {Peng}, Z.-H. and {Reguitti}, A. and {Tartaglia}, L. and {Filippenko}, A.~V. and {Pan}, Y. and {Valerin}, G. and {Kumar}, B. and {Wang}, Z. and {Fraser}, M. and {Anderson}, J.~P. and {Benetti}, S. and {Bose}, S. and {Brink}, T.~G. and {Cappellaro}, E. and {Chen}, T.-W. and {Chen}, X.-L. and {Elias-Rosa}, N. and {Esamdin}, A. and {Gal-Yam}, A. and {Gonz{\'a}lez-Ba{\~n}uelos}, M. and {Gromadzki}, M. and {Guti{\'e}rrez}, C.~P. and {Inserra}, C. and {Iskandar}, A. and {Kangas}, T. and {Kankare}, E. and {Kravtsov}, T. and {Kuncarayakti}, H. and {Li}, L.-P. and {Liu}, C.-X. and {Liu}, X.-K. and {Lundqvist}, P. and {Matilainen}, K. and {Mattila}, S. and {Moran}, S. and {M{\"u}ller-Bravo}, T.~E. and {Nagao}, T. and {Petrushevska}, T. and {Pignata}, G. and {Salmaso}, I. and {Smartt}, S.~J. and {Sollerman}, J. and {Srivastav}, S. and {Stritzinger}, M.~D. and {Wang}, L.-T. and {Yan}, S.-Y. and {Yang}, Y. and {Yang}, Y.-P. and {Zheng}, W. and {Zou}, X.-Z. and {Chen}, L.-Y. and {Du}, X.-L. and {Fang}, Q.-L. and {Fiore}, A. and {Ragosta}, F. and {Zha}, S. and {Zhang}, J.-J. and {Liu}, X.-W. and {Bai}, J.-M. and {Wang}, B. and {Wang}, X.-F.},
       title = "{Massive stars exploding in a He-rich circumstellar medium: XII. SN 2024acyl: A fast, linearly declining Type Ibn supernova with early flash-ionisation features}",
     journal = {\aap},
        year = 2026,
       month = mar,
      volume = {707},
         eid = {A157},
       pages = {A157},
         doi = {10.1051/0004-6361/202558014},
archivePrefix = {arXiv},
      eprint = {2511.04337},
primaryClass = {astro-ph.SR},
      adsurl = {https://ui.adsabs.harvard.edu/abs/2026A&A...707A.157C}
}

@ARTICLE{2025MNRAS.536.3588W,
       author = {{Warwick}, B. and {Lyman}, J. and {Pursiainen}, M. and {Coppejans}, D.~L. and {Galbany}, L. and {Jones}, G.~T. and {Killestein}, T.~L. and {Kumar}, A. and {Oates}, S.~R. and {Ackley}, K. and {Anderson}, J.~P. and {Aryan}, A. and {Breton}, R.~P. and {Chen}, T.~W. and {Clark}, P. and {Dhillon}, V.~S. and {Dyer}, M.~J. and {Gal-Yam}, A. and {Galloway}, D.~K. and {Guti{\'e}rrez}, C.~P. and {Gromadzki}, M. and {Inserra}, C. and {Jim{\'e}nez-Ibarra}, F. and {Kelsey}, L. and {Kotak}, R. and {Kravtsov}, T. and {Kuncarayakti}, H. and {Magee}, M.~R. and {Matilainen}, K. and {Mattila}, S. and {M{\"u}ller-Bravo}, T.~E. and {Nicholl}, M. and {Noysena}, K. and {Nuttall}, L.~K. and {O'Brien}, P. and {O'Neill}, D. and {Pall{\'e}}, E. and {Pessi}, T. and {Petrushevska}, T. and {Pignata}, G. and {Pollacco}, D. and {Ragosta}, F. and {Ramsay}, G. and {Sahu}, A. and {Sahu}, D.~K. and {Singh}, A. and {Sollerman}, J. and {Stanway}, E. and {Starling}, R. and {Steeghs}, D. and {Teja}, R.~S. and {Ulaczyk}, K.},
        title = "{SN 2023tsz: a helium-interaction-driven supernova in a very low-mass galaxy}",
      journal = {\mnras},
         year = 2025,
        month = feb,
       volume = {536},
       number = {4},
        pages = {3588-3600},
          doi = {10.1093/mnras/stae2784},
archivePrefix = {arXiv},
       eprint = {2409.14147},
 primaryClass = {astro-ph.HE},
       adsurl = {https://ui.adsabs.harvard.edu/abs/2025MNRAS.536.3588W}
}

@ARTICLE{2022Natur.601..201G,
       author = {{Gal-Yam}, A. and {Bruch}, R. and {Schulze}, S. and {Yang}, Y. and {Perley}, D.~A. and {Irani}, I. and {Sollerman}, J. and {Kool}, E.~C. and {Soumagnac}, M.~T. and {Yaron}, O. and {Strotjohann}, N.~L. and {Zimmerman}, E. and {Barbarino}, C. and {Kulkarni}, S.~R. and {Kasliwal}, M.~M. and {De}, K. and {Yao}, Y. and {Fremling}, C. and {Yan}, L. and {Ofek}, E.~O. and {Fransson}, C. and {Filippenko}, A.~V. and {Zheng}, W. and {Brink}, T.~G. and {Copperwheat}, C.~M. and {Foley}, R.~J. and {Brown}, J. and {Siebert}, M. and {Leloudas}, G. and {Cabrera-Lavers}, A.~L. and {Garcia-Alvarez}, D. and {Marante-Barreto}, A. and {Frederick}, S. and {Hung}, T. and {Wheeler}, J.~C. and {Vink{\'o}}, J. and {Thomas}, B.~P. and {Graham}, M.~J. and {Duev}, D.~A. and {Drake}, A.~J. and {Dekany}, R. and {Bellm}, E.~C. and {Rusholme}, B. and {Shupe}, D.~L. and {Andreoni}, I. and {Sharma}, Y. and {Riddle}, R. and {van Roestel}, J. and {Knezevic}, N.},
        title = "{A WC/WO star exploding within an expanding carbon-oxygen-neon nebula}",
      journal = {\nat},
         year = 2022,
        month = jan,
       volume = {601},
       number = {7892},
        pages = {201-204},
          doi = {10.1038/s41586-021-04155-1},
archivePrefix = {arXiv},
       eprint = {2111.12435},
 primaryClass = {astro-ph.SR},
       adsurl = {https://ui.adsabs.harvard.edu/abs/2022Natur.601..201G}
}

@ARTICLE{2013ApJ...769...39S,
       author = {{Sanders}, N.~E. and {Soderberg}, A.~M. and {Foley}, R.~J. and {Chornock}, R. and {Milisavljevic}, D. and {Margutti}, R. and {Drout}, M.~R. and {Moe}, M. and {Berger}, E. and {Brown}, W.~R. and {Lunnan}, R. and {Smartt}, S.~J. and {Fraser}, M. and {Kotak}, R. and {Magill}, L. and {Smith}, K.~W. and {Wright}, D. and {Huang}, K. and {Urata}, Y. and {Mulchaey}, J.~S. and {Rest}, A. and {Sand}, D.~J. and {Chomiuk}, L. and {Friedman}, A.~S. and {Kirshner}, R.~P. and {Marion}, G.~H. and {Tonry}, J.~L. and {Burgett}, W.~S. and {Chambers}, K.~C. and {Hodapp}, K.~W. and {Kudritzki}, R.~P. and {Price}, P.~A.},
        title = "{PS1-12sk is a Peculiar Supernova from a He-rich Progenitor System in a Brightest Cluster Galaxy Environment}",
      journal = {\apj},
         year = 2013,
        month = may,
       volume = {769},
       number = {1},
          eid = {39},
        pages = {39},
          doi = {10.1088/0004-637X/769/1/39},
archivePrefix = {arXiv},
       eprint = {1303.1818},
 primaryClass = {astro-ph.CO},
       adsurl = {https://ui.adsabs.harvard.edu/abs/2013ApJ...769...39S}
}

@ARTICLE{2014MNRAS.443..671G,
       author = {{Gorbikov}, Evgeny and {Gal-Yam}, Avishay and {Ofek}, Eran O. and {Vreeswijk}, Paul M. and {Nugent}, Peter E. and {Chotard}, Nicolas and {Kulkarni}, Shrinivas R. and {Cao}, Yi and {De Cia}, Annalisa and {Yaron}, Ofer and {Tal}, David and {Arcavi}, Iair and {Kasliwal}, Mansi M. and {Cenko}, S. Bradley and {Sullivan}, Mark and {Chen}, Juncheng},
        title = "{iPTF13beo: the double-peaked light curve of a Type Ibn supernova discovered shortly after explosion}",
      journal = {\mnras},
         year = 2014,
        month = sep,
       volume = {443},
       number = {1},
        pages = {671-677},
          doi = {10.1093/mnras/stu1184},
archivePrefix = {arXiv},
       eprint = {1312.0012},
 primaryClass = {astro-ph.SR},
       adsurl = {https://ui.adsabs.harvard.edu/abs/2014MNRAS.443..671G}
}

@ARTICLE{2024ApJ...967L..45W,
       author = {{Wu}, Chengyuan and {Zha}, Shuai and {Cai}, Yongzhi and {Zhang}, Zhengyang and {Yang}, Yi and {Xiang}, Danfeng and {Lin}, Weili and {Wang}, Xiaofeng and {Wang}, Bo},
        title = "{Light Curves of the Explosion of ONe White Dwarf + CO White Dwarf Merger Remnant and Type Icn Supernovae}",
      journal = {\apjl},
         year = 2024,
        month = jun,
       volume = {967},
       number = {2},
          eid = {L45},
        pages = {L45},
          doi = {10.3847/2041-8213/ad4a7a},
       adsurl = {https://ui.adsabs.harvard.edu/abs/2024ApJ...967L..45W}
}

@ARTICLE{1980ApJ...237..541A,
       author = {{Arnett}, W.~D.},
        title = "{Analytic solutions for light curves of supernovae of Type II}",
      journal = {\apj},
         year = 1980,
        month = apr,
       volume = {237},
        pages = {541-549},
          doi = {10.1086/157898},
       adsurl = {https://ui.adsabs.harvard.edu/abs/1980ApJ...237..541A}
}

@ARTICLE{2015ApJ...814...63M,
       author = {{Morozova}, Viktoriya and {Piro}, Anthony L. and {Renzo}, Mathieu and {Ott}, Christian D. and {Clausen}, Drew and {Couch}, Sean M. and {Ellis}, Justin and {Roberts}, Luke F.},
        title = "{Light Curves of Core-collapse Supernovae with Substantial Mass Loss Using the New Open-source SuperNova Explosion Code (SNEC)}",
      journal = {\apj},
         year = 2015,
        month = nov,
       volume = {814},
       number = {1},
          eid = {63},
        pages = {63},
          doi = {10.1088/0004-637X/814/1/63},
archivePrefix = {arXiv},
       eprint = {1505.06746},
 primaryClass = {astro-ph.HE},
       adsurl = {https://ui.adsabs.harvard.edu/abs/2015ApJ...814...63M}
}

@ARTICLE{1999ApJ...510..379M,
       author = {{Matzner}, Christopher D. and {McKee}, Christopher F.},
        title = "{The Expulsion of Stellar Envelopes in Core-Collapse Supernovae}",
      journal = {\apj},
         year = 1999,
        month = jan,
       volume = {510},
       number = {1},
        pages = {379-403},
          doi = {10.1086/306571},
archivePrefix = {arXiv},
       eprint = {astro-ph/9807046},
 primaryClass = {astro-ph},
       adsurl = {https://ui.adsabs.harvard.edu/abs/1999ApJ...510..379M}
}

@ARTICLE{2007ARA&A..45..177C,
       author = {{Crowther}, Paul A.},
        title = "{Physical Properties of Wolf-Rayet Stars}",
      journal = {\araa},
         year = 2007,
        month = sep,
       volume = {45},
       number = {1},
        pages = {177-219},
          doi = {10.1146/annurev.astro.45.051806.110615},
archivePrefix = {arXiv},
       eprint = {astro-ph/0610356},
 primaryClass = {astro-ph},
       adsurl = {https://ui.adsabs.harvard.edu/abs/2007ARA&A..45..177C}
}

@INPROCEEDINGS{1986SPIE..627..733T,
       author = {{Tody}, Doug},
        title = "{The IRAF Data Reduction and Analysis System}",
    booktitle = {Instrumentation in astronomy VI},
         year = 1986,
       editor = {{Crawford}, David L.},
       series = {SPIE Conference Series},
       volume = {627},
        month = jan,
        pages = {733},
          doi = {10.1117/12.968154},
       adsurl = {https://ui.adsabs.harvard.edu/abs/1986SPIE..627..733T}
}

@INPROCEEDINGS{1993ASPC...52..173T,
       author = {{Tody}, Doug},
        title = "{IRAF in the Nineties}",
    booktitle = {Astronomical Data Analysis Software and Systems II},
         year = 1993,
       editor = {{Hanisch}, R.~J. and {Brissenden}, R.~J.~V. and {Barnes}, J.},
       series = {Astronomical Society of the Pacific Conference Series},
       volume = {52},
        month = jan,
        pages = {173},
       adsurl = {https://ui.adsabs.harvard.edu/abs/1993ASPC...52..173T}
}

@misc{2012ascl.soft07011S,
       author = {{Science Software Branch at STScI}},
        title = "{PyRAF: Python alternative for IRAF}",
 howpublished = {Astrophysics Source Code Library, record ascl:1207.011},
         year = 2012,
        month = jul,
          eid = {ascl:1207.011},
archivePrefix = {ascl},
       eprint = {1207.011},
       adsurl = {https://ui.adsabs.harvard.edu/abs/2012ascl.soft07011S}
}

@ARTICLE{2019PASP..131g5004F,
       author = {{Fabricant}, Daniel and {Fata}, Robert and {Epps}, Harland and {Gauron}, Thomas and {Mueller}, Mark and {Zajac}, Joseph and {Amato}, Stephen and {Barberis}, Jack and {Bergner}, Henry and {Brennan}, Patricia and {Brown}, Warren and {Chilingarian}, Igor and {Geary}, John and {Kradinov}, Vladimir and {McLeod}, Brian and {Smith}, Matthew and {Woods}, Deborah},
        title = "{Binospec: A Wide-field Imaging Spectrograph for the MMT}",
      journal = {\pasp},
         year = 2019,
        month = jul,
       volume = {131},
       number = {1001},
        pages = {075004},
          doi = {10.1088/1538-3873/ab1d78},
archivePrefix = {arXiv},
       eprint = {1905.03320},
 primaryClass = {astro-ph.IM},
       adsurl = {https://ui.adsabs.harvard.edu/abs/2019PASP..131g5004F}
}

@ARTICLE{1972AJ.....77..312W,
       author = {{Walborn}, N.~R.},
        title = "{Spectral classification of OB stars in both hemispheres and the absolute-magnitude calibration.}",
      journal = {\aj},
         year = 1972,
        month = may,
       volume = {77},
        pages = {312-318},
          doi = {10.1086/111285},
       adsurl = {https://ui.adsabs.harvard.edu/abs/1972AJ.....77..312W}
}

@ARTICLE{1945CoPri..20....1M,
       author = {{Moore}, Charlotte},
        title = "{A Multiplet Table of Astrophysical Interest. Revised Edition. Part I - Table of Multiplets}",
      journal = {Contributions from the Princeton University Observatory},
         year = 1945,
        month = jan,
       volume = {20},
        pages = {1-110},
       adsurl = {https://ui.adsabs.harvard.edu/abs/1945CoPri..20....1M}
}

@ARTICLE{2025A&A...703A.177T,
       author = {{Tartaglia}, L. and {Valerin}, G. and {Pastorello}, A. and {Reguitti}, A. and {Benetti}, S. and {Tomasella}, L. and {Ochner}, P. and {Brocato}, E. and {Cond{\`o}}, L. and {De Luise}, F. and {Onori}, F. and {Salmaso}, I.},
        title = "{Signatures of anti-social mass loss in the ordinary Type II SN 2024bch: A non-interacting supernova with early high-ionisation features}",
      journal = {\aap},
         year = 2025,
        month = nov,
       volume = {703},
          eid = {A177},
        pages = {A177},
          doi = {10.1051/0004-6361/202452359},
archivePrefix = {arXiv},
       eprint = {2409.15431},
 primaryClass = {astro-ph.HE},
       adsurl = {https://ui.adsabs.harvard.edu/abs/2025A&A...703A.177T}
}

@ARTICLE{1998MNRAS.296..367C,
       author = {{Crowther}, P.~A. and {De Marco}, Orsola and {Barlow}, M.~J.},
        title = "{Quantitative classification of WC and WO stars}",
      journal = {\mnras},
         year = 1998,
        month = may,
       volume = {296},
       number = {2},
        pages = {367-378},
          doi = {10.1046/j.1365-8711.1998.01360.x},
       adsurl = {https://ui.adsabs.harvard.edu/abs/1998MNRAS.296..367C}
}

@ARTICLE{2015ApJ...806..213S,
       author = {{Shivvers}, Isaac and {Groh}, Jose H. and {Mauerhan}, Jon C. and {Fox}, Ori D. and {Leonard}, Douglas C. and {Filippenko}, Alexei V.},
        title = "{Early Emission from the Type IIn Supernova 1998S at High Resolution}",
      journal = {\apj},
         year = 2015,
        month = jun,
       volume = {806},
       number = {2},
          eid = {213},
        pages = {213},
          doi = {10.1088/0004-637X/806/2/213},
archivePrefix = {arXiv},
       eprint = {1408.1404},
 primaryClass = {astro-ph.HE},
       adsurl = {https://ui.adsabs.harvard.edu/abs/2015ApJ...806..213S}
}

@ARTICLE{2020A&A...635A..39T,
       author = {{Tartaglia}, L. and {Pastorello}, A. and {Sollerman}, J. and {Fransson}, C. and {Mattila}, S. and {Fraser}, M. and {Taddia}, F. and {Tomasella}, L. and {Turatto}, M. and {Morales-Garoffolo}, A. and {Elias-Rosa}, N. and {Lundqvist}, P. and {Harmanen}, J. and {Reynolds}, T. and {Cappellaro}, E. and {Barbarino}, C. and {Nyholm}, A. and {Kool}, E. and {Ofek}, E. and {Gao}, X. and {Jin}, Z. and {Tan}, H. and {Sand}, D.~J. and {Ciabattari}, F. and {Wang}, X. and {Zhang}, J. and {Huang}, F. and {Li}, W. and {Mo}, J. and {Rui}, L. and {Xiang}, D. and {Zhang}, T. and {Hosseinzadeh}, G. and {Howell}, D.~A. and {McCully}, C. and {Valenti}, S. and {Benetti}, S. and {Callis}, E. and {Carracedo}, A.~S. and {Fremling}, C. and {Kangas}, T. and {Rubin}, A. and {Somero}, A. and {Terreran}, G.},
        title = "{The long-lived Type IIn SN 2015da: Infrared echoes and strong interaction within an extended massive shell}",
      journal = {\aap},
         year = 2020,
        month = mar,
       volume = {635},
          eid = {A39},
        pages = {A39},
          doi = {10.1051/0004-6361/201936553},
archivePrefix = {arXiv},
       eprint = {1908.08580},
 primaryClass = {astro-ph.HE},
       adsurl = {https://ui.adsabs.harvard.edu/abs/2020A&A...635A..39T}
}

@ARTICLE{2018MNRAS.475.1261H,
       author = {{Huang}, Chenliang and {Chevalier}, Roger A.},
        title = "{Electron scattering wings on lines in interacting supernovae}",
      journal = {\mnras},
         year = 2018,
        month = mar,
       volume = {475},
       number = {1},
        pages = {1261-1273},
          doi = {10.1093/mnras/stx3163},
archivePrefix = {arXiv},
       eprint = {1712.01237},
 primaryClass = {astro-ph.HE},
       adsurl = {https://ui.adsabs.harvard.edu/abs/2018MNRAS.475.1261H}
}

@ARTICLE{2014ApJ...797..118F,
       author = {{Fransson}, Claes and {Ergon}, Mattias and {Challis}, Peter J. and {Chevalier}, Roger A. and {France}, Kevin and {Kirshner}, Robert P. and {Marion}, G.~H. and {Milisavljevic}, Dan and {Smith}, Nathan and {Bufano}, Filomena and {Friedman}, Andrew S. and {Kangas}, Tuomas and {Larsson}, Josefin and {Mattila}, Seppo and {Benetti}, Stefano and {Chornock}, Ryan and {Czekala}, Ian and {Soderberg}, Alicia and {Sollerman}, Jesper},
        title = "{High-density Circumstellar Interaction in the Luminous Type IIn SN 2010jl: The First 1100 Days}",
      journal = {\apj},
         year = 2014,
        month = dec,
       volume = {797},
       number = {2},
          eid = {118},
        pages = {118},
          doi = {10.1088/0004-637X/797/2/118},
archivePrefix = {arXiv},
       eprint = {1312.6617},
 primaryClass = {astro-ph.HE},
       adsurl = {https://ui.adsabs.harvard.edu/abs/2014ApJ...797..118F}
}

@ARTICLE{2020A&A...638A..92T,
       author = {{Taddia}, F. and {Stritzinger}, M.~D. and {Fransson}, C. and {Brown}, P.~J. and {Contreras}, C. and {Holmbo}, S. and {Moriya}, T.~J. and {Phillips}, M.~M. and {Sollerman}, J. and {Suntzeff}, N.~B. and {Ashall}, C. and {Burns}, C.~R. and {Busta}, L. and {Campillay}, A. and {Castell{\'o}n}, S. and {Corco}, C. and {Di Mille}, F. and {Gall}, C. and {Gonz{\'a}lez}, C. and {Hsiao}, E.~Y. and {Morrell}, N. and {Nyholm}, A. and {Simon}, J.~D. and {Ser{\'o}n}, J.},
        title = "{The Carnegie Supernova Project II. The shock wave revealed through the fog: The strongly interacting Type IIn SN 2013L}",
      journal = {\aap},
         year = 2020,
        month = jun,
       volume = {638},
          eid = {A92},
        pages = {A92},
          doi = {10.1051/0004-6361/201936654},
archivePrefix = {arXiv},
       eprint = {2003.09709},
 primaryClass = {astro-ph.SR},
       adsurl = {https://ui.adsabs.harvard.edu/abs/2020A&A...638A..92T}
}

@ARTICLE{2022A&A...658A.130D,
       author = {{Dessart}, L. and {Hillier}, D. John and {Kuncarayakti}, H.},
        title = "{Helium stars exploding in circumstellar material and the origin of Type Ibn supernovae}",
      journal = {\aap},
         year = 2022,
        month = feb,
       volume = {658},
          eid = {A130},
        pages = {A130},
          doi = {10.1051/0004-6361/202142436},
archivePrefix = {arXiv},
       eprint = {2111.13360},
 primaryClass = {astro-ph.SR},
       adsurl = {https://ui.adsabs.harvard.edu/abs/2022A&A...658A.130D}
}

@ARTICLE{2013MNRAS.434.1636T,
       author = {{Tomasella}, L. and {Cappellaro}, E. and {Fraser}, M. and {Pumo}, M.~L. and {Pastorello}, A. and {Pignata}, G. and {Benetti}, S. and {Bufano}, F. and {Dennefeld}, M. and {Harutyunyan}, A. and {Iijima}, T. and {Jerkstrand}, A. and {Kankare}, E. and {Kotak}, R. and {Magill}, L. and {Nascimbeni}, V. and {Ochner}, P. and {Siviero}, A. and {Smartt}, S. and {Sollerman}, J. and {Stanishev}, V. and {Taddia}, F. and {Taubenberger}, S. and {Turatto}, M. and {Valenti}, S. and {Wright}, D.~E. and {Zampieri}, L.},
        title = "{Comparison of progenitor mass estimates for the Type IIP SN 2012A}",
      journal = {\mnras},
         year = 2013,
        month = sep,
       volume = {434},
       number = {2},
        pages = {1636-1657},
          doi = {10.1093/mnras/stt1130},
archivePrefix = {arXiv},
       eprint = {1305.5789},
 primaryClass = {astro-ph.SR},
       adsurl = {https://ui.adsabs.harvard.edu/abs/2013MNRAS.434.1636T}
}

@ARTICLE{2014ARA&A..52..487S,
       author = {{Smith}, Nathan},
        title = "{Mass Loss: Its Effect on the Evolution and Fate of High-Mass Stars}",
      journal = {\araa},
         year = 2014,
        month = aug,
       volume = {52},
        pages = {487-528},
          doi = {10.1146/annurev-astro-081913-040025},
archivePrefix = {arXiv},
       eprint = {1402.1237},
 primaryClass = {astro-ph.SR},
       adsurl = {https://ui.adsabs.harvard.edu/abs/2014ARA&A..52..487S}
}

@ARTICLE{2007Natur.447..829P,
       author = {{Pastorello}, A. and {Smartt}, S.~J. and {Mattila}, S. and {Eldridge}, J.~J. and {Young}, D. and {Itagaki}, K. and {Yamaoka}, H. and {Navasardyan}, H. and {Valenti}, S. and {Patat}, F. and {Agnoletto}, I. and {Augusteijn}, T. and {Benetti}, S. and {Cappellaro}, E. and {Boles}, T. and {Bonnet-Bidaud}, J.-M. and {Botticella}, M.~T. and {Bufano}, F. and {Cao}, C. and {Deng}, J. and {Dennefeld}, M. and {Elias-Rosa}, N. and {Harutyunyan}, A. and {Keenan}, F.~P. and {Iijima}, T. and {Lorenzi}, V. and {Mazzali}, P.~A. and {Meng}, X. and {Nakano}, S. and {Nielsen}, T.~B. and {Smoker}, J.~V. and {Stanishev}, V. and {Turatto}, M. and {Xu}, D. and {Zampieri}, L.},
        title = "{A giant outburst two years before the core-collapse of a massive star}",
      journal = {\nat},
         year = 2007,
        month = jun,
       volume = {447},
       number = {7146},
        pages = {829-832},
          doi = {10.1038/nature05825},
archivePrefix = {arXiv},
       eprint = {astro-ph/0703663},
 primaryClass = {astro-ph},
       adsurl = {https://ui.adsabs.harvard.edu/abs/2007Natur.447..829P}
}

@ARTICLE{2007ApJ...657L.105F,
       author = {{Foley}, Ryan J. and {Smith}, Nathan and {Ganeshalingam}, Mohan and {Li}, Weidong and {Chornock}, Ryan and {Filippenko}, Alexei V.},
        title = "{SN 2006jc: A Wolf-Rayet Star Exploding in a Dense He-rich Circumstellar Medium}",
      journal = {\apjl},
         year = 2007,
        month = mar,
       volume = {657},
       number = {2},
        pages = {L105-L108},
          doi = {10.1086/513145},
archivePrefix = {arXiv},
       eprint = {astro-ph/0612711},
 primaryClass = {astro-ph},
       adsurl = {https://ui.adsabs.harvard.edu/abs/2007ApJ...657L.105F}
}

@ARTICLE{2016MNRAS.458.2094D,
       author = {{Dessart}, Luc and {Hillier}, D. John and {Audit}, Edouard and {Livne}, Eli and {Waldman}, Roni},
        title = "{Models of interacting supernovae and their spectral diversity}",
      journal = {\mnras},
         year = 2016,
        month = may,
       volume = {458},
       number = {2},
        pages = {2094-2121},
          doi = {10.1093/mnras/stw336},
archivePrefix = {arXiv},
       eprint = {1602.02977},
 primaryClass = {astro-ph.SR},
       adsurl = {https://ui.adsabs.harvard.edu/abs/2016MNRAS.458.2094D}
}

@ARTICLE{2009ARA&A..47..481A,
       author = {{Asplund}, Martin and {Grevesse}, Nicolas and {Sauval}, A. Jacques and {Scott}, Pat},
        title = "{The Chemical Composition of the Sun}",
      journal = {\araa},
         year = 2009,
        month = sep,
       volume = {47},
       number = {1},
        pages = {481-522},
          doi = {10.1146/annurev.astro.46.060407.145222},
archivePrefix = {arXiv},
       eprint = {0909.0948},
 primaryClass = {astro-ph.SR},
       adsurl = {https://ui.adsabs.harvard.edu/abs/2009ARA&A..47..481A}
}

@ARTICLE{2003ApJ...591..288H,
       author = {{Heger}, A. and {Fryer}, C.~L. and {Woosley}, S.~E. and {Langer}, N. and {Hartmann}, D.~H.},
        title = "{How Massive Single Stars End Their Life}",
      journal = {\apj},
         year = 2003,
        month = jul,
       volume = {591},
       number = {1},
        pages = {288-300},
          doi = {10.1086/375341},
archivePrefix = {arXiv},
       eprint = {astro-ph/0212469},
 primaryClass = {astro-ph},
       adsurl = {https://ui.adsabs.harvard.edu/abs/2003ApJ...591..288H}
}

@ARTICLE{1997ARA&A..35..309F,
       author = {{Filippenko}, Alexei V.},
        title = "{Optical Spectra of Supernovae}",
      journal = {\araa},
         year = 1997,
        month = jan,
       volume = {35},
        pages = {309-355},
          doi = {10.1146/annurev.astro.35.1.309},
       adsurl = {https://ui.adsabs.harvard.edu/abs/1997ARA&A..35..309F}
}

@ARTICLE{2019MNRAS.482.1545S,
       author = {{Shivvers}, Isaac and {Filippenko}, Alexei V. and {Silverman}, Jeffrey M. and {Zheng}, WeiKang and {Foley}, Ryan J. and {Chornock}, Ryan and {Barth}, Aaron J. and {Cenko}, S. Bradley and {Clubb}, Kelsey I. and {Fox}, Ori D. and {Ganeshalingam}, Mohan and {Graham}, Melissa L. and {Kelly}, Patrick L. and {Kleiser}, Io K.~W. and {Leonard}, Douglas C. and {Li}, Weidong and {Matheson}, Thomas and {Mauerhan}, Jon C. and {Modjaz}, Maryam and {Serduke}, Franklin J.~D. and {Shields}, Joseph C. and {Steele}, Thea N. and {Swift}, Brandon J. and {Wong}, Diane S. and {Yuk}, Heechan},
        title = "{The Berkeley sample of stripped-envelope supernovae}",
      journal = {\mnras},
         year = 2019,
        month = jan,
       volume = {482},
       number = {2},
        pages = {1545-1556},
          doi = {10.1093/mnras/sty2719},
archivePrefix = {arXiv},
       eprint = {1810.03650},
 primaryClass = {astro-ph.SR},
       adsurl = {https://ui.adsabs.harvard.edu/abs/2019MNRAS.482.1545S}
}

@ARTICLE{2014Natur.509..471G,
       author = {{Gal-Yam}, Avishay and {Arcavi}, I. and {Ofek}, E.~O. and {Ben-Ami}, S. and {Cenko}, S.~B. and {Kasliwal}, M.~M. and {Cao}, Y. and {Yaron}, O. and {Tal}, D. and {Silverman}, J.~M. and {Horesh}, A. and {De Cia}, A. and {Taddia}, F. and {Sollerman}, J. and {Perley}, D. and {Vreeswijk}, P.~M. and {Kulkarni}, S.~R. and {Nugent}, P.~E. and {Filippenko}, A.~V. and {Wheeler}, J.~C.},
        title = "{A Wolf-Rayet-like progenitor of SN 2013cu from spectral observations of a stellar wind}",
      journal = {\nat},
         year = 2014,
        month = may,
       volume = {509},
       number = {7501},
        pages = {471-474},
          doi = {10.1038/nature13304},
archivePrefix = {arXiv},
       eprint = {1406.7640},
 primaryClass = {astro-ph.HE},
       adsurl = {https://ui.adsabs.harvard.edu/abs/2014Natur.509..471G}
}

\begin{appendix}
\onecolumn
\section{Observations and data reduction} \label{sec:obsredu}
\subsection{Photometry} \label{sec:photodata}
Photometric followup started soon after the ATLAS discovery (see Sect.~\ref{sec:intro}), with early Near ultra-violet (UV; $\sim1600-4700\,\text{\AA}$) and $UBV$ observations carried out using the $0.3\,\text{m}$ Ultra-Violet Optical Telescope (UVOT) on board the Swift Gamma Ray Burst Explorer \citep{2004AAS...20511601G}.
The corresponding apparent UV/optical magnitudes were obtained following the prescription of \citet{2009AJ....137.4517B} with {\sc HEASoft v. 6.33} \citep{2014ascl.soft08004N} on pre-processed images retrieved from the Swift archive\footnote{\url{https://swift.gsfc.nasa.gov/archive/}}.
Optical photometry was also obtained with facilities of the Las Cumbres Observatory network \citep{2013PASP..125.1031B} within the Global Supernova Project (GSP), as well as the $2.56\,\rm{m}$ Nordic Optical Telescope (NOT) equipped with the ``Alhambra Faint Object Spectrograph and Camera" (ALFOSC\footnote{\url{https://www.not.iac.es/instruments/alfosc/}}), the $2\,\rm{m}$ Liverpool Telescope (LT) with the ``Infrared-Optical: Optical" (IO:O) camera, the $1.82\,\rm{m}$ Copernico telescope with the ``Asiago Faint Object Spectrograph and Camera" (AFOSC) and the $69/92\,\rm{cm}$ Schmidt with a G4-16000LC Moravian camera of the Asiago observatory (Mount Ekar), all within the NOT Unbiased Transient Survey 2 (NUTS2\footnote{\url{https://nuts.sn.ie/}}).
Additional $ugriz$ photometry was obtained with the $1.3\,\rm{m}$ Devasthal Fast Optical Telescope \citep[DFOT;][]{2022JAI....1140004J} and the $3.6\,\rm{m}$ Devasthal Optical Telescope \citep[DOT;][]{2018BSRSL..87...29K}.
Publicly available ZTF $gri$ frames obtained with the Samuel Oschin $1.2\,\rm{m}$ ($48\,\rm{inches}$) Telescope (Palomar 48 - P48) were retrieved from the NASA/IPAC Infrared Science Archive\footnote{https://irsa.ipac.caltech.edu/Missions/ztf.html}.
All of these data were reduced using the superNOva photometry ({\sc ecsnoopy}\footnote{\url{https://sngroup.oapd.inaf.it/ecsnoopy.html}}) pipeline, which is based on point spread function (PSF) fitting and zero-point calibration using stars from the Sloan Digital Sky Survey (SDSS) catalog \citep[Data Release 17 - DDR17;][]{2022ApJS..259...35A}. 
Additional details on the reduction steps within {\sc ecsnoopy} are given in \citet{2016PhDT.......149T}.
ATLAS light curves obtained in $orange$ and $cyan$ bands ($o$ and $c$, respectively) were retrieved using the ATLAS Forced Photometry server for public use \citep{2021TNSAN...7....1S}.
UV and $UBV$ magnitudes were calibrated in the Vega photometric system, applying the transformations derived by \citet{2008AJ....135..264C} to the $ugriz$ magnitudes of the stars in the field to compute the proper zero points.
$ugcroiz$ magnitudes were calibrated in the AB photometric system.

\subsection{Spectroscopy} \label{sec:spectrodata}
The spectroscopic follow-up started on 2020 December 8.5~UT ($\rm{JD}=2459192.0$), approximately $7\,\rm{days}$ after the estimated explosion (considering the rest-frame phase corrected for cosmological redshift).
Data were collected within the GSP, using the $2\,\rm{m}$ Faulkes north and south telescopes of the Las Cumbres Observatory, located at  the Haleakala Observatories (Faulkes Telescope North, Hawaii - USA; FTN) and the Siding Spring Observatory (Faulkes Telescope South, Australia; FTS) using the cross-dispersed, low-resolution spectrographs FLOYDS.
These were reduced using their dedicated pipeline\footnote{\url{https://lco.global/documentation/data/floyds-pipeline/}}.
Spectra were also obtained using the $2.56\,\rm{m}$ NOT with ALFOSC and the $1.82\,\rm{m}$ Copernico telescope with AFOSC, both within the NUTS2 collaboration.
Data obtained with ALFOSC and AFOSC were reduced using the {\sc foscgui} pipeline\footnote{\url{https://sngroup.oapd.inaf.it/foscgui.html}}, which performs standard {\sc iraf} reduction steps \citep{1986SPIE..627..733T,1993ASPC...52..173T} through {\sc pyraf} \citep{2012ascl.soft07011S}.
Additional low- and medium-resolution spectra (see Table~\ref{tab:speclog}) were obtained using DOT with ADFOSC \citep{2019BSRSL..88...31O,2023JATIS...9a8002P} and the $2\,\rm{m}$ Himalayan Chandra Telescope with the Hanle Faint Object Spectrograph Camera \citep[HFOSC;][]{2014PINSA..80..887P}, both processed through standard \textsc{iraf} reduction steps.
Three mid-resolution ($R\simeq4000$) spectra were obtained using the $6.5\,\rm{m}$ MMT located at the Fred Lawrence Whipple Observatory in Arizona with Blue Channel and Binospec \citep{2019PASP..131g5004F} spectrographs.
Observations taken with Binospec \citep{Fabricant2019} on MMT were initially processed using the Binospec IDL pipeline \citep{Kansky2019} for flat-fielding, sky subtraction, and wavelength and flux calibrations. The 1D spectrum were then extracted using IRAF techniques \citep{1986SPIE..627..733T}.
Spectra obtained with BlueChannel \citep{Angel1979} on MMT are reduced using standard IRAF techniques.
An additional spectrum was obtained with the Boller \& Chivens (B\&C) spectrograph on the $2.3\,\rm{m}$ Bok telescope, located at the Kitt Peak National Observatory (KPNO) in Arizona, USA and reduced using standard {\sc iraf} routines.
A log of the observations, including additional information on the instrumental setup used, is reported in Table~\ref{tab:speclog}.
\begin{table*}[ht!]
\caption{Log of the spectroscopic observations of SN~2020able.} \label{tab:speclog}
\centering
\resizebox{0.92\textwidth}{!}{%
\begin{adjustbox}{tabular=cccccccc,center} \\
\toprule
Date & JD & Rest-frame phase & Telescope+Instrument & Grism/grating & Range & Resolution & Exposure  \\
\midrule
& (days) & (days) & & & (\AA) & ($\lambda/\Delta\lambda$) & (s) \\
\midrule
20201208 & 2459191.99 & $+7$  & OGG2m+FLOYDS  & red+blue            & $3500-10000$   & 505       & 3600          \\
20201209 & 2459193.07 & $+8$  & OGG2m+FLOYDS  & red+blue            & $3500-10000$   & 530       & 2700          \\
20201211 & 2459194.95 & $+10$ & OGG2m+FLOYDS  & red+blue            & $3500-10000$   & 555       & 2700          \\
20201212 & 2459195.74 & $+11$ & NOT+ALFOSC    & Gr4                 & $3700-9700$    & 330       & 900           \\
20201214 & 2459197.64 & $+13$ & Ekar182+AFOSC & Gr4                 & $3400-9400$    & 390       & $2\times1800$ \\
20201213 & 2459197.39 & $+13$ & MMT+BCH       & 1200GPM/LP-530      & $5700-7000$    & 4000      & $3\times600$  \\
20201215 & 2459198.65 & $+14$ & NOT+ALFOSC    & Gr4                 & $3700-9700$    & 410       & $2\times1800$ \\
20201215 & 2459199.13 & $+14$ & OGG2m+FLOYDS  & red+blue            & $3500-10000$   & 520       & 2700          \\
20201216 & 2459200.46 & $+16$ & Ekar182+AFOSC & VPH6+VPH7           & $3700-9400$    & 400       & $1800+1800$   \\
20201217 & 2459200.96 & $+16$ & OGG2m+FLOYDS  & red+blue            & $3500-10000$   & 510       & 2700          \\
20201218 & 2459201.59 & $+17$ & NOT+ALFOSC    & Gr4                 & $3700-9700$    & 400       & 1800          \\
20201221 & 2459204.65 & $+20$ & NOT+ALFOSC    & Gr4                 & $3700-9700$    & 330       & 1800          \\
20201222 & 2459206.04 & $+21$ & OGG2m+FLOYDS  & red+blue            & $3500-10000$   & 525       & 2700          \\
20201225 & 2459208.70 & $+24$ & NOT+ALFOSC    & Gr4                 & $3700-9700$    & 290       & 2400          \\
20201230 & 2459213.92 & $+29$ & OGG2m+FLOYDS  & red+blue            & $3500-10000$   & 520       & 2700          \\
20210105 & 2459220.40 & $+36$ & MMT+BCH       & 1200GPM/LP-530      & $5700-7000$    & 4000      & $3\times600$  \\
20210107 & 2459221.53 & $+36$ & DOT+ADFOSC    & 676R/RG-610         & $6100-8200$    & 1100      & $2\times900$  \\
20210111 & 2459226.45 & $+41$ & DOT+ADFOSC    & 676R                & $4000-9000$    & 620       & $3\times900$  \\
20210114 & 2459228.61 & $+43$ & DOT+ADFOSC    & 676R                & $4000-9000$    & 620       & $3\times900$  \\
20210114 & 2459228.86 & $+43$ & OGG2m+FLOYDS  & red+blue            & $3500-10000$   & 540       & 2700          \\
20210116 & 2459230.58 & $+45$ & DOT+ADFOSC    & 676R                & $4000-9000$    & 620       & $4\times900$  \\
20210118 & 2459232.49 & $+47$ & NOT+ALFOSC    & Gr4                 & $3700-9700$    & 430       & $2\times1800$ \\
20210125 & 2459239.54 & $+54$ & NOT+ALFOSC    & Gr4                 & $3700-9700$    & 430       & 3600          \\
20210204 & 2459250.24 & $+65$ & MMT+Binospec  & $\times1000$/LP3800 & $5700-7000$    & 3900      & $5\times600$ \\
20210206 & 2459251.78 & $+66$ & Bok+B\&C      & 300g/mm             & $4000-8000$   & 700       & $3\times1200$ \\
20210215 & 2459261.12 & $+75$ & HCT+HFOSC     & Gr7+Gr8             & $4000-9000$    & 1330+2190 & $2700+2700$   \\
20210216 & 2459262.47 & $+76$ & NOT+ALFOSC    & Gr4                 & $3700-9700$    & 400       & 3600          \\
20210312 & 2459285.49 & $+98$ & NOT+ALFOSC    & Gr4                 & $3700-9700$    & 420       & 3600          \\
\bottomrule
\end{adjustbox}}
\tablefoot{OGG2m: 2m0-01 telescope with FLOYDS at Las Cumbres Observatory node at Haleakala Observatory; NOT: $2.56\,\rm{m}$ Nordic Optical Telescope with ALFOSC; Ekar182: $1.82\,\rm{m}$ Copernico telescope with AFOSC; MMT: $6.5\,\rm{m}$ MMT with the Blue Channel and Binospec spectrographs; DOT: $3.6\,\rm{m}$ Devasthal Optical Telescope with the ARIES Devasthal Faint Object Spectrograph and Camera (ADFOSC); HCT: $2\,\rm{m}$ Himalayan Chandra Telescope with the Himalayan Faint Object Spectrograph and Camera (HFOSC). Rest-frame phases refer to the estimated epoch of the SN explosion.}
\end{table*}
After data reduction, absolute flux calibrations were checked against photometry obtained on the closest night, mostly by multiplying the spectral continuum by a constant.
At early phases ($t\lesssim+10\,\rm{days}$), the spectral shape was corrected using low-order (1 or 2) polynomials, since the observed spectral continuum was too red with respect to the SED computed from the photometry. 

\newpage
\section{Analysis of the early light curves of Type Ibn/Icn SNe} \label{sec:fitRise}
\begin{table*}
\caption{Data points used to produce Fig.~\ref{fig:riseMax} and \ref{fig:declines}}
\label{tab:riseMax}
\centering
\resizebox{0.82\textwidth}{!}{%
\begin{adjustbox}{tabular=lcccccccc,center} \\
\toprule
SN & MJD$_{\rm{expl}}$ (err) & t$_{\rm{rise}}$ (err) & Decline (err) & $\mu$ & m$_{\rm{abs}}$ (err) & Filter & Type/subclass & Reference \\
\midrule
   &   JD-2400000.5          & (d)                   & ($\rm{mag\,d^{-1}}$) & (mag) & (mag)                &        &               &           \\
\midrule
2010al           & 55267.5(1.5)         &  17.2(1.5)      & 0.0792(0.0020)   & 34.27(0.16)   & $-18.81(0.16)$   & $R$ & HI       & (1) \\
OGLE-2012-SN-006 & 56206.339(0.031)     & 11.276(0.061)   & 0.0477(0.0012)   & 36.94(0.18)   & $-19.65(0.18)$   & $I$ & P Cygni  & (2) \\
PTF12ldy         & 56236.190(0.047)     &  8.984(0.056)   & 0.130(0.012)     & 38.36(0.15)   & $-18.97(0.15)$   & $r$ & P Cygni  & (3) \\
PS1-12sk         & 55993.624(0.055)     & 12.559(0.064)   & 0.079(0.021)     & 36.75(0.15)   & $-18.80(0.15)$   & $z$ & emission & (4) \\
iPTF13beo        & 56430.900(0.025)     &  2.843(0.072)   & 0.0575(0.0033)   & 38.01(0.15)   & $-18.41(0.15)$   & $r$ & emission & (5) \\
LSQ13ccw         & 56534.879(0.014)     &  4.847(0.025)   & 0.134(0.012)     & 37.07(0.14)   & $-18.25(0.34)$   & $g$ & emission & (6) \\
2014av           & 56763.277(0.056)     &   6.44(0.15)    & 0.1409(0.0090)   & 35.56(0.15)   & $-19.68(0.15)$   & $R$ & Ibn      & (7) \\
ASASSN-14ms      & 57014.99(0.25)       &   6.77(0.28)    & 0.0419(0.0034)   & 36.72(0.14)   & $-20.41(0.14)$   & $V$ & Ibn      & (8) \\
iPTF14aki        & 56756.359(0.026)     &  9.615(0.032)   & 0.1205(0.0018)   & 37.13(0.15)   & $-19.01(0.15)$   & $R$ & emission & (3) \\
OGLE-2014-SN-131 & 56921.27(0.75)       &   48.5(1.6)     & 0.0089(0.0040)   & 37.85(0.15)   & $-18.08(0.15)$   & $I$ & Ibn      & (9) \\
ASASSN-15ed      & 57078.5(2.7)         &   8.5(3.8)      & 0.1159(0.0094)   & 36.59(0.14)   & $-19.98(0.34)$   & $r$ & P Cygni  & (10) \\
2015U            & 57062.598(0.073)     &   9.46(0.10)    & 0.1510(0.0020)   & 33.85(0.15)   & $-19.61(0.15)$   & $r$ & P Cygni  & (11) \\
iPTF15ul         & 57090.302(0.006)     &  4.2960(0.0078) & 0.1824(0.0070)   & 37.18(0.15)   & $-19.40(0.15)$   & $g$ & Ibn      & (3) \\
2018bcc          & 58225.355(0.018)     &  5.403(0.032)   & 0.1476(0.0036)   & 37.19(0.15)   & $-19.41(0.15)$   & $r$ & Ibn      & (12) \\
2018jmt          & 58456.709(0.080)     &   8.26(0.19)    & 0.1140(0.0081)   & 35.65(0.22)   & $-19.13(0.22)$   & $g$ & Ibn      & (13) \\
2019cj           & 58483.54(0.18)       &   8.22(0.26)    & 0.0666(0.0072)   & 36.38(0.16)   & $-18.84(0.16)$   & $g$ & HI       & (13) \\
2019hgp          & 58640.982(0.035)     &   8.06(0.12)    & 0.0904(0.0028)   & 37.21(0.15)   & $-18.49(0.15)$   & $r$ & Icn      & (14) \\
2019jc           & 58489.42(0.13)       &   5.17(0.13)    & 0.106(0.025)     & 34.55(0.15)   & $-17.06(0.15)$   & $r$ & Icn      & (15) \\
2019uo           & 58499.578(0.067)     &  9.455(0.071)   & 0.1710(0.0023)   & 34.74(0.15)   & $-18.10(0.15)$   & $r$ & HI       & (16) \\
2020able         & 59183.96(0.43)       & 15.289(0.054)   & 0.0505(0.0016)   & 35.16(0.15)   & $-18.79(0.15)$   & $r$ & HI       & This work \\
2020bqj          & 58861.69(0.98)       &   37.7(1.1)     & 0.0068(0.0020)   & 37.92(0.15)   & $-19.90(0.15)$   & $r$ & Ibn      & (17) \\
2020nxt          & 59032.4830(0.0090)   &  6.482(0.012)   & 0.1638(0.0028)   & 34.80(0.03)   & $-18.999(0.030)$ & $o$ & Ibn      & (18) \\
2020taz          & 59098.11(0.15)       &  13.55(0.28)    & 0.0475(0.0060)   & 36.62(0.02)   & $-17.858(0.024)$ & $o$ & Ibn      & (18) \\
2021ckj          & 59251.88(0.48)       &   7.44(0.86)    & 0.1256(0.0082)   & 39.07(0.15)   & $-19.88(0.16)$   & $r$ & Icn      & (19,20) \\
2021csp          & 59251.43(0.32)       &   8.34(0.33)    & 0.0760(0.0028)   & 37.82(0.15)   & $-19.99(0.16)$   & $g$ & Icn      & (21) \\
2022ablq         & 59905.613(0.060)     & 10.951(0.066)   & 0.0940(0.0031)   & 33.87(0.15)   & $-19.46(0.15)$   & $g$ & Ibn      & (22) \\
2023emq          & 60033.315(0.030)     &  7.582(0.037)   & 0.236(0.020)     & 35.77(0.15)   & $-19.00(0.15)$   & $o$ & HI       & (23) \\
2023fyq          & 60147.913(0.049)     & 6.139(0.050)    & 0.14939(0.00038) & 31.28(0.45)   & $-18.25(0.45)$   & $r$ & Ibn      & (24) \\
2023utc          & 60227.31(0.10)       &   7.15(0.11)    & 0.0707(0.0039)   & 33.80(0.46)   & $-16.35(0.46)$   & $r$ & Ibn      & (18) \\
2023tsz          & 60199.7(1.7)         &  13.8(1.7)      & 0.1201(0.0084)   & 35.35(0.15)   & $-19.45(0.21)$   & $g$ & Ibn      & (25) \\
2023xgo          & 60256.925(0.023)     &  4.954(0.080)   & 0.1531(0.0021)   & 33.678(0.056) & $-17.661(0.057)$ & $r$ & Icn      & (26) \\
2024acyl         & 60644.78(0.23)       &   5.85(0.26)    & 0.0880(0.0054)   & 35.23(0.15)   & $-18.18(0.15)$   & $V$ & Ibn      & (27) \\
2024aej          & 60321.47(0.13)       &   7.10(0.14)    & 0.1417(0.0067)   & 37.17(0.11)   & $-19.22(0.11)$   & $r$ & Ibn      & (18) \\
\bottomrule
\end{adjustbox}}
\tablefoot{HI: Type Ibn SNe showing high-ionisation features in their early spectra. The other labels refer to the sub-classification proposed by \citet{2017ApJ...836..158H}}
\tablebib{
(1)~\cite{2015MNRAS.449.1921P};
(2) \cite{2015MNRAS.449.1941P};
(3) \cite{2017ApJ...836..158H};
(4) \cite{2013ApJ...769...39S};
(5) \cite{2014MNRAS.443..671G};
(6) \cite{2015MNRAS.449.1954P};
(7) \cite{2016MNRAS.456..853P};
(8) \cite{2021ApJ...917...97W};
(9) \cite{2017A&A...602A..93K};
(10) \cite{2015MNRAS.453.3649P};
(11) \cite{2015MNRAS.454.4293P};
(12) \cite{2021A&A...649A.163K};
(13) \cite{2024A&A...691A.156W};
(14) \cite{2022Natur.601..201G};
(15) \cite{2024ApJ...967L..45W};
(16) \cite{2020ApJ...889..170G};
(17) \cite{2021A&A...652A.136K};
(18) \cite{2025A&A...700A.156W};
(19) \cite{2022ApJ...938...73P};
(20) \cite{2023A&A...673A..27N};
(21) \cite{2022ApJ...927..180P};
(22) \cite{2024ApJ...977....2P};
(23) \cite{2023ApJ...959L..10P};
(24) \cite{2024ApJ...977..254D};
(25) \cite{2025MNRAS.536.3588W};
(26) \cite{2026MNRAS.547f1517G};
(27) \cite{2026A&A...707A.157C}}
\end{table*}

\newpage
\section{Modelling of the light curves of SNe~2010al and 2020able with {\sc MOSFiT}} \label{sec:mosfit}

\begin{figure*}
\centering
\resizebox{\hsize}{!}{\includegraphics{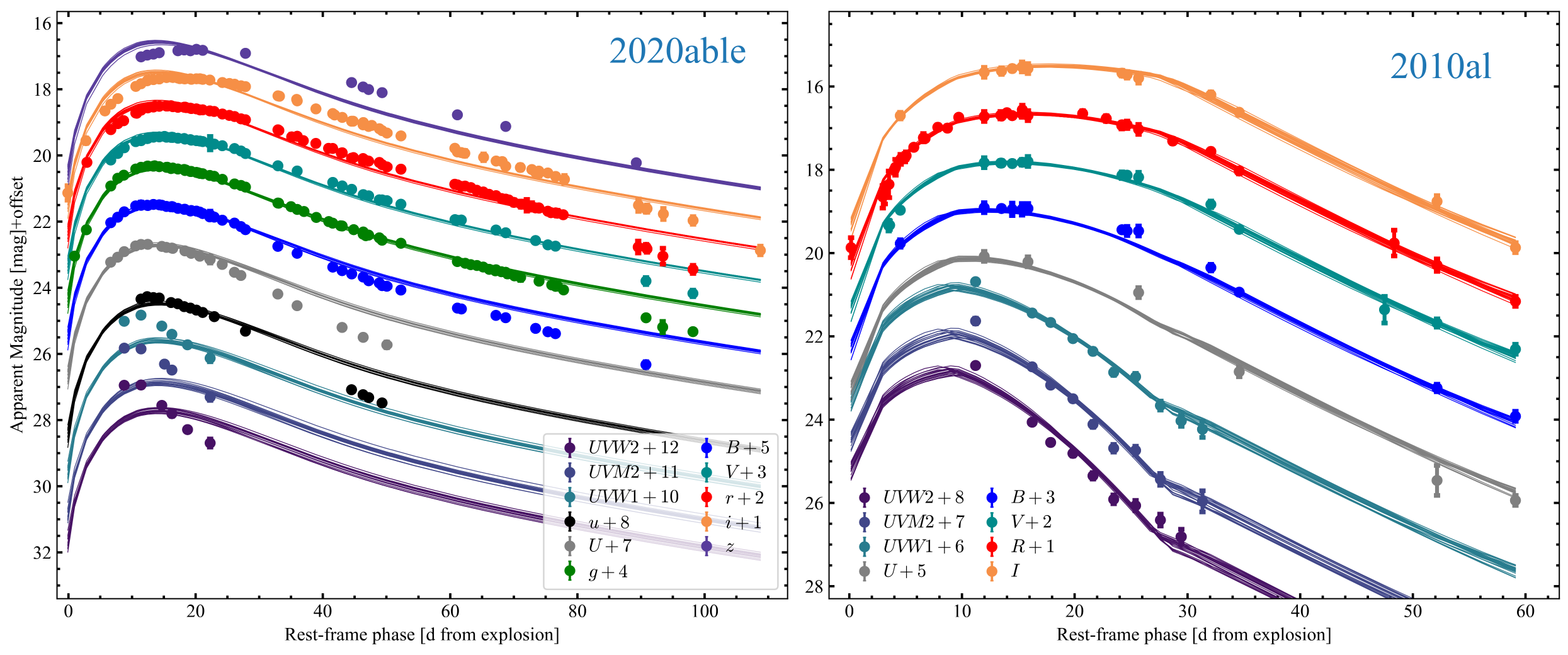}}
\caption{Multi-band light curves of SN~2020able (left) and SN~2010al (right) compared with an ensemble of {\sc MOSFiT} models generated using the {\sc default} module, which implements a standard nickel-cobalt ($^{56}$Ni $\rightarrow$ $^{56}$Co) radioactive decay power source without accounting for ejecta-CSM interaction. The model fails to reproduce the overall light curve evolution, particularly the early-time luminous peaks observed in the UV bands. Light curves have been vertically displaced by the indicated offsets for clarity. Rest-frame phases are relative to the estimated explosion epoch of each event.}
\label{fig:lcmodeldefault}
\end{figure*}

\begin{table*}[ht!]
\caption{Parameters of the {\sc default} analytical model obtained with {\sc MOSFiT} for SNe~2010al and 2020able}
\label{tab:mosfitPardefault}
\centering
\begin{tabular}{ccccc}
\toprule
Parameter & Prior range & 2010al & 2020able & Unit \\
\midrule
$M_{\rm{ej}}$      & $0.1-10.0$                                      & $0.474_{-0.026}^{+0.030}$    & $17.95_{-1.88}^{+1.15}$            & $\rm{M_\odot}$    \\
$M_{\rm{^{56}Ni}}$ & $10^{-3}\times M_{\rm{ej}}-1.0\times M_{\rm{ej}}$ & $0.461_{-0.030}^{+0.031}$    & $15.67_{-2.95}^{+1.84}$            & $\rm{M_\odot}$    \\
$v_{\rm{ej}}$      & $1.0\times10^3-1.0\times10^5$                   & $8280_{-370}^{+390}$         & $(9.42_{-0.71}^{+0.40})\times10^4$ & $\rm{km\,s^{-1}}$ \\
$A_V$              & $4.5\times10^{-6}-45$                           & $0.0002_{-0.0002}^{+0.0021}$ & $0.79_{-0.06}^{+0.04}$             & $\rm{mag}$        \\
$T_{\rm{min}}$     & $1.0-10^4$                                      & $2100_{-800}^{+1000}$        & $22800_{-1300}^{+1000}$            & $\rm{K}$          \\
$\rm{MJD_{expl}}$    & $-10-0^{\rm{a}}$                                & $55265.76_{-0.373}^{+0.353}$ & $59182.961_{-0.177}^{+0.163}$      & -                 \\
\bottomrule
\end{tabular}
\tablefoot{$^{a}$ The prior range refers to the phase with respect to the first detection. Corresponding light curves and corner plots showing the posterior probability distributions are available at \url{https://zenodo.org/records/21678623}}
\end{table*}

\begin{table*}
\caption{Parameters of the {\sc csmni} analytical model obtained with {\sc MOSFiT} for SNe~2010al and 2020able}
\label{tab:mosfitParcsmni}
\centering
\begin{tabular}{ccccc}
\toprule
Parameter & Prior range & 2010al & 2020able & Unit \\
\midrule
$E_k$              & $0.5-10.0$                                         & $0.38_{-0.49}^{+0.56}$                  & $0.29_{-0.013}^{+0.012}$                 & $10^{51}\,\rm{erg}$    \\
$M_{\rm{ej}}$      & $0.1-10.0$                                         & $1.15_{-0.17}^{+0.20}$                  & $1.315_{-0.068}^{+0.062}$                 & $\rm{M_{\sun}}$        \\
$n$                & $6-10$                                             & $7.92^{+0.48}_{-0.44}$                  & $6.595^{+0.080}_{-0.068}$                & -                      \\
$M_{\rm{^{56}Ni}}$ & $10^{-3}\times\,M_{\rm{ej}}-1.0\times M_{\rm{ej}}$ & $0.0065_{-0.0022}^{+0.0028}$            & $0.0315_{-0.0038}^{+0.0043}$             & $\rm{M_{\sun}}$        \\
$R_0$              & $2.0\times10^{-3}-25$                              & $0.110_{-0.098}^{+0.74}$                 & $0.25_{-0.19}^{+0.32}$                   & AU                     \\
$M_{\rm{CSM}}$     & $0.1-10.0$                                         & $0.490^{+0.023}_{-0.023}$               & $0.573^{+0.013}_{-0.014}$                & $\rm{M_{\sun}}$        \\
$\rho_{CSM}$       & $10^{-16}-10^{-6}$                                 & $(4.17^{+367.37}_{-4.10})\times10^{-8}$ & $(9.33^{+122.80}_{-7.50})\times10^{-9}$  & $\rm{g}\,\rm{cm^{-3}}$ \\
$A_V$              & $4.5\times10^{-6}-45$                              & $0.442_{-0.043}^{+0.036}$               & $0.147_{-0.013}^{+0.012}$                & mag                    \\
$T_{min}$          & $1.0-10^4$                                         & $7762_{-18}^{+18}$                      & $7499_{-52}^{+52}$                       & K                      \\
$\rm{MJD_{expl}}$   & $-10-0^{\rm{a}}$                                   & $55266.99^{+0.13}_{-0.15}$              & $59183.433^{+0.083}_{-0.086}$            & -                      \\
\bottomrule
\end{tabular}
\tablefoot{$^{a}$ The prior range refers to the phase with respect to the first detection.}
\end{table*}

\begin{figure*}
\centering
\resizebox{\hsize}{!}{\includegraphics{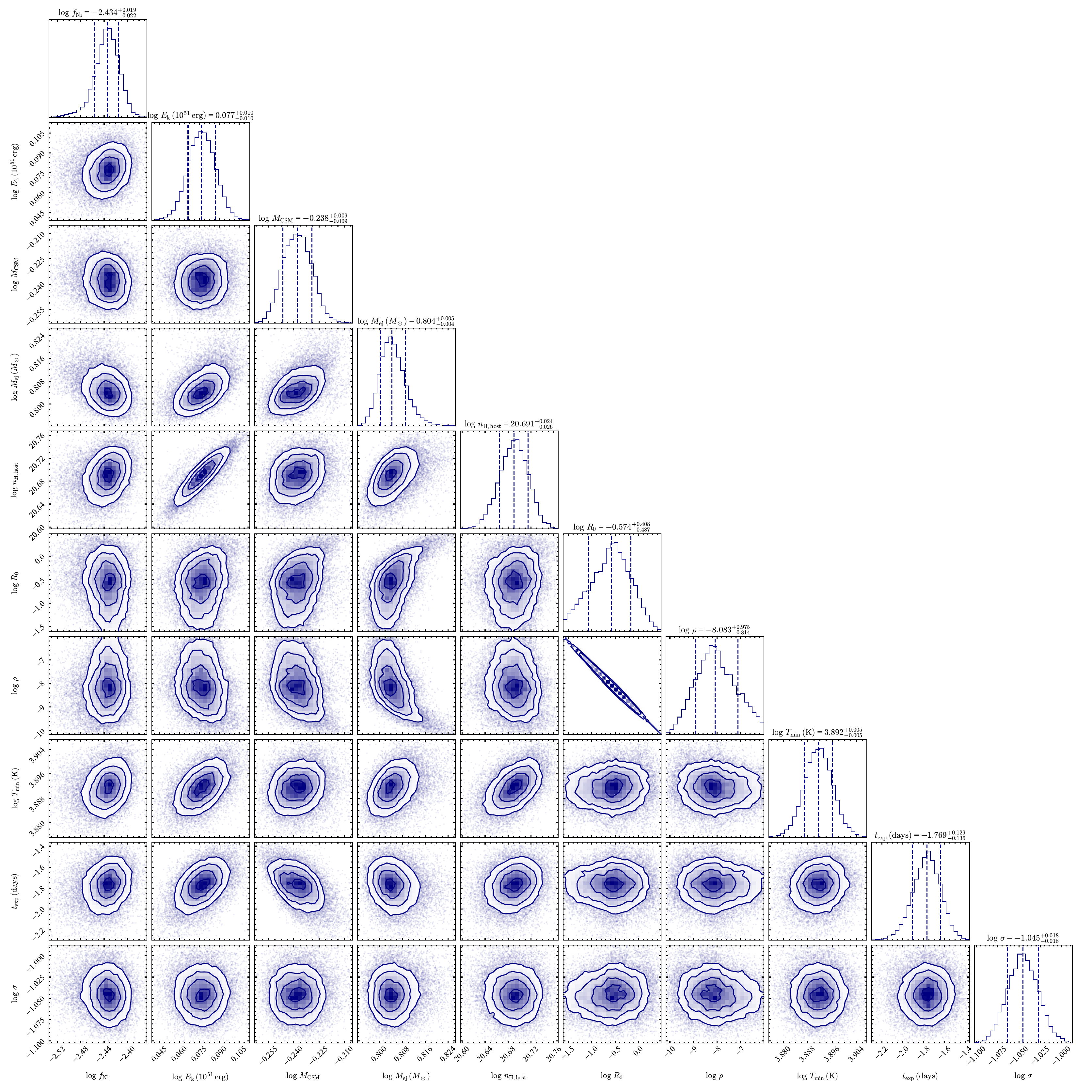}}
\caption{Posterior probability distributions for SN~2020able, generated by {\sc MOSFiT} using the {\sc csmni} model. The corner plot displays the multidimensional parameter space. Vertical dashed lines in the 1D histograms represent the median values and the $1\sigma$ confidence intervals.}
\label{fig:corner20able}
\end{figure*}

\begin{figure*}
\centering
\resizebox{\hsize}{!}{\includegraphics{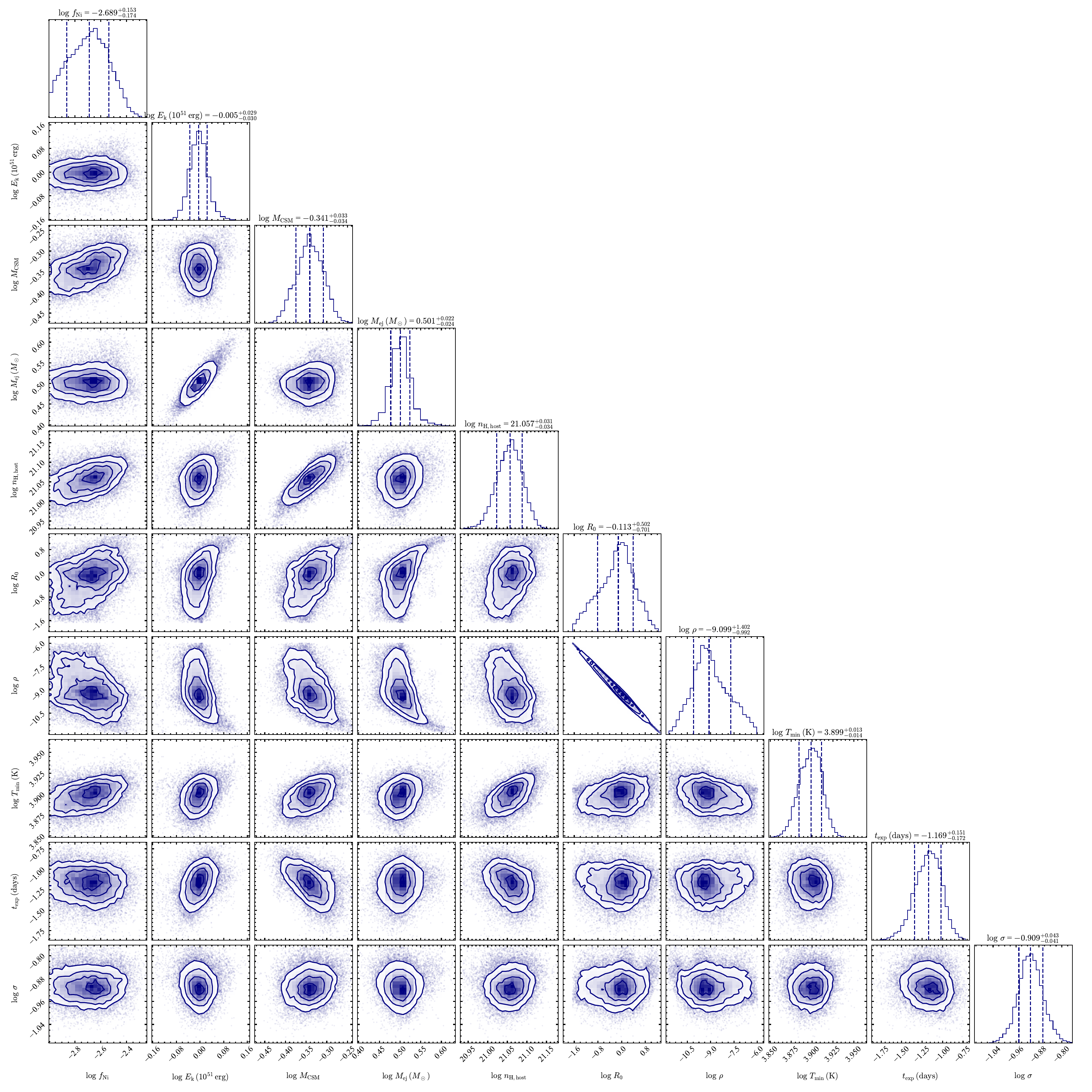}}
\caption{Same as Fig.~\ref{fig:corner20able} for SN~2010al.}
\label{fig:corner10al}
\end{figure*}

\begin{table*}
\caption{Parameters of the {\sc csm} analytical model obtained with {\sc MOSFiT} for SNe~2010al and 2020able}
\label{tab:mosfitParcsm}
\centering
\begin{tabular}{ccccc}
\toprule
Parameter & Prior range & 2010al & 2020able & Unit \\
\midrule
$M_{\rm{ej}}$      & $0.1-10.0$                                         & $1.32_{-0.18}^{+0.17}$               & $1.9055_{-0.0044}^{+0.0016}$              & $\rm{M_{\sun}}$        \\
$v_{ej}$           & $1.0\times10^3-1.0\times10^5$                      & $7520^{+140}_{-137}$                 & $5.5\times10^3$ (fixed)                  & $\rm{km}\,\rm{s^{-1}}$ \\
$n$                & $6-10$                                             & $9.29^{+0.42}_{-0.62}$               & $6.79^{+0.07}_{-0.07}$                   & -                      \\
$R_0$              & $2.0\times10^{-3}-25$                              & $6.9_{-3.6}^{+5.4}$                  & $35.5_{-3.1}^{+3.4}$                     & AU                     \\
$M_{\rm{CSM}}$     & $0.1-10.0$                                         & $2.64^{+0.19}_{-0.16}$               & $0.6918^{+0.0016}_{-0.0016}$             & $\rm{M_{\sun}}$        \\
$\rho_{CSM}$       & $10^{-16}-10^{-6}$                                 & $(7.6^{+25.0}_{-5.0})\times10^{-12}$ & $(5.62^{+0.83}_{-0.73})\times10^{-13}$   & $\rm{g}\,\rm{cm^{-3}}$ \\
$A_V$              & $4.5\times10^{-6}-45^{\rm{a}}$                     & $0.412^{+0.034}_{-0.034}$            & $0.14$ (fixed)                           & mag                    \\
$T_{min}$          & $1.0-10^4$                                         & $7710_{-140}^{+143}$                 & $7460_{-250}^{+190}$                     & K                      \\
$\rm{MJD_{expl}}$   & $-10-0^{a}$                                        & $55266.94^{+0.13}_{-0.14}$           & $59182.6^{+0.13}_{-0.13}$                & -                      \\
\bottomrule
\end{tabular}
\tablefoot{$^{a}$ The prior range refers to the phase with respect to the first detection. The corresponding light curves and corner plots showing the posterior probability distributions are available at \url{https://zenodo.org/records/21678623}}
\end{table*}

\end{appendix}
\end{document}